\documentclass[prx,aps,superscriptaddress,twocolumn,preprintnumbers,footinbib,longbibliography]{revtex4-1}

\usepackage{amsmath,amssymb,amsfonts,mathrsfs,bm,dsfont}
\usepackage[all]{xy} % commutative diagrams
\usepackage{array}
\usepackage{graphicx}
\usepackage[space]{grffile} % to avoid compilation problem with the figures name with spaces, see https://tex.stackexchange.com/questions/8422/how-to-include-graphics-with-spaces-in-their-path

\usepackage{url}
\usepackage{float}
\usepackage{bibentry}
\usepackage{tcolorbox}
\usepackage[scr=boondoxo]{mathalfa}
\usepackage{xcolor}
\usepackage{braket}
\usepackage{enumitem} % for customized enumerate env
\usepackage{cleveref}
\usepackage{ragged2e}
\usepackage{diagbox} % diagonal box in table construction
\definecolor{mypink}{rgb}{0.858, 0.188, 0.478}
\definecolor{myred}{cmyk}{0, 0.7808, 0.4429, 0.1412}

\bibpunct{[}{]}{,}{n}{}{}
\allowdisplaybreaks[4] % all `align' env to go across pages

\makeatletter
\DeclareFontFamily{OMX}{MnSymbolE}{}
\DeclareSymbolFont{MnLargeSymbols}{OMX}{MnSymbolE}{m}{n}
\SetSymbolFont{MnLargeSymbols}{bold}{OMX}{MnSymbolE}{b}{n}
\DeclareFontShape{OMX}{MnSymbolE}{m}{n}{
    <-6>  MnSymbolE5
   <6-7>  MnSymbolE6
   <7-8>  MnSymbolE7
   <8-9>  MnSymbolE8
   <9-10> MnSymbolE9
  <10-12> MnSymbolE10
  <12->   MnSymbolE12
}{}
\DeclareFontShape{OMX}{MnSymbolE}{b}{n}{
    <-6>  MnSymbolE-Bold5
   <6-7>  MnSymbolE-Bold6
   <7-8>  MnSymbolE-Bold7
   <8-9>  MnSymbolE-Bold8
   <9-10> MnSymbolE-Bold9
  <10-12> MnSymbolE-Bold10
  <12->   MnSymbolE-Bold12
}{}

\let\llangle\@undefined
\let\rrangle\@undefined
\DeclareMathDelimiter{\llangle}{\mathopen}%
                     {MnLargeSymbols}{'164}{MnLargeSymbols}{'164}
\DeclareMathDelimiter{\rrangle}{\mathclose}%
                     {MnLargeSymbols}{'171}{MnLargeSymbols}{'171}
\makeatother

\begin{document}

\title{Hyperdeterminants and Composite fermion States in Fractional Chern Insulators}
\author{Xiaodong Hu}
\affiliation{Department of Physics, Boston College, Chestnut Hill, MA 02467}
\author{Di Xiao}
\affiliation{Department of Materials Science and Engineering, University of Washington, Seattle, WA 98195}
\affiliation{Department of Physics, University of Washington, Seattle, WA 98195}
\author{Ying Ran}
\affiliation{Department of Physics, Boston College, Chestnut Hill, MA 02467}

\begin{abstract}
Fractional Chern insulators (FCI) were proposed theoretically about a decade ago. These exotic states of matter are fractional quantum Hall states realized when a nearly flat Chern band is partially filled, even in the absence of an external magnetic field. Recently, exciting experimental signatures of such states have been reported in twisted MoTe$_2$ bilayer systems. Motivated by these experimental and theoretical progresses, in this paper, we develop a projective construction for the composite fermion states (either the  Jain's sequence or the composite Fermi liquid) in a partially filled Chern band with Chern number $C=\pm1$, which is capable of capturing the microscopics, e.g., symmetry fractionalization patterns and magnetoroton excitations. On the mean-field level, the ground states' and excitated states' composite fermion wavefunctions are found self-consistently in an enlarged Hilbert space. Beyond the mean-field, these wavefunctions can be projected back to the physical Hilbert space to construct the electronic wavefunctions, allowing direct comparison with FCI states from exact diagonalization on finite lattices. We find that the projected electronic wavefunction corresponds to the \emph{combinatorial hyperdeterminant} of a tensor. When applied to the traditional Galilean invariant Landau level context, the present construction exactly reproduces Jain's composite fermion wavefunctions. We apply this projective construction to the twisted bilayer MoTe$_2$ system. Experimentally relevant properties are computed, such as the magnetoroton band structures and quantum numbers.
\end{abstract}
\maketitle

\tableofcontents

\section{Introduction}
Fractional Chern insulators (FCI) were theoretically proposed about a decade ago \cite{neupert2011fractional,tang2011high,sheng2011fractional,sun2011nearly,regnault2011fractional,xiao2011interface} as fractional quantum Hall states in the absence of the external magnetic field. Different from the traditional fractional quantum Hall (FQH) states realized in Landau Levels (LL), in FCI the electrons partially fill a nearly flat Chern band, and the Berry's curvature from the Chern band plays the role of the magnetic field. When Coulomb interactions are strong enough compared with the bandwidth of the Chern band, fractional quantum Hall states may be realized, which may host Abelian or non-Abelian anyon excitations. Although the theoretical possibility of such fascinating correlated states of matter in realistic materials has been known for quite some time, and intensive experimental efforts have been made in various candidate materials \cite{xie2021fractional}, only recently the experimental signatures of FCI have been reported in twisted MoTe$_2$ bilayer systems \cite{cai2023signatures,zeng2023thermodynamic,park2023observation,xu2023observation} and rhombohedral pentalayer graphene/hBN moiré superlattice \cite{lu2023fractional}.

In traditional FQH states, the energy scale of the excitations is determined by the Coulomb energy scale $\frac{e^2}{\epsilon l_B}$, where $l_B$ is the magnetic length and $\epsilon$ is the dielectric constant. In FCI, however, $l_B$ should be essentially replaced by the lattice constant $a$ of the crystalline order. This suggests that FCI states, as a matter of principle, may host dynamics with much larger energy scales, and could be ideal experimental platforms to investigate quantum phenomena like anyon statistics. The ongoing theoretical development mainly focus on clarifying the criterion to realize FCI phases, from ideal flatband condition to vortexability \cite{roy2014band,wang2021exact,wang2023origin,ledwith2023vortexability}, and on constructing analytic ground state wavefunctions in certain limits \cite{ledwith2020fractional,dong2023many,wang2023origin}. However, the microscopic theoretical tools suitable to study FCI states are limited: the main theoretical tools currently available include exact diagonalization (ED) and density matrix renormalization group (DMRG) numerics \cite{haldane1985finite,feiguin2008density,dong2023anomalous,dong2023composite,yu2023fractional,goldman2023zero,dong2023theory}. Several outstanding issues are directly related to the ongoing experimental efforts, yet they are challenging to answer using the available theoretical tools. Below, we remark on some of them.

Some of these issues concern the ground state properties of FCI systems. One crucial question is whether the experimental FCI states realize entirely new states of matter, that are not adiabatically connected to the traditional LL FQH states. Theoretically, from the classification viewpoint, such new states of matter could exist from two perspectives. First, the topological order, namely the anyon contents of the FCI states may not be realized in traditional FQH states. Second, even if the topological orders of the FCI states is identical to the traditional FQH states, the presence of the crystalline symmetry may enrich the topological orders, giving rise to different symmetry-enriched topological (SET) states \cite{mesaros2013classification,lu2016classification}. One such SET phenomenon that has been discussed in the literature is the analog of the Wen-Zee shift \cite{wen1992shift} for the discrete crystalline rotation symmetry group \cite{manjunath2023rotational,han2019generalized,li2020fractional,zhang2022fractional}, which is related to the spin angular momentum carried per quasiparticle. Nontrivial Wen-Zee shift would lead to, for instance, fractionally quantized charges at lattice disclinations in the bulk \cite{li2020fractional,zhang2022fractional}.

In the traditional FQH context, the composite fermion states \cite{jain1989composite} are associated with a simple mean-field picture. After the flux attachment \cite{jain1990theory}, the electron in a physical LL becomes a composite fermion (CF) that sees an effective magnetic field -- a fraction of the physical magnetic field. Consequently, the CF fills an integer number of effective composite fermion LLs. The Jain's sequence at filling $\nu=\frac{p}{2ps+1}$ corresponds to attaching $2s$-unit of flux to the electron, and the CF fills $p$ CF LLs. Note that the CF wavefunction is a free fermion state in this mean-field picture -- a Slater determinant. The physical electronic wavefunction, e.g., the Laughlin's wavefunctions, obviously, is not a Slater determinant. 

In the FCI context, this mean-field picture is expected to be modified naturally: the electron LL is replaced by a Chern band, while the CF Chern bands also replace the CF LLs. Physically, the different CF LLs are characterized by the spin angular momentum carried per CF: for the $n$-th ($n=0,1,2...$) CF LL, the CF carries spin angular momentum $l=n$. Although continuous rotation symmetry is absent in the FCI context, the angular momentum mod $m$ is still sharply defined for  $C_m$ crystalline rotation symmetry. 

% [Illustration figure of CF LL's.]
\begin{figure}[!htp]
    \centering
    \includegraphics[width=0.45\textwidth]{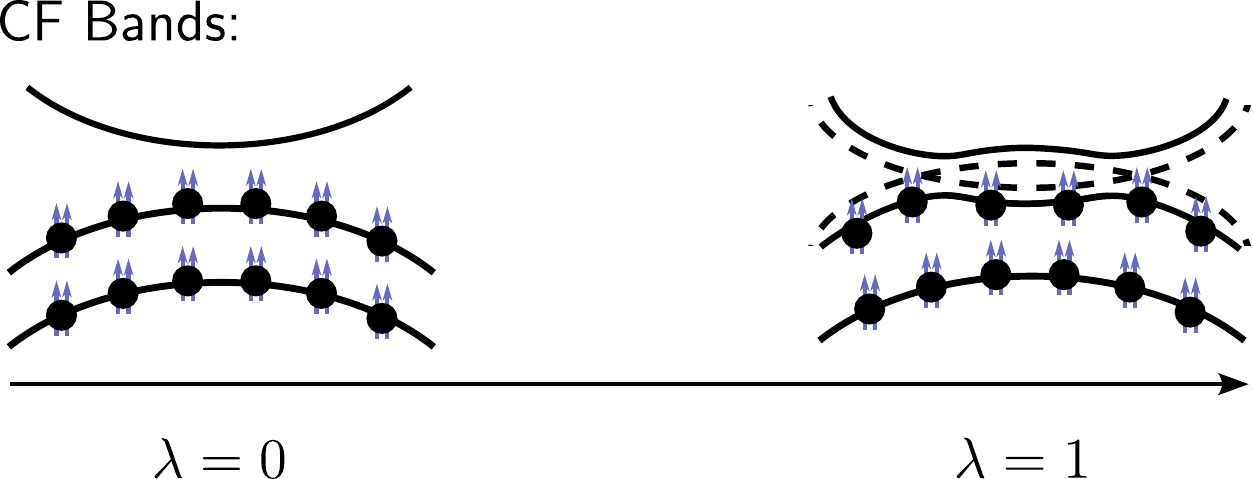}
    \caption{Illustration for the possible change of Wen-Zee shift due to the CF band inversion when tunning the parameter $\lambda$, so that the CF states at $\lambda=0$ and $\lambda=1$ belongs to different SET phases.}
    \label{fig:CF band inversion}
\end{figure}

As a thought experiment, one may imagine smoothly deforming the electronic Hamiltonian $H(\lambda)$ with a parameter $\lambda$ while preserving the physical symmetries so that a LL CF state at $\lambda=0$ is connected with an FCI CF state at $\lambda=1$. The question is whether or not the two states are in the same quantum phase.  The physics of topological insulators teaches us that band inversion may give rise to new states of matter. Indeed, when the CF Chern bands have a full band inversion from $\lambda=0$ to $\lambda=1$, the system would have a corresponding change of the Wen-Zee shift, in which case the two states are in different SET phases. See FIG.\ref{fig:CF band inversion} for illustration.

Some other theoretical issues are about the dynamical properties of the FCI states. For instance, the magnetorotons are the charge-neutral bulk excitations and have been experimentally probed using Raman scattering in the traditional FQH systems \cite{pinczuk1998light,kang2000inelastic,kukushkin2009dispersion}. In the presence of the Galilean invariance, the magnetoroton at wavevector $q=0$ has been recently interpreted by Haldane as the collective mode of the geometry fluctuations \cite{haldane2011geometrical,haldane2009hall}, analogous to the graviton, carrying angular momentum $l=\pm 2$ \cite{yang2012model}.  In the FCI systems, there is no reason these magnetorotons necessarily carry angular momentum  $l=\pm 2$. What are the crystalline quantum numbers carried by the magnetorotons in FCI systems? How to theoretically compute the magnetoroton spectra in FCI systems? These questions are also relevant to the quantum phase transitions involving FCI states. For instance, when magnetorotons become gapless at certain momenta, the system is expected to break translational symmetry and develop charge density wave order.

Due to the limitation of the small system sizes for ED and the difficulty of implementing DMRG on sizable torus samples, answering questions about the crystalline quantum numbers has been challenging for FCI systems. Developing new theoretical tools to investigate these important questions would be desirable. 

On the other hand, a different class of quantum systems hosting topologically ordered phases is the quantum spin liquids. There, a nice theoretical tool is available: projective constructions such as the Schwinger-boson and Abrikosov-fermion methods \cite{arovas1988functional,read1991large,sachdev1991large,sachdev1992kagome,wang2010schwinger,wen1991mean,wen2002quantum,gros1989physics,tay2011variational}. These projective constructions are very helpful: they provide mean-field theories for the topologically ordered states by enlarging the physical Hilbert space. The mean-field wavefunctions can be improved by projection back to the physical space, leading to physical wavefunctions that can be directly compared with other numerical methods, e.g., ED and DMRG. The detailed microscopic information, such as the crystalline symmetry quantum numbers carried by the ground states and excited states, is accessible in these methods. However, in FCI systems, similar projective construction has been lacking. 

Motivated by these issues, we establish a projective construction for the composite fermion states in fractional Chern insulators in this paper. Our main results can be summarized as a general procedure. The procedure input is the Hamiltonian $\mathbf H_{\text{CB}}$ describing a partially filled Chern band with Chern number $C=\pm 1$, which we want to investigate. The procedure output is two-fold. First, on the mean-field level, the procedure outputs a Hartree-Fock (HF) mean-field Hamiltonian for the CF states in an enlarged Hilbert space, whose ground state is the CF wavefunction $|\psi^{MF}_{CF}\rangle$ and is a Slater determinant. The excitations of the system (e.g., the magnetoroton collective modes) can be calculated within the time-dependent Hartree-Fock (TDHF) framework. Second, beyond the mean-field level, the CF wavefunction can be projected into the physical electronic wavefunction $|\psi_{e}(\psi^{MF}_{CF})\rangle=\mathbf P |\psi^{MF}_{CF}\rangle$ ($\mathbf P$ is a projector), which turns out to be a so-called hyperdeterminant of a tensor and can be compared with wavefunctions obtained from ED. 

The paper is organized as follows. Because we will discuss both $|\psi^{MF}_{CF}\rangle$ and $|\psi_{e}(\psi^{MF}_{CF})\rangle$, to avoid confusion, below we will denote the former wavefunction as the mean-field (MF) CF state, while the latter wavefunction as the electronic (or projected) CF state.  To present a self-contained discussion, in Sec.\ref{sec:review} we briefly review several related pieces of previous works, including Jain's CF construction \cite{jain1990theory}, Murthy-Shankar's Hamiltonian formalism \cite{murthy2003hamiltonian}, the construction of $\nu=1$ bosonic composite Fermi liquid developed by Pasquier-Haldane \cite{pasquier1998dipole} and Read \cite{read1998lowest}. In Sec.\ref{sec:M_S_finite} we discuss the general projective construction on finite-size crystalline systems for composite fermion states (either in the Jain's sequence or the composite Fermi liquid), which is based on Murthy-Shankar's construction.  This projective construction leads to the MF CF ground states and excited states on the mean-field level, as well as a projection operation $\mathbf P$ to go beyond the mean-field. In Sec.\ref{sec:proj}, we study the mathematical details of the projection $\mathbf P$ operation and show that the general projected CF states are hyperdeterminants of tensors. We then connect our results with previous works including Jain's construction in the traditional FQH context and the parton construction in the FCI context. Interestingly, despite the current construction reproduces Jain's wavefunctions in the traditional Galilean invariant Landau level context, in the absence of the Galilean invariance (e.g. in a FCI system), the present construction and the naive generalization of Jain's prescription are \emph{different} in general. In Sec.\ref{sec:benchmark}, we apply this general procedure to two microscopic FCI models: a toy model of mixed Landau levels introduced by Murthy and Shankar \cite{murthy2012hamiltonian}, and the realistic model for the twisted bilayer MoTe$_2$ \cite{wang2023fractional}. Experimentally relevant properties of the FCI states are computed, such as the magnetoroton quantum numbers and spectra. Finally, we discuss possible future developments of our construction and conclude in Sec.\ref{sec:conclusion}

\section{A brief review of related previous works}\label{sec:review}
\subsection{Jain's composite-fermion construction}
Jain's wavefunctions for composite fermion states \cite{jain1989composite,jain1989incompressible,jain1990theory,jain2007composite,balram2013state} are based on the seminal idea of the flux attachment. To describe the fractional quantum Hall states at filling $\nu=\frac{p}{2ps+1}$ where $p,s\in\mathbb Z$ are integers, Jain proposed the following wavefunctions in the symmetric gauge of the lowest Landau level with the open boundary condition \cite{kamilla1996composite}:
\begin{align}\label{eq1. Jain's wave function}
\psi_{\frac{p}{2ps+1}}=\mathcal P_{\text{LLL}} \prod_{i<j} (z_i-z_j)^{2s}\cdot \chi_p(z,\bar z)
\end{align}
Here $\chi_p(z,\bar z)$ is the Jain's composite fermion wavefunction with $p$-filled Landau levels. The flux attachment in this scheme is achieved by the Jastrow factor $(z_i-z_j)^{2s}$: when one electron moves around another electron by a circle, this factor gives a $4\pi s$ phase shift, similar to when an electron moves around $2s$-flux tube. The projection operation $\mathcal P_{\text{LLL}}$ ensures the final wavefunction is within the lowest Landau level (LLL). Precisely, Jain proposes the prescription to replace $\bar z$ by the derivative $\mathcal P_{\text{LLL}}: \bar z\rightarrow 2 l_e^2 \frac{\partial}{\partial z}$, where $l_e$ is the electron's magnetic length. By moving all the derivatives to the left, the obtained wavefunction is holomorphic as required by the LLL. Jain's wavefunctions, after adapted to appropriated boundary conditions, have been demonstrated to have excellent overlap with those obtained from the exact diagonalization. 

Jain's composite fermion wavefunction $\chi_p(z,\bar z)$ is a single Slater determinant. In the simplest $p=1$ case, it is the Vandermonde determinant together with the Gaussian factor:
\begin{align}
\chi_{p=1}&=\det\left|\begin{array}{cccc} 
        z_1^0 & z_2^0 & z_3^0 & \cdots\\
        z_1^1 & z_2^1 & z_3^1 & \cdots\\
        z_1^2 & z_2^2 & z_3^2 & \cdots\\
        \vdots & \vdots &\vdots &\ddots
    \end{array}\right|\exp\left[-\sum_{i}\frac{|z_i|^2}{4l_e^2}\right]\notag\\
&=\prod_{i<j}(z_i-z_j)~\exp\left[-\sum_{i}\frac{|z_i|^2}{4l_e^2}\right].\label{eq.2 Jain's CF wave function}
\end{align}
In this particular case, the projection $\mathcal P_{\text{LLL}}$ is unnecessary since $\bar z$ is not present, and Jain's wavefunctions becomes Laughlin's wavefunctions \cite{laughlin1983anomalous}.

Despite the success of Jain's wavefunctions in the FQHE, how to generalize them to the context of FCI remains unclear. In fact, we want to mention two conceptual puzzles in Jain's original construction, which motivated us to develop the new construction. First, the physical meaning of the composite fermion Landau levels needs to be clarified. For instance, how many composite fermion Landau levels are there? In a finite-size system, the dimension of the physical electronic Hilbert space is finite. It would be unphysical to have a construction involving an infinite number of composite fermion Landau levels. So, if this number is finite for a finite-size sample, what is it? 

Second, let us pay attention to the Gaussian factor $\exp\left[-\sum_{i}\frac{|z_i|^2}{4l_e^2}\right]$ in the composite fermion state Eq.\eqref{eq.2 Jain's CF wave function}. The puzzle is the appearance of the \emph{electronic} magnetic length $l_e$. On the one hand, this is required by Jain's prescription to obtain a wavefunction $\psi_{\frac{p}{2ps+1}}$ within the LLL of the electrons. On the other hand, physically, if the composite fermion sees a weaker magnetic field with an effective magnetic length $l_{CF}>l_e$, wouldn't $l_{CF}$ be appearing in the Gaussian?

We will come back to these two puzzles in Sec.\ref{sec:connection_jain}, where we demonstrate that the new construction solves both puzzles naturally.

\subsection{Murthy-Shankar's Hamiltonian theory}
Focusing on the composite fermion states, Murthy and Shankar developed a Hamiltonian theory for FQHE \cite{murthy2003hamiltonian}. Let us first set up some basic notations. The electron's full position operator $\bm r_e$ can be separated into the mutually commuting guiding-center $\mathcal R_e$ and cyclotron $\eta_e$ degrees of freedom:
\begin{align}
\bm r_e=\mathcal R_e+\eta_e,
\end{align}
satisfying the algebra:
\begin{align}\label{eq:e_guiding_center}
[\mathcal R_{e,x},\mathcal R_{e,y}]&=-il_e^2,&[\eta_{e,x},\eta_{e,y}]&=il_e^2
\end{align}

For the dynamics within the LLL, the $\eta_e$ degrees of freedom are frozen and one needs to focus on the guiding-center part of the density operator ($i$ labels the particle) 
\begin{align}
\pmb{\boldsymbol\rho}_e(\mathbf q_e)=\sum_{i}e^{i\mathbf q_e\cdot \mathcal R_{e(i)}},\label{eq:e_density}
\end{align}
which satisfies the Girvin-MacDonald-Platzman (GMP) algebra \cite{girvin1986magneto}
\begin{align}
\pmb{\boldsymbol\rho}_e(\mathbf q_e)\pmb{\boldsymbol\rho}_e(\mathbf q_e')&=e^{i\frac{\mathbf q_e\times \mathbf q_e'}{2}l_e^2}\pmb{\boldsymbol\rho}_e(\mathbf q_e+\mathbf q_e')\notag\\
[\pmb{\boldsymbol\rho}_e(\mathbf q_e),\pmb{\boldsymbol\rho}_e(\mathbf q_e')]&=2i\sin\left[\frac{\mathbf q_e\times \mathbf q_e'}{2}l_e^2\right]\pmb{\boldsymbol\rho}_e(\mathbf q_e+\mathbf q_e')\label{eq:GMP}
\end{align}
The electron Hamiltonian constrained within the LLL can be represented using this density operator. For instance, for the Coulomb interaction,
\begin{align}
\mathbf H_e=\frac{1}{2 A}\sum_{\mathbf q_e} e^{-\frac{\mathbf q_e^2l_e^2}{2}}v(\mathbf q_e):\pmb{\boldsymbol\rho}_e(\mathbf q_e)\pmb{\boldsymbol\rho}_e(-\mathbf q_e):,
\end{align}
where $v(\mathbf q_e)=\frac{2\pi e^2}{\epsilon |\mathbf q_e|}$ and $A$ is the real-space sample size.

To achieve the flux attachment, Murthy and Shankar introduced auxiliary degrees of freedom, the vortex guiding-center $\mathcal R_v$, to enlarge the Hilbert space:
\begin{align}
[\mathcal R_{v,x},\mathcal R_{v,y}]&=il_v^2.\label{eq:v_guiding_center}
\end{align}
Here the vortex magnetic length $l_v=\frac{l_e}{c}$ with $c=\sqrt{\frac{2ps}{2ps+1}}$. Physically, $\mathcal R_v$ describes the vortex that carries an electric charge $q_v$ that has an opposite sign of the electron's electric charge $q_e$:  $q_v=-\frac{2ps}{2ps+1}q_e$. With these auxiliary degrees of freedom, the full composite fermion degrees of freedom can be constructed, including the mutually commuting guiding-center and the cyclotron components:
\begin{align}
\mathcal R_{CF}&=\frac{\mathcal R_e-c^2\mathcal R_v}{1-c^2},&\eta_{CF}&=\frac{c}{1-c^2}(\mathcal R_v-\mathcal R_e), \label{eq:CF_substitution}
\end{align}
satisfying the algebra:
\begin{align}
[\mathcal R_{CF,x},\mathcal R_{CF,y}]&=-il_{CF}^2,&[\eta_{CF,x},\eta_{CF,y}]&=il_{CF}^2.\label{eq:CF_algebra}
\end{align}
Here the CF magnetic length $l_{CF}=\frac{l_e}{\sqrt{1-c^2}}=\sqrt{2ps+1}l_e$, which can be interpreted as the CF electric charge $q_{CF}=\frac{q_e}{2ps+1}$. We also list the inverse transformation of Eq.(\ref{eq:CF_substitution}):
\begin{align}
\mathcal R_e&=\mathcal R_{CF}+c\cdot\eta_{CF},&\mathcal R_v&=\mathcal R_{CF}+\frac{1}{c}\eta_{CF}\label{eq:CF_substitution_inverse}
\end{align}

If we denote the electron's and the vortex's single-particle Hilbert spaces as $\mathcal H_e$ and $\mathcal H_v$,  the composite fermion lives in an enlarged Hilbert space $\mathcal H_{CF}=\mathcal H_e\otimes \mathcal H_v$, that can be decomposed into CF's guiding-center and the cyclotron components: $\mathcal H_{CF}=\mathcal H_{\mathcal R_{CF}}\otimes \mathcal H_{\eta_{CF}}$. 

Any physical operator $\hat O_e$ acting in the electronic Fock space, including the Hamiltonian $\mathbf H_e$, then can be mapped to the composite fermion Fock space. As a fundamental example, the electron's density operator can be identified with:
\begin{align}
\pmb{\boldsymbol\rho}_e(\mathbf q_e)=\sum_{i}e^{i\mathbf q_e\cdot \mathcal R_{e(i)}} \rightarrow \sum_{i}e^{i\mathbf q_e\cdot\left[\mathcal R_{CF(i)}+c\cdot\eta_{CF(i)}\right]}  \label{eq:rho_CF_substitution}
\end{align}
The composite fermion states with $p$-filled CF LLs can be viewed as the Hartree-Fock mean-field ground states of $\mathbf H_e$. In addition, the bulk excitations such as the magnetorotons spectra can be computed within the time-dependent Hartree-Fock framework \cite{murthy1999hamiltonian, murthy2001hamiltonian}. These mean-field results are qualitatively consistent with other calculation methods.

More recently, Murthy and Shankar generalized this Hamiltonian approach to the context of FCI \cite{murthy2012hamiltonian}. This generalization is based on two important observations. First, the bloch states in a Chern band with Chern number $C=\pm 1$ can be mapped to the wavefunctions in the LLL on a torus \cite{haldane1985periodic}. Accordingly, an FCI Hamiltonian with $C=\pm 1$ can be exactly mapped to an electronic Hamiltonian in the LLL, with the presence of a crystalline potential. Second, the density operators as $\rho_e(\mathbf q_e)$ in Eq.(\ref{eq:e_density}) on a finite-size torus actually form a complete basis for any fermion bilinears (i.e., single-body operators). Therefore, any density-density interactions can also be straightforwardly mapped into the LLL problem based on Eq.(\ref{eq:rho_CF_substitution}).

In Murthy-Shankar's Hamiltonian construction, the physical origin of the CF LL is clear as it is a consequence of the enlarged Hilbert space. The relation with Jain's wavefunctions, however, remains a puzzle. It is also unclear how to improve beyond the mean-field treatment, a challenge related to the gauge structure of the construction that was first studied by Read \cite{read1998lowest} in the context of the bosonic $\nu=1$ composite Fermi liquid, as we will discuss shortly. 

\subsection{Pasquier-Haldane-Read construction for the bosonic $\nu=1$ composite fermion liquid}
The Pasquier-Haldane's work \cite{pasquier1998dipole} considered bosonic charged particles at $\nu=1$. Here, one may argue that after attaching one unit flux, the boson becomes a composite fermion that sees no effective magnetic field, which forms a composite Fermi sea. The boson's Fock space is enlarged by introducing fermions with two indices $c_{mn}$ satisfying the usual algebra:
\begin{align}
\{c_{mn},c^\dagger_{m'n'}\}=\delta_{mm'}\delta_{nn'}.
\end{align}
Here $m,n\in 1,2,...N$, and $N$ is the number of bosonic particles and the number of orbitals in the LLL. The basis states of the physical Fock space of bosons are then constructed as
\begin{align}
|m_1,m_2,...,m_N\rangle=\epsilon^{n_1,n_2,...,n_N} c_{n_1m_1}^\dagger c_{n_2m_2}^\dagger ... c_{n_N m_N}^\dagger|0\rangle,\label{eq:Read_projection}
\end{align}
where $|0\rangle$ is the vacuum of the $c_{mn}$ fermion's Fock space, and $\epsilon$ is the fully antisymmetric Levi-Civita symbol, and we have used the Einstein notation. Read \cite{read1998lowest} studied the mean-field theory and gauge fluctuations of this theory. As any theory involving an enlarged Hilbert space, the physical state is obtained only when the gauge redundancy is removed. In the present case, the constraint that the physical states need to satisfy is exactly the invariance under the $SU(N)_R$ transformation generated by (apart from the trace):
\begin{align}
\pmb{\boldsymbol\rho}^R_{nn'}=c^\dagger_{nm} c_{mn'}
\end{align}
On the other hand, the physical density operators are:
\begin{align}
\pmb{\boldsymbol\rho}^L_{mm'}=c^\dagger_{nm'} c_{mn}
\end{align}
Note that $\pmb{\boldsymbol\rho}_R$ and $\pmb{\boldsymbol\rho}_L$ commute. The constraint can then be implemented as the identity on the operator level: $\pmb{\boldsymbol\rho}^R_{nn'}=\delta_{nn'}$, which is treated using the Hartree-Fock and time-dependent Hartree-Fock approximation (also called the conserving approximation) in Ref.\cite{read1998lowest}.

The Pasquier-Haldane-Read construction, although only applicable to the bosonic $\nu=1$ CFL, is closely related to the Murthy-Shankar Hamiltonian theory, which we will explain below.

\section{The projective construction on a finite-size system}\label{sec:M_S_finite}
In this section, step by step, we present a general projective construction of CF states applicable for both traditional FQHE and FCI systems. Several steps of this construction are based on Murthy-Shankar's Hamiltonian theory but on finite-size systems. 

\emph{In this paper, to avoid confusion, we will always use the regular font for operators in the single-particle Hilbert spaces and the bold font for corresponding operators in many-particle Fock spaces.}

\subsection{Mapping a Chern band to the lowest Landau level}
Firstly, let us introduce the single-particle bloch basis in the LLL. The mutually commuting guiding-center and cyclotron degrees of freedom in the LLL are ($e>0$):
\begin{align}
\mathcal R_e&=\bm r_e-\eta_e, & \eta_e=\frac{l_e^2}{\hbar}\hat z\times (\mathbf p_e+e\mathbf A_e),
\end{align}
where the magnetic length $l_e\equiv \sqrt{\frac{\hbar}{eB_e}}$. The usual kinetic Hamiltonian only depends on the $\eta_e$:
\begin{align}
H_K= \frac{(\mathbf p_e+e\mathbf A_e)^2}{2m_e}=\frac{\hbar^2}{2m_e l_e^4}\eta_e^2.
\end{align}
Without loss of generality, we choose the Landau gauge $\mathbf A_e=(B_e y,0)$ in this section. The subscript $e$ highlights the objects for physical electrons, because, in the next step, we will introduce similar objects for vortices and composite fermions.

\emph{Throughout this paper, unless explicitly stated otherwise, we focus on the case with $B_e>0$, whose LLL has a Chern number $C=-1$.} For Chern band systems whose Chern number $C=+1$, one needs to perform a time-reversal transformation to map to the LLL discussed here.

\emph{To save notation, we will interchangeably use the complex number $z\equiv x+iy$ to represent a vector $\bm r\equiv(x,y)$}. The single-particle magnetic translation operator is
\begin{align}
D_e(z_0)\equiv U_{T,e}(z_0)T_e(z_0),
\end{align}
where $T_e(z_0)$ is the usual translation operator: $T_e(z_0)\psi_e(z)=\psi_e(z-z_0)$, and $U_{T,e}(z_0)$ is the associated gauge transformation that is fixed up to a $U(1)$ phase factor. One choice to fix this $U(1)$ phase ambiguity is to define $D_e(z_0)$ as the density operator:
\begin{align}
D_e(z_0)\equiv\rho_e(\mathbf q_{e, z_0})=e^{i \mathbf q_{e, z_0} \cdot \mathcal R_e}, \quad\text{where }\mathbf q_{e, z_0}\equiv i\frac{z_0}{l_e^2}.\label{eq:real_momentum}
\end{align}

We will fix this definition in the discussion below. One may straightforwardly check that the explicit form of $U_{T,e}(z_0)$ is now
\begin{align}
U_{T,e}(z_0)=e^{\frac{i}{2l_e^2}(x_0y_0-2y_0 x)}\label{eq:U_T}
\end{align}
The magnetic translations satisfy the algebra:
\begin{align}
D_e(z_0)D_e(z_1)&=e^{\frac{i}{2l_e^2}(x_0 y_1-y_0 x_1)}D_e(z_0+z_1),
\end{align}
and consequently, they satisfy the commutation relation
\begin{align}
[D_e(z_0),D_e(z_1)]=2i\sin\frac{x_0 y_1-y_0 x_1}{2l_e^2} D_e(z_0+z_1).
\end{align}
This is just another way to write down the GMP algebra Eq.(\ref{eq:GMP}).

Note that, although on the single-particle level we have Eq.(\ref{eq:real_momentum}),  the many-particle versions of the density operator and the magnetic translation operator in the Fock spaces are defined differently. In the first-quantization language:
\begin{align}
\pmb{\boldsymbol\rho}_e(\mathbf q_e)&\equiv \sum_i \rho_e(\mathbf q_e)_{(i)}\notag\\
\pmb{\bm D}_e(z_0)&\equiv \prod_i D_e(z_0)_{(i)},
\end{align}
where the subscript-$i$ means the operator is acting on the $i$-th particle. They satisfy:
\begin{align}
\pmb{\bm D}_e(z_0)\pmb{\boldsymbol\rho}_e(\mathbf q_e)\pmb{\bm D}_e(z_0)^\dagger=e^{i\mathbf q_e\cdot z_0}\pmb{\boldsymbol\rho}_e(\mathbf q_e)\label{eq:D_e_rho_e}
\end{align}
We will come back to these many-particle operators later.

On a finite-size torus, the boundary conditions can be described by the operator identities:
\begin{align}
D_e(L_1)&=e^{-i\varphi_{e,1}},& D_e(L_1\tau)&=e^{-i\varphi_{e,2}},\label{eq:boundary_conditions}
\end{align}
where $L_1>0$, $\tau$ is a complex number with positive imaginary part capturing the shape of the sample, and $L_1$ and $|L_1\tau|$ specify the real-space sample size. Note that $D_e(L_1)$ and $D_e(L_1\tau)$ must commute to apply the boundary conditions, leading to the flux quantization condition: the total number of fluxes through the sample is an integer $N_{\phi,e}$.

Haldane and Rezayi \cite{haldane1985periodic} pointed out that the orbital wavefunctions in the LLL, in the present gauge, can be compactly written in terms of the odd Jacobi-$\vartheta$ function, parameterized by $N_{\phi,e}$ zeros $z_\nu$ ($\nu=1,2,...,N_{\phi,e}$) and a complex number $k$ (Note that the convention for the magnetic translation in the present work differs from that in Ref.\cite{haldane1985periodic} by a minus sign.):
\begin{align}
\psi_e(z)&\propto f(z)e^{-\frac{y^2}{2l_e^2}},\notag\\
\text{with } f(z)&=e^{ikz}\prod_{\nu=1}^{N_{\phi,e}}\vartheta_1(\pi(z-z_\nu)/L_1 | \tau), \label{eq:Haldane_theta}
\end{align}
where the value of $k$ and the sum of zeros $z_{sum}\equiv\sum_{\nu} z_\nu$ need to be consistent with the boundary conditions:
\begin{align}
e^{ikL_1}&=(-1)^{N_{\phi,e}}\cdot e^{i\varphi_{e,1}},\notag\\
e^{i 2\pi z_{sum}/L_1 }&=(-1)^{N_{\phi,e}}\cdot e^{i\varphi_{e,2}-ikL_1\tau}. \label{eq:k_z_nu}
\end{align}
Although it appears that the wavefunction can be smoothly tuned, there are only $N_{\phi,e}$ linearly independent wavefunctions. Let $(k_0,z_{sum,0})$ be one solution of Eq.(\ref{eq:k_z_nu}),
the other solutions have the form:
\begin{align}
k=k_0+\frac{2\pi l_1}{L_1},\quad z_{sum}=z_{sum,0}+l_2 L_1-l_1 L_1\tau, \quad l_i\in\mathbb Z \label{eq:diff_z_sum}
\end{align}

In order to study a Chern-band sample with $N_{1,e}\cdot N_{2,e}$ unit cells, one can construct the corresponding bloch basis in the LLL. Namely, we consider the two real-space basis vectors $\mathbf a_{1,e},\mathbf a_{2,e}$ with $L_1=N_1 \mathbf a_{1,e}$, and $L_1\tau=N_2 \mathbf a_{2,e}$. $D_e(\mathbf a_{1,e})$ and $D_e(\mathbf a_{2,e})$ need to commute as the usual lattice translations. To have a one-to-one mapping between the Chern band and the LLL, one further chooses the area spanned by $\mathbf a_{1,e},\mathbf a_{2,e}$ contains exactly one flux unit, so that $N_{1,e}\cdot N_{2,e}=N_{\phi,e}$. With this setup, the minimal magnetic translations along $\mathbf a_{1,e},\mathbf a_{2,e}$ directions allowed by the boundary conditions are $\frac{\mathbf a_{1,e}}{N_{2,e}},\frac{\mathbf a_{2,e}}{N_{1,e}}$, respectively. 

The bloch basis in the LLL is formed by $N_{1,e}\cdot N_{2,e}$ simultaneous eigenstates of $\mathbf a_{1,e},\mathbf a_{2,e}$ magnetic translations:
\begin{align}
D_e(\mathbf a_{1,e})|\mathbf k_e\rangle_{\text{LLL}}&=e^{-i\mathbf k_e\cdot \mathbf a_1}|\mathbf k_e\rangle_{\text{LLL}},\notag\\
D_e(\mathbf a_{2,e})|\mathbf k_e\rangle_{\text{LLL}}&=e^{-i\mathbf k_e\cdot \mathbf a_2}|\mathbf k_e\rangle_{\text{LLL}},\label{eq:bloch_condition}
\end{align}
where 
\begin{align}
&\mathbf k_{e,(m_{1,e},m_{2,e})}\notag\\
&\quad\equiv \frac{m_{1,e}+\varphi_{1,e}/(2\pi)}{N_{1,e}} \mathbf G_{1,e} + \frac{m_{2,e}+\varphi_{2,e}/(2\pi)}{N_{2,e}} \mathbf G_{2,e},\label{eq:k_e_vals}\\
&\mathbf G_{1,e}=\frac{-i\mathbf a_{2,e}}{l_e^2}, \;\;\mathbf G_{2,e}=\frac{i\mathbf a_{1,e}}{l_e^2}.\notag
\end{align}
Here, $\mathbf G_{i,e}$ are the reciprocal basis vectors of $\mathbf a_{i,e}$, and one may choose a Brillouin Zone (BZ) with $m_{i,e}\in [0,N_{i,e}-1]$ being integers. These bloch states can be written in terms of infinite sums, as performed in Ref.[Murthy, Qi] for the case of a square lattice. Here, instead, we simply represent them using the Haldane-Rezayi's Jacobi-$\vartheta$ function via paramters $k$ and $z_\nu$.

\begin{figure}
    \centering
    \includegraphics[width=0.48\textwidth]{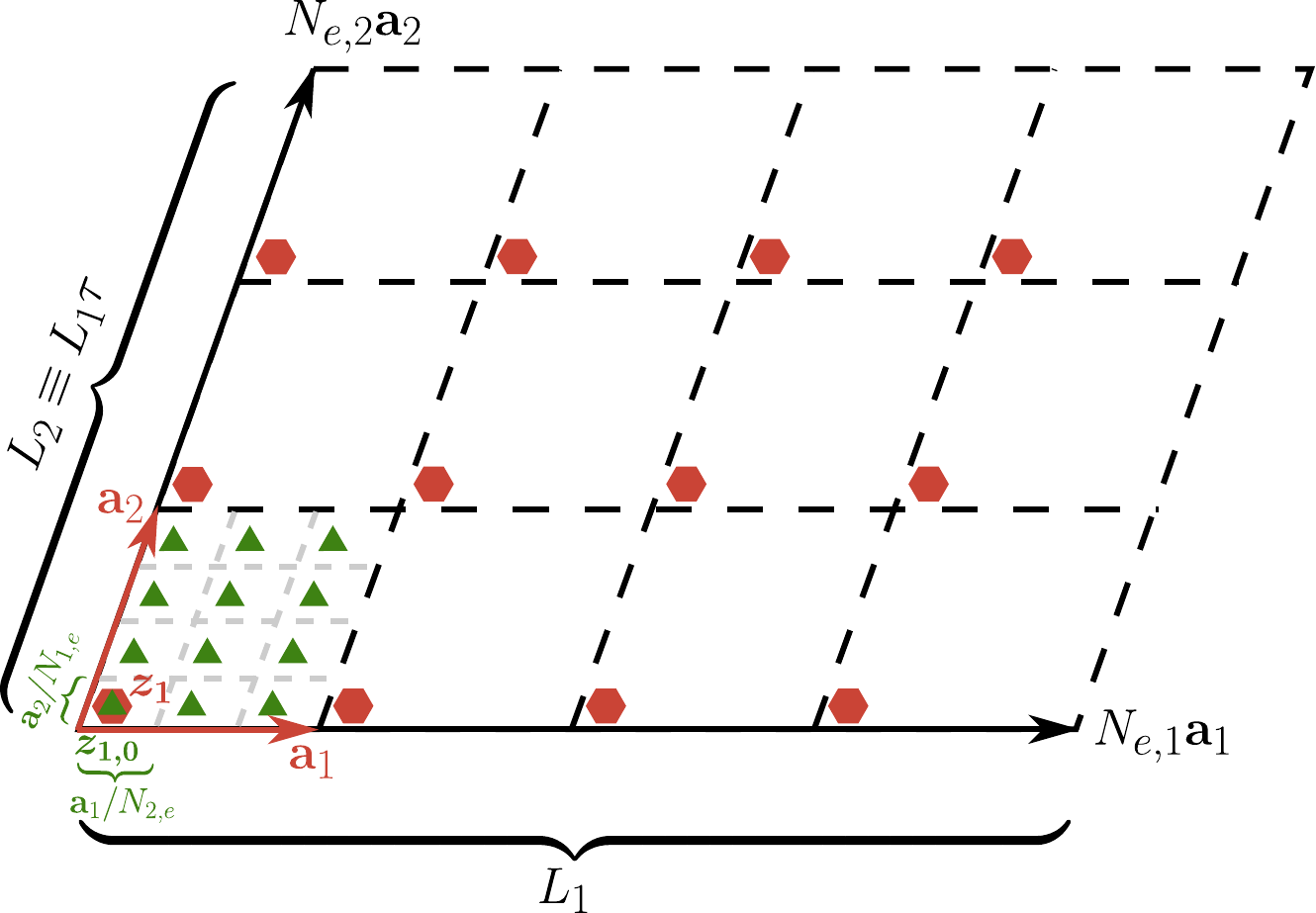}
    \caption{Geometry Setup of the LLL Bloch Basis: The torus sample, equivalent to a parallelogram, is parameterized by a real length $L_1$ and a modular parameter $\tau$. The $N_{1,e}\cdot N_{2,e}$ zeros $\{z_\nu\}$ (red hexagons) of the Haldane-Rezayi wave function are equally distributed on the torus sample by construction, see Eq.\eqref{eq:bloch_pattern_of_zeros}. There are also $N_{1,e}\cdot N_{2,e}$ independent groups of $\{k, z_{sum}\}$ satisfying the boundary condition Eq.\eqref{eq:k_z_nu}, i.e. $N_{1,e}\cdot N_{2,e}$ independent basis parameterized by different values of $z_{1,0}$, which also forms a $N_{2,e}\cdot N_{1,e}$ grids as green triangles, see Eq.\eqref{eq:z_1_vals}.}
    \label{fig:LLL_bloch_basis}
\end{figure}

To satisfy the eigen-condition Eq.(\ref{eq:bloch_condition}), obviously the zeros $z_\nu$ of $|\mathbf k_e\rangle_{\text{LLL}}$ need to form a $N_{1,e}\cdot N_{2,e}$ grid in the real-space:
\begin{align}
z_\nu=z_1+n_1 \mathbf a_{1,e}+ n_2 \mathbf a_{2,e},\label{eq:bloch_pattern_of_zeros}
\end{align}
where $n_i \in [0,N_{i,e}-1]$ are integers, and $z_1$ can be completely determined by $z_{sum}$ (see FIG.\ref{fig:LLL_bloch_basis} for an illustration).
Different $|\mathbf k_e\rangle_{\text{LLL}}$'s correspond to different values of $z_1$. According to Eq.(\ref{eq:diff_z_sum}), one finds that the possible values of $z_1$ are related as:
\begin{align}
z_1=z_{1,0}+l_2\frac{\mathbf a_{1,e}}{N_{2,e}}-l_1\frac{\mathbf a_{2,e}}{N_{1,e}}, \label{eq:z_1_vals}
\end{align}
where $z_{1,0}$ is determined by $z_{sum,0}$. Since the pattern of zeros returns to itself after $l_i\rightarrow l_i+N_{i,e}$ in Eq.(\ref{eq:bloch_pattern_of_zeros}), the linearly independent choices of $z_1$ correspond to $l_i\in [0,N_{i,e}-1]$. These allowed values of $z_{1,0}$ also form a grid (see FIG.\ref{fig:LLL_bloch_basis} for an illustration), related by magnetic translations $D_e(l_2\frac{\mathbf a_{1,e}}{N_{2,e}}-l_1\frac{\mathbf a_{2,e}}{N_{1,e}})$. There is a one-to-one mapping between the values of $\mathbf k_e$ in Eq.(\ref{eq:k_e_vals}) and the values of $z_1$ in Eq.(\ref{eq:z_1_vals}).

At this point, an instructive observation is that the magnetic translations $D_e(l_2\frac{\mathbf a_{1,e}}{N_{2,e}}-l_1\frac{\mathbf a_{2,e}}{N_{1,e}})$ are exactly the density operators in Eq.(\ref{eq:e_density}) for the finite-size sample. The relation Eq.(\ref{eq:real_momentum}) leads to the correspondence
\begin{align}
D_e\left(l_2\frac{\mathbf a_{1,e}}{N_{2,e}}-l_1\frac{\mathbf a_{2,e}}{N_{1,e}}\right)=\rho_e\left(\mathbf q_e= \frac{l_1}{N_{1,e}} \mathbf G_{1,e} + \frac{l_2}{N_{2,e}} \mathbf G_{2,e}\right)\label{eq:mag_transl_vs_density}
\end{align}

Due to the GMP algebra, we know that for $\mathbf q_e= \frac{l_1}{N_{1,e}} \mathbf G_{1,e} + \frac{l_2}{N_{2,e}} \mathbf G_{2,e}$, 
\begin{align}
\rho_e(\mathbf q_e)|\mathbf k_e\rangle_{\text{LLL}} \propto |\mathbf q_e+\mathbf k_e\rangle_{\text{LLL}}.
\end{align}
Therefore, if one chooses $z_{1,0}$ corresponding to $\mathbf k_{e,(0,0)}$ in Eq.(\ref{eq:k_e_vals}), we have the identification $l_i=m_{i,e}$ between Eq.(\ref{eq:z_1_vals}) and Eq.(\ref{eq:k_e_vals}), as expected.

To have a concrete discussion, we still need to fix a gauge for these bloch states. In this paper, we choose the gauge of $|\mathbf k_e\rangle_{\text{LLL}}$ so that:
\begin{align}
&\rho_e\left(\mathbf q_e=\frac{\mathbf G_{1,e}}{N_{1,e}}\right)|\mathbf k_{e,(m_{1,e},m_{2,e})}\rangle_{\text{LLL}}\notag\\
&\quad\quad=e^{\frac{2\pi i}{N_{\phi,e}}\big(m_{2,e}+\varphi_{2,e}/(2\pi)\big)}|\mathbf k_{e,(m_{1,e}+1,m_{2,e})}\rangle_{\text{LLL}},\notag\\
&\rho_e\left(\mathbf q_e=\frac{\mathbf G_{2,e}}{N_{2,e}}\right)|\mathbf k_{e,(m_{1,e},m_{2,e})}\rangle_{\text{LLL}}\notag\\
&\quad\quad=|\mathbf k_{e,(m_{1,e},m_{2,e}+1)}\rangle_{\text{LLL}}.\label{eq:rho_matrix_element}
\end{align}
The phase factor in the first line is to satisfy the GMP algebra. Applying the GMP algebra, the matrix elements of any density operator are analytically known in this LLL bloch basis. In particular, as noted in Ref.\cite{murthy2012hamiltonian}, \emph{the $N_{\phi,e}^2$ density operators with $l_1,l_2\in [0,N_{\phi,e}-1]$ form a complete basis of single-body operators in the LLL}. In fact, one can show that for any single-body operator $\hat{\mathbf A}_e$, one can expand it by the density operators:
\begin{align}
\hat{\mathbf A}_e=\sum_{l_1,l_2\in [0,N_{\phi,e}-1]} a_{l_1 l_2}\cdot\pmb{\boldsymbol\rho}_e\left(\mathbf q_e=l_1\frac{\mathbf G_{1,e}}{N_{1,e}}+l_2\frac{\mathbf G_{1,e}}{N_{2,e}}\right),
\end{align}
where
\begin{align}
a_{l_1 l_2}=\frac{1}{N_{\phi,e}}\mathop{\mathrm{Tr}}\left[\hat A_e\rho_e\left(\mathbf q_e=-l_1\frac{\mathbf G_{1,e}}{N_{1,e}}-l_2\frac{\mathbf G_{1,e}}{N_{2,e}}\right)\right],
\end{align}
which follows the GMP algebra and the fact that $\rho_e(\mathbf q_e)$ is traceless unless $\mathbf q_e$ is a linear superposition of $N_{2,e}\mathbf G_{1,e}$ and $N_{1,e}\mathbf G_{2,e}$ with integer coefficients.

One could extend the smooth gauge Eq.(\ref{eq:rho_matrix_element}) of the LLL bloch states beyond the BZ specified by $m_{i,e}\in [0,N_{i,e}-1]$, leading to the BZ boundary condition:
\begin{align}
|\mathbf k_e+\mathbf G_{1,e}\rangle_{\text{LLL}}&=|\mathbf k_e\rangle_{\text{LLL}}\notag\\
|\mathbf k_e+\mathbf G_{2,e}\rangle_{\text{LLL}}&=e^{-i \mathbf k_e\cdot\mathbf a_{1,e}}|\mathbf k_e\rangle_{\text{LLL}}\label{eq:BZ_BC}
\end{align}

It is known that the bloch wavefunctions in a $C=\pm 1$ Chern band (CB) can be mapped to the orbitals in the LLL, preserving the rotation and translation symmetries \cite{jian2013crystal}. To this end we need to discuss the magnetic rotation operation by an angle $\theta$ in the LLL:
\begin{align}
\psi_e(z)\rightarrow U_{R,e}(\theta)R_e(\theta)\psi_e(z)=U_{R,e}(\theta)\psi_e(e^{-i\theta}z),
\end{align}
where $R_e(\theta): \psi_e(z)\rightarrow \psi_e(e^{-i\theta}z)$ is the usual rotation, $U_{R,e}(\theta)$ is the associated gauge transformation, determined up to a $U(1)$ phase factor. In this paper, we fix this phase ambiguity by choosing
\begin{align}
U_{R,e}(\theta)=\exp\left[\frac{-i}{2l_e^2}\Big[\frac{\sin2\theta}{2}(x^2-y^2)+(1-\cos2\theta)xy\Big]\right],\label{eq:U_R}
\end{align}
satisfying $U_{R,e}(2\pi)R_e(2\pi)=\mathbf 1$ and
\begin{align}
 [U_{R,e}(\theta_1)R_e(\theta_1)]&[U_{R,e}(\theta_2)R_e(\theta_2)]\notag\\
 &\quad=U_{R,e}(\theta_1+\theta_2)R_e(\theta_1+\theta_2),\label{eq:U_R_identity}
\end{align}
As long as the modular parameter $\tau$ and the boundary conditions are consistent with the rotation angle (e.g., there exists $n_i\in\mathbb Z$ such that $e^{i\theta}\tau\equiv n_1+n_2\tau$), the magnetic rotation in the LLL is legitimate. Generally speaking, the magnetic rotation sends $|\mathbf k_e\rangle_{\text{LLL}}$ to a linear superposition of the bloch basis. 

If one further requires $\mathbf a_{i,e}$ to be consistent with the rotation, e.g., when $\mathbf a_{i,e}$ generates a square lattice and $\theta=\frac{\pi}{2}$, the magnetic rotation does send $|\mathbf k_e\rangle_{\text{LLL}}$ to a single bloch state:
\begin{align}
U_{R,e}(\theta)R_e(\theta)|\mathbf k_e\rangle_{\text{LLL}}=e^{i \xi(\theta,\mathbf k_e)}|R_{\theta}\mathbf k_e\rangle_{\text{LLL}}.\label{eq:R_trans}
\end{align}
It turns out that, generally speaking, the rotation should be interpreted as about the $[\pi,\pi]$-point, i.e., $R_{\theta}\mathbf k_e=e^{i\theta}(\mathbf k_e-\mathbf K_e)+\mathbf K_e$ where $\mathbf K_e=\frac{\mathbf G_{1,e}}{2}+\frac{\mathbf G_{2,e}}{2}$. This is the consequence of the magnetic translation algebra (see Appendix \ref{app:mag_rot} for details).

The phase factor $e^{i \xi(\theta,\mathbf k_e)}$ is fixed by the gauge choice in Eq.(\ref{eq:rho_matrix_element}). One way to compute it is to realize the magnetic symmetry group compatibility condition:
\begin{align}
[U_{R,e}(\theta)R_e(\theta)]D_e(z_0)[U_{R,e}(\theta)R_e(\theta)]^{-1}=D_e(e^{i\theta}z_0),\label{eq:R_T_identity}
\end{align}
which can be established using Eq.(\ref{eq:U_T},\ref{eq:U_R}). Choosing $z_0=\frac{\mathbf a_{1,e}}{N_{2,e}},\frac{\mathbf a_{2,e}}{N_{1,e}}$ and applying this identity to the bloch gauge condition Eq.(\ref{eq:rho_matrix_element}), an equation determining $\xi(\theta,\mathbf k_e)$ can be obtained and solved (see Appendix \ref{app:mag_rot} for details and explicit forms of $e^{i \xi(\theta,\mathbf k_e)}$).

In a Chern-band (CB), we will have the usual rotation $R^{\text{CB}}_{\theta}$ and usual translations $T^{\text{CB}}(\mathbf a_{i,e})$. Generally, one can show that the following correspondence can be made:
\begin{align}
T_e^{\text{CB}}(\mathbf a_{i,e})\leftrightarrow (-1) D_e(\mathbf a_{i,e}),\quad R^{\text{CB}}_{\theta}\leftrightarrow U_{R,e}(\theta)R_e(\theta),\label{eq:CB_map_LLL_symmetry}
\end{align}
because the algebra satisfied by the corresponding operators is identical. The minus sign in the first relation is not required for $C_2$ and $C_4$ systems but is required for the $C_3$ and $C_6$ systems. To have a uniform discussion, \emph{we choose this minus sign as a convention even for $C_2$ and $C_4$ systems}. Namely, the crystal momentum for the CB system will be shifted by $(\pi,\pi)$ when mapping into the LLL: 
\begin{align}
|\mathbf k_e\rangle_{\text{CB}}\leftrightarrow |\mathbf k_e+\mathbf K_e\rangle_{\text{LLL}}.\label{eq:CB_LLL_map}
\end{align}

Precisely, one needs to choose a smooth gauge in the CB satisfying the same BZ boundary condition as Eq.(\ref{eq:BZ_BC}) \cite{jian2013crystal}:
\begin{align}
|\mathbf k_e+\mathbf G_{1,e}\rangle_{\text{CB}}&=|\mathbf k_e\rangle_{\text{CB}}\notag\\
|\mathbf k_e+\mathbf G_{2,e}\rangle_{\text{CB}}&=e^{-i \mathbf k_e\cdot\mathbf a_{1,e}}|\mathbf k_e\rangle_{\text{CB}},\label{eq:CB_BZ_BC}
\end{align}
and the physical rotation $R^{\text{CB}}_e(\theta)$ needs to satisfy the same rule as Eq.(\ref{eq:R_trans}):
\begin{align}
R^{\text{CB}}_e(\theta)|\mathbf k_e\rangle_{\text{CB}}=e^{i \xi(\theta,\mathbf k_e+\mathbf K_e)}|e^{i\theta}\mathbf k_e\rangle_{\text{CB}}\label{eq:CB_rotation}
\end{align}
Under conditions Eq.(\ref{eq:CB_BZ_BC},\ref{eq:CB_rotation}), the identification Eq.(\ref{eq:CB_LLL_map}) allows one to map the original Hamiltonian $\mathbf H_{\text{CB}}$ in the CB into a Hamiltonian $\mathbf H_e$ in the LLL, preserving the rotation symmetry. If $\mathbf H_{\text{CB}}$ has the form:
\begin{align}
\mathbf H_{\text{CB}}=\sum_{\mathbf G_e} h(\mathbf G_e) \pmb{\boldsymbol\rho}_{\text{CB}}(\mathbf G_e)+\frac{1}{2}\sum_{\mathbf q_e} V(\mathbf q_e)\pmb{\boldsymbol\rho}_{\text{CB}}(\mathbf q_e)\pmb{\boldsymbol\rho}_{\text{CB}}(-\mathbf q_e),\label{eq:CB_H_e}
\end{align}
then $\mathbf H_e$ is \cite{murthy2012hamiltonian}:
\begin{align}
\mathbf H_e&=\sum_{\mathbf G_e} h(\mathbf G_e) \Big[\sum_{\mathbf G_e'}c(\mathbf G_e,\mathbf G_e')\pmb{\boldsymbol\rho}_{e}(\mathbf G_e+\mathbf G_e')\Big]\notag\\
+&\frac{1}{2}\sum_{\mathbf q_e} V(\mathbf q_e)\Big[\sum_{\mathbf G_e'}c(\mathbf q_e,\mathbf G_e')\pmb{\boldsymbol\rho}_{e}(\mathbf q_e+\mathbf G_e')\Big]\cdot\Big[h.c.\Big].\label{eq:H_e}
\end{align}
Here, $\pmb{\boldsymbol\rho}_{\text{CB}}(\mathbf q_e)$ is the density operator $\sum_i e^{i\mathbf q_e\cdot \bm{r}_i}$ projected into the CB. The first term in $\mathbf H_{\text{CB}}$ represents the CB dispersion, and the second term is the density-density interaction. Because the LLL density operators $\pmb{\boldsymbol\rho}_e(\mathbf q_e)$ form a complete basis for single-body operators, one has the expansion:
\begin{align}
\pmb{\boldsymbol\rho}_{\text{CB}}(\mathbf q_e)=\sum_{\mathbf G_e'}c(\mathbf q_e,\mathbf G_e')\pmb{\boldsymbol\rho}_{e}(\mathbf q_e+\mathbf G_e'),\label{eq:rho_CB}
\end{align}
where the summation is over $N_{2,e}\cdot N_{1,e}$ reciprocal lattice vectors.

Finally we comment on the conditions Eq.(\ref{eq:CB_BZ_BC},\ref{eq:CB_rotation}). One may wonder whether certain obstruction is present in the CB so that these conditions cannot be satisfied in a smooth gauge. The BZ boundary condition Eq.(\ref{eq:CB_BZ_BC}) can always be satisfied provided the CB has $C=-1$ that is identical to the LLL. The rotation condition Eq.(\ref{eq:CB_rotation}) requires further discussion. It is known that the Chern number of a band gives a constraint to the rotation eigenvalues at the high-symmetry points in the momentum space \cite{fang2012bulk}. We list these constraints in Eq.\eqref{eq:C_constraint} in Appendix \ref{app:mag_rot}.

These eigenvalues are preserved in the CB to LLL mapping. What if the CB and the LLL have different rotation eigenvalues? As computed in Appendix \ref{app:mag_rot}, the magnetic rotation eigenvalue for the LLL is $e^{-i\theta}$ at $\mathbf K_e$ point, corresponding to the $\Gamma$-point in the CB, while it is trivial for all other high-symmetry points. It turns out that one can always redefine the physical rotation operator, after which exactly the same eigenvalues are realized in the CB, and the conditions Eq.(\ref{eq:CB_BZ_BC},\ref{eq:CB_rotation}) can be satisfied in a smooth gauge following the prescription in Ref.\cite{jian2013crystal}. This redefinition is a source of the possible nontrivial Wen-Zee shift. We leave details in Appendix \ref{app:mag_rot}.

\subsection{Composite fermion substitution}\label{sec:CF_substitution}
From the previous section, we have the Hamiltonian $\mathbf H_e$ in the Fock space constructed from with the single-particle Hilbert space $\mathcal H_e$ in the LLL, which is exactly mapped from the CB problem. In this section, following the Murthy-Shankar construction \cite{murthy2003hamiltonian}, we need to enlarge the single-particle Hilbert space and construct the composite fermion single-particle Hilbert space for the finite-size systems:
\begin{align}
\mathcal H_e\otimes \mathcal H_v=\mathcal H_{CF}=\mathcal H_{\mathcal R_{CF}}\otimes \mathcal H_{\eta_{CF}}\label{eq:Hilbert_decomposition}
\end{align}

First, we introduce the vortex single-particle Hilbert space $\mathcal H_v$. $\mathcal H_v$ describes the guiding-center degrees of freedom of a particle carrying charge $q_v=-c^2q_e=-\frac{2ps}{2ps+1}q_e$ in \emph{the same} sample size specified by $L_1$ and $\tau$ as the electron. Consequently, the number of flux quanta seen by the vortex, i.e., the dimension of $\mathcal H_v$, is $N_{\phi,v}=c^2 N_{\phi,e}=\frac{2ps}{2ps+1}N_{\phi,e}$. One cannot define guiding-center operators as in Eq.(\ref{eq:v_guiding_center}) for a finite system. However, the density operators (magnetic translation operators) are well-defined for discrete momentum points (discrete displacements). We define them as:
\begin{align}
D_v(z_0)=\rho_v(\mathbf q_{v,z_0}) \equiv e^{-i\mathbf q_{v,z_0}\cdot \mathcal R_v},\text{ where }\mathbf q_{v,z_0}\equiv i\frac{z_0}{l_v^2}.\label{eq:D_v}
\end{align}
Here, the additional minus sign in the exponent is due to the sign of the vortex's charge. The periodic boundary condition is specified as:
\begin{align}
D_v(L_1)&=e^{i\varphi_{1,v}},&D_v(L_1\tau)&=e^{i\varphi_{2,v}}.
\end{align}
A simple way to understand the vortex's density operator $\rho_v(\mathbf q_v)\equiv e^{-i\mathbf q_{v}\cdot \mathcal R_v}$ and $\mathcal H_v$ is to consider the antilinear complex conjugate operator $K$. $K$ sends the $\bar z$ in a wavefunction in $\mathcal H_v$ to $z$, and consequently sends $\mathcal H_v$ to the Hilbert space $\bar{\mathcal H}_v$ of LLL wavefunctions of a particle carrying $-q_v$, with the same sign of the electrons' charge. At the same time:
\begin{align}
K \rho_v(\mathbf q_v) K=e^{i\mathbf q_v\cdot \bar{\mathcal R}_v},
\end{align}
where $\bar{\mathcal R}_v\equiv K\mathcal R_v K$ also satisfies the guiding-center algebra the charge $-q_v$. Namely, our results for the density operator of electrons, e.g., Eq.(\ref{eq:rho_matrix_element}), can be directly reused for the vortex case after the caution is made for the antilinear nature of $K$:
\begin{align}
\langle v_1|\rho_v(\mathbf q_v)|v_2\rangle=\langle \bar{v}_1|e^{i\mathbf q_v\cdot \bar{\mathcal R}_v}|\bar{v}_2 \rangle^*,
\end{align}
for any $|v_i\rangle\in\mathcal H_v$ and $|\bar{v}_i \rangle\equiv K|v_i\rangle\in \bar{\mathcal H}_v$.

Next, we decompose the tensor product of the enlarged Hilbert space $\mathcal H_e\otimes \mathcal H_v$ by introducing the full composite fermion with both the guiding-center and cyclotron degrees of freedom. We consider two cases separately: the Jain's sequence for $\nu=\frac{p}{2ps+1}$, and the composite Fermi liquid (CFL) case for $\nu=\frac{1}{2s}$. In the main text below, we focus on the Jain's sequence, and the CF substitution for CFL can be found in Appendix \ref{app:CFL_substitution}.

$\bullet$ \emph{Jain's sequence}.
The CF carries an electric charge $q_{CF}=\frac{1}{2ps+1} q_e$ as dictated by the algebra Eq.(\ref{eq:CF_algebra}). \emph{To save notation, we neglect the subscripts for $\mathcal R_{CF}$ and $\eta_{CF}$ from now on}. We similarly define the density operators (magnetic translation operators) for the CF degrees of freedom on finite-size systems:
\begin{align}
D_{\mathcal R}(z_0)=\rho_{\mathcal R}(\mathbf q_{\mathcal R,z_0}) \equiv e^{i\mathbf q_{\mathcal R,z_0}\cdot \mathcal R},\text{ where }\mathbf q_{\mathcal R,z_0}\equiv i\frac{z_0}{l_{CF}^2},\notag\\
D_{\eta}(z_0)=\rho_{\eta}(\mathbf q_{\eta,z_0}) \equiv e^{-i\mathbf q_{\eta,z_0}\cdot \eta},\text{ where }\mathbf q_{\eta,z_0}\equiv i\frac{z_0}{l_{CF}^2}.\label{eq:D_R_D_eta}
\end{align}
The CF guiding-center $\mathcal R$ lives on a real-space sample with \emph{the same} size as $\mathcal R_e$ and $\mathcal R_v$, specified by the boundary condition:
\begin{align}
D_{\mathcal R}(L_1)&=e^{-i\varphi_{1,\mathcal R}},&D_{\mathcal R}(L_1\tau)&=e^{-i\varphi_{2,\mathcal R}}.
\end{align}
For reasons that will be clear shortly, the CF cyclotron coordinates $\eta$, however, should be viewed as living on a sample whose linear size is enlarged by a factor $\frac{c}{1-c^2}=\sqrt{2ps(2ps+1)}$ where $c=\sqrt{\frac{2ps}{2ps+1}}$, satisfying the boundary condition:
\begin{align}
D_{\eta}\left(\frac{c}{1-c^2}L_1\right)=e^{i\varphi_{1,\eta}},\quad D_{\eta}\left(\frac{c}{1-c^2}L_1\tau\right)=e^{i\varphi_{2,\eta}}.
\end{align}

Consequently, the total number of flux quanta seen by $\mathcal R$ is $N_{\phi,\mathcal R}=\frac{1}{2ps+1} N_{\phi,e}$, while that seen by $\eta$ is $N_{\phi,\eta}=\frac{1}{2ps+1}\cdot\big(\frac{c}{1-c^2}\big)^2 N_{\phi,e}=2psN_{\phi,e}$. The Hilbert space dimensions must be consistent with the decomposition relation Eq.(\ref{eq:Hilbert_decomposition}): 
\begin{equation}
N_{\phi,e}\cdot N_{\phi,v}=N_{\phi,\mathcal R}\cdot N_{\phi,\eta}    
\end{equation}
Note that states in the space $\mathcal H_{\eta}$ label the CF LL indices. Namely, \emph{on a finite-size system, the number of CF LLs is finite and is equal to $N_{\phi,\eta}$.} We list these results in Table.\ref{tb:dimensions} for the convenience of readers. 

\begin{table}
\centering
\begin{tabular}{||c | c | c | c | c||} 
 \hline
  & $\mathcal R_e$ & $\mathcal R_v$ & $\mathcal R$ & $\eta$ \\ 
 \hline\hline
 \# of particles & $N$ & $N$ & $N$ & $N$ \\ \hline
 sample-size & $A$ & $A$ & $A$ & $2ps(2ps+1) A$\\\hline
 charge $q/q_e$ & 1 & $-\frac{2ps}{2ps+1}$ & $\frac{1}{2ps+1}$ & $-\frac{1}{2ps+1}$\\ \hline
 \# of fluxes & $\frac{2ps+1}{p} N$ & $2s N$  & $\frac{1}{p} N $ & $2s(2ps+1) N$ \\ \hline
 filling-fraction & $\frac{p}{2ps+1}$ & $\frac{1}{2s}$ & $p$ & $\frac{1}{2s(2ps+1)}$ \\ \hline
 \hline
\end{tabular}
\caption{Counting of the electronic guiding-center degrees of freedom $\mathcal R_e$, vortice's guiding-center degrees of freedom $\mathcal R_v$, and composite-fermion's both guiding-center $\mathcal R$ and cyclotron $\eta$ degrees of freedom on a finite-size sample for Jain's sequences $\nu=\frac{p}{2sp+1}$.}
\label{tb:dimensions}
\end{table}

After taking the exponential, the linear superposition Eq.(\ref{eq:CF_substitution},\ref{eq:CF_substitution_inverse}) in an infinite system becomes the operator identities together with boundary condition relations:
\begin{align}
D_e(z_0)&=D_{\mathcal R}\left(\frac{1}{1-c^2}z_0\right) D_{\eta}\left(-\frac{c}{1-c^2}z_0\right),\notag\\
D_v(z_0)&=D_{\mathcal R}\left(-\frac{c^2}{1-c^2}z_0\right) D_{\eta}\left(\frac{c}{1-c^2}z_0\right),\notag\\
e^{-i\varphi_{i,e}}&=e^{-i(2ps+1)\varphi_{i,\mathcal R}-i\varphi_{i,\eta}},\notag\\ e^{i\varphi_{i,v}}&=e^{i(2ps)\varphi_{i,\mathcal R}+i\varphi_{i,\eta}}\label{eq:D_e_D_v_to_D_R_D_eta}
\end{align}
and their inverse:
\begin{align}
D_{\mathcal R}(z_0)&=D_e(z_0)D_v(z_0),& D_{\eta}(z_0)&=D_e(cz_0)D_v(z_0/c),\notag\\
e^{-i\varphi_{i,\mathcal R}}&=e^{-i\varphi_{i,e}+i\varphi_{i,v}},& e^{i\varphi_{i,\eta}}&=e^{i(2ps+1)\varphi_{i,v}-i(2ps)\varphi_{i,e}}.\label{eq:D_R_D_eta_to_D_e_D_v}
\end{align}

These identities can be translated as identities of the density operators.
\begin{align}
\pmb{\boldsymbol\rho}_e(\mathbf q_e)&=\pmb{\boldsymbol\rho}_{\mathcal R}(\mathbf q_e)\pmb{\boldsymbol\rho}_{\eta}(-c\mathbf q_e),\notag\\
\pmb{\boldsymbol\rho}_v(\mathbf q_v)&=\pmb{\boldsymbol\rho}_{\mathcal R}(-\mathbf q_v)\pmb{\boldsymbol\rho}_{\eta}(\mathbf q_v/c),\label{eq:CF_substitution_density}
\end{align}
and the inverse
\begin{align}
\pmb{\boldsymbol\rho}_{\mathcal R}\left(\mathbf q_{\mathcal R}\right)&=\pmb{\boldsymbol\rho}_e\left(\frac{1}{1-c^2}\mathbf q_{\mathcal R}\right)\pmb{\boldsymbol\rho}_v\left(\frac{c^2}{1-c^2}\mathbf q_{\mathcal R}\right),\notag\\
\pmb{\boldsymbol\rho}_{\eta}\left(\mathbf q_{\eta}\right)&=\pmb{\boldsymbol\rho}_e\left(\frac{c}{1-c^2}\mathbf q_{\eta}\right)\pmb{\boldsymbol\rho}_v\left(\frac{c}{1-c^2}\mathbf q_{\eta}\right)
\end{align}
One can check that the dictionary above indeed provides a one-to-one mapping between the (finite number of) single-body operators in the electron/vortex spaces and those in the CF space. 

Importantly, after choosing the bloch bases in $\mathcal H_v$, $\mathcal H_{\mathcal R}$ and $\mathcal H_{\eta}$ similar to that in $\mathcal H_e$ (see Eq.(\ref{eq:bloch_condition},\ref{eq:k_e_vals})), this dictionary also specifies the fusion coefficients $\langle  \mathbf k_{\mathcal R},\mathbf k_{\eta}  |\mathbf k_e,\mathbf k_v\rangle$ (up to an unimportant overall phase factor):
\begin{align}
|\mathbf k_e\rangle\otimes |\mathbf k_v\rangle = \sum_{\mathbf k_{\mathcal R},\mathbf k_{\eta}} \langle  \mathbf k_{\mathcal R},\mathbf k_{\eta}  |\mathbf k_e,\mathbf k_v\rangle |\mathbf k_{\mathcal R}\rangle\otimes |\mathbf k_{\eta}\rangle.\label{eq:CG_coeff}
\end{align}

The composite fermion substitution can now be performed. The number of electrons, vortices and composite fermions are all the same. Following Eq.(\ref{eq:CF_substitution_density}), the Hamiltonian $\mathbf H_e$ in Eq.(\ref{eq:H_e}) is then mapped to the composite fermion Hamiltonian $\mathbf H_e\rightarrow \mathbf H_{CF}$ by substituting $\pmb{\boldsymbol\rho}_e(\mathbf q_e)=\pmb{\boldsymbol\rho}_{\mathcal R}(\mathbf q_e)\pmb{\boldsymbol\rho}_{\eta}(-c\mathbf q_e)$. One can perform the Hartree-Fock as well as time-dependent Hartree-Fock analysis for $\mathbf H_{CF}$, after caution is taken with respect to the constraint. However, before that, we need to discuss the consequence of the auxiliary vortex space.

\subsection{Gauge redundancy and projective symmetries}
The Fock spaces for electrons $\mathcal K_e$ and composite fermions $\mathcal K_{CF}$ (including both the CF guiding-center and cyclotron degrees of freedom) are fermionic spaces. The vortex Fock space $\mathcal K_v$ should be bosonic: fusing a fermionic electron with a bosonic vortex gives rise to a composite fermion. On the other hand, the construction above leads to a vast gauge redundancy. Given a state $|\psi_e\rangle\in \mathcal K_e$, one can tensor with an arbitrary state $|\psi_v\rangle\in \mathcal K_v$ and obtains a state $|\psi_e\rangle\otimes|\psi_v\rangle \in \mathcal K_{CF}$. Note that the reverse is not true: it is generally impossible to write a state in $\mathcal K_{CF}$ as a linear superposition of states in $\mathcal K_e\otimes \mathcal K_v$. Namely, $\mathcal K_e\otimes \mathcal K_v\subsetneq \mathcal K_{CF}$.

In order to go back to the physical Fock space $\mathcal K_e$, one needs to constrain states in $\mathcal K_{CF}$ by explicitly choosing some state $|\psi_v^g\rangle\in \mathcal K_v$, and only consider states in $\mathcal K_{CF}$ with the form $|\psi_e\rangle\otimes |\psi_v^g\rangle$. This choice of $|\psi_v^g\rangle$ is a gauge choice (hence the superscript $g$), and in principle, it can be arbitrary. One may write the constraint as the projector
\begin{align}
\mathbf P_g\equiv |\psi_v^g\rangle\langle\psi_v^g|,
\end{align}
and the partition function of $\mathbf H_{CF}$ and the original $\mathbf H_e$ are identical after the projection:
\begin{align}
\mathcal Z=\mathop{\mathrm{Tr}}[e^{-\beta \mathbf H_e}]=\mathop{\mathrm{Tr}}[ e^{-\beta \mathbf H_{CF}} \cdot \mathbf P_g]\label{eq:full_constraint}
\end{align}

A physical electronic state can be obtained from any state $|\psi_{CF}\rangle\in K_{CF}$ using this projector:
\begin{align}
|\psi^g_e(\psi_{CF})\rangle\otimes|\psi_v^g\rangle\equiv \mathbf P_g |\psi_{CF}\rangle =|\psi_v^g\rangle\langle\psi_v^g|\psi_{CF}\rangle.\label{eq:projection_ket}
\end{align}
Here $|\psi^g_e(\psi_{CF})\rangle\in \mathcal K_e$ is the projected wavefunction in the physical Fock space. We may view $\psi_{CF}$ as a "label" for the physical state $|\psi^g_e(\psi_{CF})\rangle$. As in all projective constructions, this is a many-to-one labeling, and caution needs to be taken when considering the symmetry and low energy fluctuations of $|\psi^g_e(\psi_{CF})\rangle$.

In terms of the first quantization, this projection is implemented as follows. The bosonic state $|\psi_v^g\rangle$ can be expanded in the bloch basis using the wavefunction:
\begin{align}
|\psi_v^g\rangle=\sum_{\{\mathbf k_{v_i}\}} \psi_v^g(\mathbf k_{v_1},\mathbf k_{v_2},...,\mathbf k_{v_N}) |\mathbf k_{v_1},\mathbf k_{v_2},...,\mathbf k_{v_N}\rangle.
\end{align}
We may expand the fermionic state $|\psi_{CF}\rangle$ in the bloch bases in $\mathcal H_e$ and $\mathcal H_v$:
\begin{align}
|\psi_{CF}\rangle=\sum_{\{\mathbf k_{e_i},\mathbf k_{v_i}\}} &\psi^{e,v}_{CF}(\mathbf k_{e_1},\mathbf k_{v_1},...,\mathbf k_{e_N},\mathbf k_{v_N}) \notag\\
&\cdot|\mathbf k_{e_1},\mathbf k_{v_1},...,\mathbf k_{e_N},\mathbf k_{v_N}\rangle.
\end{align}
The electronic state is then:
\begin{align}
|\psi^g_e(\psi_{CF})\rangle=\sum_{\{\mathbf k_{e_i}\}} \psi_e^g(\mathbf k_{e_1},\mathbf k_{e_2},...,\mathbf k_{e_N}) |\mathbf k_{e_1},\mathbf k_{e_2},...,\mathbf k_{e_N}\rangle,
\end{align}
where
\begin{align}
 &\psi_e^g(\mathbf k_{e_1},\mathbf k_{e_2},...,\mathbf k_{e_N})\notag\\
 &=\sum_{\{\mathbf k_{v_i}\}}\psi_v^{g*}(\mathbf k_{v_1},\mathbf k_{v_2},...,\mathbf k_{v_N})\psi^{e,v}_{CF}(\mathbf k_{e_1},\mathbf k_{v_1},...,\mathbf k_{e_N},\mathbf k_{v_N}).\label{eq:projection}
\end{align}
The fully symmetric nature of $\psi_v^g$ and the fully antisymmetric nature of $\psi^{e,v}_{CF}$ dictate that $\psi_e^g$ is fully antisymmetric.

In the case of bosonic $\nu=1$ CFL, this projection can be exactly implemented by the $SU(N)_R$ ($N$ is the number of physical particles) singlet condition, giving rise to a field theory treatment by Read \cite{read1998lowest}. This is because in that case, electron is bosonic, and vortex is fermionic, both at filling fraction $\nu=1$. This leads to a simple and technically helpful fact: the fermionic Fock space $\mathcal K_v$ is only one-dimensional because one is filling $N$ fermionic vortices in a $N$-dimensional single-particle Hilbert space. Eq.(\ref{eq:Read_projection}) is simply the second-quantization version of Eq.(\ref{eq:projection}).

However, in the present fermionic electron case, we do not know how to implement $\mathbf P_g$ in an elegant field-theory fashion. Instead, in this paper, we will focus on the wave function perspective of the projective construction, and comment on the associated effective field theories towards the end of the paper in the Discussion Section.\ref{sec:conclusion}.

In the remaining part of this section, we implement physical symmetries in the enlarged single-particle Hilbert space $\mathcal H_{CF}$, including the magnetic translation $D_e(\mathbf a_{i,e})$ and the magnetic rotation $U_{R,e}(\theta)R_e(\theta)$. In principle, one may combine the physical symmetry operations in $\mathcal H_e$ with an \emph{arbitrary} operation in $\mathcal H_v$, as long as $|\psi_v^g\rangle$ is invariant under that operation up to a phase factor. Namely, there is a gauge choice for the symmetry operation in $\mathcal H_v$. In order to have symmetries in $\mathcal H_{CF}$ consistent with physical intuitions of composite fermions (see Eq.(\ref{eq:proj_symm}) below), we choose the projective symmetry transformation as:
\begin{align}
D_e(\mathbf a_{i,e}) &\rightarrow D_e(\mathbf a_{i,e})\otimes D_v(\mathbf a_{i,e})=D_{\mathcal R}(\mathbf a_{i,e}),\notag\\
U_{R,e}(\theta)R_e(\theta)&\rightarrow U_{R,e}(\theta)R_e(\theta)\otimes U_{R,v}(\theta)R_v(\theta)\notag\\
&\quad=U_{R,\mathcal R}(\theta)R_{\mathcal R}(\theta)\otimes U_{R,\eta}(\theta)R_{\eta}(\theta).
\end{align}
In addition, we will choose $|\psi_v^g\rangle$ to be symmetric under $D_v(\mathbf a_{i,e})$ and $U_{R,v}(\theta)R_v(\theta)$. Here, the magnetic translations in various spaces were already defined before, and the magnetic rotations $U_{R,\alpha}(\theta)R_{\alpha}(\theta)$ ($\alpha=v,\mathcal R,\eta$) are defined similarly to Eq.(\ref{eq:U_R}) after the magnetic length is replaced $l_e\rightarrow l_{\alpha}$ and complex conjugate is taken for $\alpha=v,\eta$, i.e., $i\rightarrow -i$ in Eq.(\ref{eq:U_R}) due to the negativity of charges. For an infinite system, consistent with Eq.(\ref{eq:CF_substitution}), these transformations are:
\begin{align}
\left(\begin{array}{c}
    \mathcal R_e \\
    \mathcal R_v
\end{array}\right)\rightarrow \left(\begin{array}{c}
    \mathcal R_e+\mathbf a_{i,e} \\
    \mathcal R_v+\mathbf a_{i,e}
\end{array}\right)&\Longleftrightarrow\left(\begin{array}{c}
    \mathcal R \\
    \eta
\end{array}\right)\rightarrow\left(\begin{array}{c}
    \mathcal R+\mathbf a_{i,e} \\
    \eta
\end{array}\right)\notag\\
\left(\begin{array}{c}
    \mathcal R_e \\
    \mathcal R_v
\end{array}\right)\rightarrow e^{i\theta}\left(\begin{array}{c}
    \mathcal R_e \\
    \mathcal R_v
\end{array}\right)&\Longleftrightarrow\left(\begin{array}{c}
    \mathcal R \\
    \eta
\end{array}\right)\rightarrow e^{i\theta}\left(\begin{array}{c}
    \mathcal R \\
    \eta
\end{array}\right)\label{eq:proj_symm}
\end{align}
% &(\mathcal R_e\rightarrow \mathcal R_e+\mathbf a_{i,e}, \mathcal R_v\rightarrow \mathcal R_v+\mathbf a_{i,e} )\notag\\
% &\qquad\Leftrightarrow(\mathcal R\rightarrow \mathcal R+\mathbf a_{i,e}, \eta\rightarrow \eta )\notag\\
% &(\mathcal R_e\rightarrow e^{i\theta}\mathcal R_e,\mathcal R_v\rightarrow e^{i\theta}\mathcal R_v)\notag\\
% &\qquad\Leftrightarrow(\mathcal R\rightarrow e^{i\theta}\mathcal R, \eta\rightarrow e^{i\theta}\eta )

It is convenient to choose the bloch bases in $\mathcal H_v,\mathcal H_{\mathcal R}, \mathcal H_{\eta}$ so that these projective symmetries are (partially) explicit. For instance, one can choose the real-space basis vectors and lattice sizes as:
\begin{align}
\mathbf a_{1,v}&=\frac{1+2ps}{2ps}\mathbf a_{1,e},& N_{1,v}&=\frac{2ps}{1+2ps} N_{1,e},\notag\\
\mathbf a_{2,v}&=\mathbf a_{2,e}, & N_{2,v}&=N_{2,e};\notag\\
\mathbf a_{1,\mathcal R}&=(1+2ps)\mathbf a_{1,e}, &N_{1,\mathcal R}&=\frac{1}{1+2ps} N_{1,e},\notag\\
\mathbf a_{2,\mathcal R}&=\mathbf a_{2,e}, & N_{2,\mathcal R}&=N_{2,e};\notag\\
\mathbf a_{1,\eta}&=\frac{c}{1-c^2}\frac{1}{2ps}\mathbf a_{1,e},& N_{1,\eta}&=(2ps) N_{1,e},\notag\\
\mathbf a_{2,\eta}&=\frac{c}{1-c^2} \mathbf a_{2,e}, & N_{2,\eta}&=N_{2,e}.\label{eq:bloch_bases}
\end{align}
Here we have assumed that the lattice size $N_{1,e}$ is a multiple of $(1+2ps)$. Note that the unit cell size needs to enclose one flux quantum in the corresponding space. For instance, the unit cell for $\mathcal R$ is enlarged $(1+2ps)$-times along the $\mathbf a_{1}$ direction. On the other hand, the $\mathbf a_2$ for $v$ and $\mathcal R$ are purposely chosen to be identical to $\mathbf a_{2,e}$, so that the $D_e(\mathbf a_{2,e})$-projective symmetry is explicit.  With these bloch bases, the $D_e(\mathbf a_{2,e})$-projective symmetry dictates the selection rule for the fusion coefficients: 
\begin{align}
&\langle \mathbf k_e,\mathbf k_v|\mathbf k_{\mathcal R},\mathbf k_{\eta} \rangle \neq 0 \notag\\
&\quad\text{ only if } m_{2,e}-m_{2,v}=m_{2,\mathcal R} \mod N_{2,\mathcal R}.\label{eq:CG_selection_2}
\end{align}
Notice that we choose the convention for the momentum eigenvalues as (the signs in the exponents are due to the signs of the charges):
\begin{align}
D_e(\mathbf a_{i,e})|\mathbf k_e\rangle&=e^{-i\mathbf k_e\cdot \mathbf a_{i,e}}|\mathbf k_e\rangle\notag\\
D_v(\mathbf a_{i,v})|\mathbf k_v\rangle&=e^{i\mathbf k_v\cdot \mathbf a_{i,v}}|\mathbf k_v\rangle\notag\\
D_{\mathcal R}(\mathbf a_{i,\mathcal R})|\mathbf k_{\mathcal R}\rangle&=e^{-i\mathbf k_{\mathcal R}\cdot \mathbf a_{i,\mathcal R}}|\mathbf k_{\mathcal R}\rangle\notag\\
D_{\eta}(\mathbf a_{i,\eta})|\mathbf k_{\eta}\rangle&=e^{i\mathbf k_{\eta}\cdot \mathbf a_{i,\eta}}|\mathbf k_{\eta}\rangle
\end{align}
and $m_{i,\alpha}\in \mathbb Z$ are the momentum quantum numbers defined as Eq.(\ref{eq:k_e_vals}) for the relevant spaces.

The $D_e((1+2ps)\mathbf a_{1,e})$ symmetry is also explicit, leading to the selection rule:
\begin{align}
&\langle \mathbf k_e,\mathbf k_v|\mathbf k_{\mathcal R},\mathbf k_{\eta} \rangle \neq 0 \notag\\
&\quad\text{ only if } (1+2ps)m_{1,e}-(2ps)m_{1,v}=m_{1,\mathcal R} \mod N_{1,\mathcal R}.\label{eq:CG_selection_1}
\end{align}

How about the $D_e(\mathbf a_{1,e})$-projective symmetry? For instance, in CF space $\mathcal H_{\mathcal R}$, it is implemented as $D_{\mathcal R}(\mathbf a_{1,e}=\frac{1}{1+2ps}\mathbf a_{1,\mathcal R})$. According to the bloch basis gauge choice Eq.(\ref{eq:rho_matrix_element}) (after modified for the $\mathcal H_{\mathcal R}$ space), we know that
\begin{align}\label{eq: translation fractionalization}
D_{\mathcal R}(\mathbf a_{1,e})|\mathbf k_{\mathcal R}\rangle=\left|\mathbf k_{\mathcal R}+\frac{\mathbf G_{2,\mathcal R}}{1+2ps}\right\rangle.
\end{align}
If the CF mean-field state satisfies this projective symmetry, it means that the CF band structure will have a $(1+2ps)$-fold periodicity in the CF BZ -- a well-known phenomenon for \emph{translational symmetry fractionalization}. Similarly,
\begin{align}
D_{v}(\mathbf a_{1,e})|\mathbf k_v\rangle=\left|\mathbf k_v+\frac{(2ps)\mathbf G_{2,v}}{1+2ps}\right\rangle.
\end{align}

As a remark, in order to respect $D_e(\mathbf a_{1,e})$-projective symmetry, the sample size $N_{2,e}$ must also be a multiple of $(1+2ps)$ in the current construction (we have already assumed that $N_{1,e}$ is a multiple of $(1+2ps)$ in Eq.(\ref{eq:bloch_bases}).), otherwise $D_v(\mathbf a_{1,e})$ and $D_{\mathcal R}(\mathbf a_{1,e})$ changes the boundary conditions in $\mathcal H_v$ and $\mathcal H_{\mathcal R}$. To implement $D_e(\mathbf a_{1,e})$-projective symmetry in the case that $\text{mod}(N_{2,e},(1+2ps))\neq 0$, the construction needs to be generalized which we will leave as a future project.

The projective magnetic rotation symmetry, when implemented in the $\alpha=v,\mathcal R,\eta$ spaces, sends a bloch basis state $|\mathbf k_\alpha\rangle$ to a linear superposition of bloch basis states. These transformation rules can be computed analytically using the gauge conditions similar to Eq.(\ref{eq:rho_matrix_element}) and numerically using Haldane-Rezayi wave function. 

\subsection{Hartree-Fock and time-dependent Hartree-Fock}\label{sec:HF_TDHF}
In this section, we describe how to perform the Hartree-Fock analysis for the CF mean-field ground state, and to perform the time-dependent Hartree-Fock analysis for CF excited states. 

In an exact study, one should have implemented the full constraint as in Eq.(\ref{eq:full_constraint}), and only states of the form $|\psi_e\rangle \otimes|\psi_v^g\rangle\in \mathcal K_{CF}$ are physical. In a Hartree-Fock analysis or time-dependent Hartree-Fock analysis, this constraint is implemented on a mean-field level: the variational states under consideration are free CF states $|\psi^{MF}_{CF}\rangle$ (i.e., single Slater-determinants in $\mathcal K_{CF}$) satisfying 
\begin{align}
\langle \psi^{MF}_{CF}|\pmb{\boldsymbol\rho}_v(\mathbf q_v)|\psi^{MF}_{CF}\rangle=\langle\psi^g_v|\pmb{\boldsymbol\rho}_v(\mathbf q_v)|\psi^g_v\rangle,\;\; \forall \pmb{\boldsymbol\rho}_v(\mathbf q_v).\label{eq:mf_constraint}
\end{align}
Note that since $\pmb{\boldsymbol\rho}_v(\mathbf q_v)$'s form a complete basis for single-body operators, Eq.(\ref{eq:mf_constraint}) means that the expectation value of any single-body operator in 
$|\psi^{MF}_{CF}\rangle$ is the same as in $|\psi_v^g\rangle$. 

We have not specified the bosonic vortex state $|\psi^g_v\rangle$ yet. But we know it should respect the magnetic symmetries in the vortex space, and at the filling fraction $\nu=\frac{1}{2s}$. We make a natural choice: \emph{$|\psi^g_v\rangle$ will be one of the $2s$-fold degenerate bosonic $\nu=\frac{1}{2s}$ Laughlin wavefunction $|\psi^{Laughlin}_{\nu=1/2s,v}\rangle$ on the torus in our discussion below.} We will comment on exactly which state we choose for practical simulations in the $2s$-dimensional subspace in Appendix \ref{app:ccs}.

With this choice, we know that \cite{wen1990ground}, apart from the trivial condition for $\mathbf{q}_v=0$, $\langle\psi^g_v|\rho_v(\mathbf q_v)|\psi^g_v\rangle=0$ except for a few specific values of $\mathbf q_v=\frac{i (n_1 \frac{L_1}{2s}+ n_2\frac{L_1\tau}{2s})}{l_v^2}$ ($n_1,n_2\in [0,2s-1]$ are integers). Even for these specific values of $\mathbf q_v$, $\langle\psi^g_v|\rho_v(\mathbf q_v)|\psi^g_v\rangle$ exponentially decays to zero as the system size increases (see Appendix. \ref{app:density_expectation} for a detailed discussion). 

For the simplicity of presentation, we choose the thermodynamic limit values for $\langle\psi^g_v|\rho_v(\mathbf q_v)|\psi^g_v\rangle$ for the discussion below, and require the CF mean-field state to satisfy:
\begin{align}
\langle \psi^{MF}_{CF}|\pmb{\boldsymbol\rho}_v(\mathbf q_v)|\psi^{MF}_{CF}\rangle=0,\;\; \forall \mathbf q_v\neq 0.\label{eq:mf_constraint_1}
\end{align}
Notice that $[\pmb{\boldsymbol\rho}_v(\mathbf q_v),\mathbf H_{CF}]=0$ by construction. This mean-field level constraint is imposed via Lagrangian multipliers.  In terms of second quantization, the CF mean-field Hamiltonian can be expressed as:
\begin{align}
\mathbf H_{CF}^{MF}&=\sum_{\mathbf k_{\mathcal R}} \sum_{\mathbf k_{\eta_i},\mathbf k_{\eta_j}} f^\dagger_{\mathbf k_{\mathcal R},\mathbf k_{\eta_i}} h_{\mathbf k_{\eta_i},\mathbf k_{\eta_j}}(\mathbf k_{\mathcal R}) f_{\mathbf k_{\mathcal R},\mathbf k_{\eta_j}}\notag\\
&+\sum_{\mathbf q_v}\lambda_{\mathbf q_v}\pmb{\boldsymbol\rho}_v(\mathbf G_e)
\end{align}
where we have used the bloch bases defined in Eq.(\ref{eq:bloch_bases}), and $f^\dagger_{\mathbf k_{\mathcal R},\mathbf k_{\eta_i}}$ are the corresponding CF creation operators. $\pmb{\boldsymbol\rho}_v(\mathbf q_v)$ operators can be expressed using the CF operators as described in Eq.(\ref{eq:CF_substitution_density}). When the sample size is consistent with $D_e(\mathbf a_{i,e})$-projective symmetry (i.e., a multiple of $(1+2ps)$ along both directions), these projective symmetry can be exactly implemented in $\mathbf H_{CF}^{MF}$, and consequently expectation values $\langle \psi^{MF}_{CF}|\pmb{\boldsymbol\rho}_v(\mathbf q_v)|\psi^{MF}_{CF}\rangle$ may be nonzero only when $\mathbf q_v=\mathbf G_e$ is a reciprocal lattice vector of electrons. In this situation only lagrangian multipliers $\lambda_{\mathbf G_e}$ are needed.

$\mathbf H_{CF}^{MF}$ and its ground state $|\psi^{MF}_{CF}\rangle$ are determined self-consistently as in a standard Hartree-Fock calculation, during which the projective symmetries can be implemented exactly (as long as the sample size is consistent with them.). The mean-field energy does have a variational meaning despite the fact that the Hilbert space is enlarged.

After $|\psi^{MF}_{CF}\rangle$ is determined, one may proceed to perform the time-dependent Hartree-Fock (TDHF) calculation for the excitations. TDHF is an approximation scheme to compute excited states (e.g., particle-hole excitations or collective modes) in quantum systems. We are not aware of a systematic TDHF treatment in the presence of constraints and lagrangian multipliers in the literature. We briefly present the main procedure and leave the details in Appendix \ref{app:TDHF}. 

TDHF is known to be a conserving approximation. Similar to static Hartree-Fock, in TDHF one considers the Slater determinant states, which are completely determined by their single-body density matrix $\boldsymbol{\mathcal P}$. Let the static Hartree-Fock self-consistent solution be $\boldsymbol{\mathcal P}_0$, the perturbed state can be parameterized by $\boldsymbol{\mathcal P}=\mathbf U\boldsymbol{\mathcal P}\mathbf U^\dagger$, where $\mathbf U=e^{i\boldsymbol{\phi}}$ is a unitary rotation generated by a small composite fermion bilinear operator $\boldsymbol{\phi}$. The time-evolution of $\boldsymbol{\phi}$ can be computed self-consistently: $\boldsymbol{\mathcal L}\cdot \boldsymbol{\phi}= i\hbar \dot{\boldsymbol{\phi}}$, where $\boldsymbol{\mathcal L}$ is a linear operator acting in a space $\boldsymbol{\mathcal W}$, spanned by fermion bilinears having nontrivial commutator with $\boldsymbol{\mathcal P}_0$. The eigenvalues of $\boldsymbol{\mathcal L}$ are the energies of the excitation modes.

In the present situation, constraints Eq.(\ref{eq:mf_constraint_1}) need to be imposed on both $\boldsymbol{\mathcal P}_0$ and $\boldsymbol{\mathcal P}$. This reduces the dimension of $\boldsymbol{\mathcal W}$ by $N_c$, the number of nontrivial constraints. $N_c=N_{\phi,v}^2-1$, since the $\mathbf q_v=0$ constraint is trivial. In addition, each symmetry generator $\boldsymbol{\rho}_v(\mathbf q_v)$ (except for $\mathbf q_v=0$) leads to an exact zero mode (the Goldstone mode) in TDHF. There are also totally $N_c$ exact zero modes. The nonzero eigenmodes can be found in the remaining subspace $\boldsymbol{\mathcal V}\subset \boldsymbol{\mathcal W}$, whose dimension is $2N_c$ smaller than the dimension of $\boldsymbol{\mathcal W}$. In the subspace $\boldsymbol{\mathcal V}$, the eigenproblem of $\boldsymbol{\mathcal L}$ can be mapped to the diagonalization problem of a free boson Hamiltonian using the symplectic Bogoliubov transformation: The eigenvalues are real and form $\pm\hbar\omega$ pairs.

A single composite fermion transforms projectively under the symmetry group, as mentioned before. The fermion bilinear $\boldsymbol{\phi}$, however, transform as a regular representation of the symmetry group. Namely, $\boldsymbol{\phi}$ carries well-defined crystalline momentum under $D_e(\mathbf a_{i,e})$: $\boldsymbol{\phi}_{\mathbf q_e}$, where $\mathbf q_e$ is inside the Brillouin Zone (BZ) of the electronic Chern band. The magnetoroton collective modes in the FCI phase form a gapped band structure $\hbar\omega_a(\mathbf q_e)$, where $a$ labels the bands. At the high symmetry points in the BZ, the crystalline rotation eigenvalues of the magnetorotons can be computed explicitly using the TDHF eigenstate $\boldsymbol{\phi}_{\mathbf q_e}$. 

Special consideration needs to be made for $\mathbf q_e=0$. Here, there is a $\pm\hbar\omega_0$ pair of approximate zero modes in the TDHF calculation, and $\omega_0$ goes to zero in the thermodynamic limit. This is again related to the gauge redundancy in the projective construction. In the infinite system, the guiding center $\mathcal R_v$ is a well-defined operator at $\mathbf q=0$, corresponding to $\mathcal R_v=\mathcal R+\frac{1}{c}\eta$ on the CF side. Just like $\boldsymbol\rho_v(\mathbf q_v)$, this operator does not act in physical Hilbert space and commutes with $\mathbf H_{CF}$, giving a pair of exact zero modes. 

To appreciate the physical picture of $\mathcal R_v$, let us consider the CF state in the traditional LL case. It is convenient to write $\mathcal R_v$ in terms of the ladder operators as in Eq.(\ref{eq:ladder_op}): $a^\dagger_{v}\propto a_{\mathcal R}+\frac{1}{c}a_{\eta}^\dagger$. For instance, Laughlin's $\nu=1/3$ state is represented as a single filled CF LLL: $|\psi_{CF}\rangle=|\psi_{\mathcal R}\rangle\otimes \prod_i |0_{\eta}\rangle_i$, where $|\psi_{\mathcal R}\rangle$ is the fully filled state in the $\mathcal R$-space, and all composite fermions have the same wavefunction in the $\eta$-space: the coherent state $|0_{\eta}\rangle$ that can be annihilated by $a_{\eta}$. It is then easy to see that of $\sum_i a^\dagger_{v,i}|\psi_{CF}\rangle\propto\sum_i a^\dagger_{\eta,i}|\psi_{CF}\rangle $ since $|\psi_{\mathcal R}\rangle$ is annihilated by $\sum_i a_{\mathcal R,i}$. Because $a^\dagger_{\eta}|0_{\eta}\rangle=|1_{\eta}\rangle$, $\mathcal R_v$ is creating the $\mathbf q_e=0$ particle-hole excitations between the CF LLL and the first LL. This excitation was known to be a zero mode in an infinite-system TDHF calculation previously \cite{murthy2001hamiltonian}. Here we have shown the physical origin of this zero mode. 

The exact zero modes $\boldsymbol{\rho}_v(\mathbf q_v)$ as well as the zero mode due to $\mathcal R_v$ are gauge modes and do not correspond to physical excitations. One way to see this is via the projective construction: the electronic ground state is given by $|\psi^{\text{GS}}_e\rangle\otimes |\psi_v^g\rangle=\mathbf P_g|\psi_{CF}\rangle$, and the would-be excited state corresponding to $\boldsymbol{\rho}_v(\mathbf q_v)$ is $|\psi^{\text{EX}}_e\rangle\otimes |\psi_v^g\rangle=\mathbf P_g \boldsymbol{\rho}_v(\mathbf q_v)|\psi_{CF}\rangle$. The latter one can be equivalently obtained via $\mathbf{\tilde{ P}}_g\equiv |\tilde\psi_v^g\rangle\langle\tilde\psi_v^g|$: $|\psi^{\text{EX}}_e\rangle\otimes |\tilde\psi_v^g\rangle=\mathbf{\tilde{ P}}_g |\psi_{CF}\rangle$, where $|\tilde\psi^g_v\rangle\equiv \mathbf\rho_v(\mathbf q_v)^\dagger |\psi^g_v\rangle$. But the difference between $\mathbf P_g$ and $\mathbf{\tilde{ P}}_g$ is really a gauge choice: In an exact study, $|\psi_{CF}\rangle=|\psi_v\rangle\otimes|\psi_e^{\text{GS}}\rangle$ where $|\psi_v\rangle$ can be an arbitrary state in $\mathcal K_v$. Therefore $|\psi^{\text{EX}}_e\rangle\propto |\psi^{\text{GS}}_e\rangle$ is the same wavefunction in the exact study (unless $|\psi_v\rangle$ being orthogonal to $|\tilde\psi^g_v\rangle$, in which case $|\psi^{\text{EX}}_e\rangle=0$ is annihilated by the projection).

\subsection{Projection to physical states: Hyperdeterminant}\label{sec:proj}
We have demonstrated the procedure to perform static Hartree-Fock calculations to obtain the mean-field CF ground state $|\psi_{CF}^{MF}\rangle$ and to perform TDHF calculations for excited states. The physical electronic state is obtained via the projection $\mathbf P_g$ as in Eq.(\ref{eq:projection_ket},\ref{eq:projection}). In this section, we study the mathematical structure of $|\psi^g_e(\psi_{CF})\rangle$.

We will focus on a single slater determinant $|\psi_{CF}\rangle$ before the projection ($|\psi_{CF}\rangle$ could be either the mean-field ground state $|\psi_{CF}^{MF}\rangle$ or the unitary rotated states $e^{i\boldsymbol{\phi}}|\psi_{CF}^{MF}\rangle$ related to excitations). It turns out that $|\psi^g_e(\psi_{CF})\rangle$ is mathematically represented as the combinatorial hyperdeterminant of a tensor. 

Given a rank $m$ tensor $T_{i_1,i_2,...,i_m}$, with each index $i_s = 1,2,..,N$ ($N$ is the dimension of the tensor), the combinatorial hyperdeterminant \cite{gelfand1994hyperdeterminants} is a direct generalization of the determinant of a matrix:
\begin{align}
&\text{Hyperdet}(T)\equiv \sum_{P_1,P_2,...P_{m-1}\in S_N} (-1)^{P_1}(-1)^{P_2}...(-1)^{P_{m-1}} \notag\\
&\cdot T_{1,P_1(1),P_2(1),..,P_{m-1}(1)}T_{2,P_1(2),P_2(2),..,P_{m-1}(2)}\cdot ...\notag\\
&\cdot T_{N,P_1(N),P_2(N),..,P_{m-1}(N)},
\end{align}
where $S_N$ is the permutation group and $(-1)^P$ is the signature of the permutation.

We will demonstrate the FCI states in Jain's sequence at $\nu=\frac{p}{2ps+1}$ with $s=1$ as an example. To perform the projection, we need an expression for the bosonic Laughlin's state at $\nu=1/2$ since $|\psi^g_v\rangle=|\psi^{Laughlin}_{\nu=1/2,v}\rangle$. It turns out that, the state $|\psi^{Laughlin}_{\nu=1/2,v}\rangle$ can be constructed via the same projective construction mentioned before, but for $s=1/2$. We will prove this later in Sec.\ref{sec:connection_jain}. Precisely, one views the bosonic $v$-particle as the ``electron'', and then attaches a single unit of flux ($2s=1$) to form a composite fermion. The corresponding vortices and composite fermions for $v$-particles will be denoted as $v-v$ and $v-CF$ respectively, both are fermionic. Following the projective construction:
\begin{align}
|\psi^{Laughlin}_{\nu=1/2,v}\rangle | \psi^g_{v-v}\rangle \equiv|\psi^g_{v-v}\rangle\langle \psi^g_{v-v}|\psi^{MF}_{v-CF}\rangle
\end{align}
But here, both $|\psi^g_{v-v}\rangle$ and $|\psi^{MF}_{v-CF}\rangle$ are single slater determinants: $|\psi^g_{v-v}\rangle$ is the full filled state in the Fock space $\mathcal K_{v-v}$, which is only one-dimensional. $|\psi^{MF}_{v-CF}\rangle$ is the filled $v-CF$ LLL state. In terms of the first quantization, We may represent them using the filled orbitals as:
\begin{align}
|\psi^g_{v-v}\rangle&=\sum_{P\in S_N} (-1)^P |\phi^{v-v}_{P(1)}\phi^{v-v}_{P(2)}...\phi^{v-v}_{P(N)}\rangle\notag\\
|\psi^{MF}_{v-CF}\rangle&=\sum_{P\in S_N} (-1)^P |\phi^{v-CF}_{P(1)}\phi^{v-CF}_{P(2)}...\phi^{v-CF}_{P(N)}\rangle\notag\\
\end{align}
It is easy to see that, if one chooses a basis $\{\phi^v_{\alpha}\}$ in $\mathcal H_v$, the wavefunction of $|\psi^{Laughlin}_{\nu=1/2,v}\rangle$ will be the hyperdeterminant of a rank-3 tensor $C$ formed by the fusion coefficients similar to Eq.(\ref{eq:CG_coeff}):
\begin{align}
&\langle \phi^v_{\alpha_1}\phi^v_{\alpha_2}...\phi^v_{\alpha_N}|\psi^{Laughlin}_{\nu=1/2,v}\rangle=\text{Hyperdet}(C), \text { where }\notag\\
& C_{ijk}=\langle \phi^v_{\alpha_i}|\langle \phi^{v-v}_{j}|\phi^{v-CF}_{k}\rangle.
\end{align}

Next, we will perform the projection $|\psi^{Laughlin}_{\nu=1/2,v}\rangle\langle\psi^{Laughlin}_{\nu=1/2,v}|\psi_{CF}^{MF}\rangle$, where the CF mean-field state $|\psi_{CF}^{MF}\rangle$ is a single slater determinant filling $p$-CF bands:
\begin{align}
|\psi_{CF}^{MF}\rangle=\sum_{P\in S_N} (-1)^P |\phi^{CF}_{P(1)}\phi^{CF}_{P(2)}...\phi^{CF}_{P(N)}\rangle
\end{align}
Similarly, after choosing a basis $\{\phi^e_{\alpha}\}$ in $\mathcal H_e$, the wavefunction of $|\psi^g_e(\psi_{CF})\rangle$ will be the hyperdeterminant of a rank-4 tensor $T$ formed by the fusion coefficients in the projective construction:
\begin{align}
&\langle \phi^e_{\alpha_1}\phi^e_{\alpha_2}...\phi^e_{\alpha_N}|\psi^g_e(\psi_{CF})\rangle=\text{Hyperdet}(T), \text { where }\notag\\
& T_{ijkl}=\langle \phi^e_{\alpha_i}|\langle \phi^{v-CF}_{j}|\phi^{v-v}_{k}\rangle |\phi^{CF}_{l}\rangle.\label{eq:hyperdet_T}
\end{align}

A very special situation is when the tensor $T$ can be represented as a product of matrices: $T_{i_1,i_2,...,i_m}=A^{(1)}_{i_1,i_2}\cdot A^{(2)}_{i_1,i_3}\cdot...\cdot A^{(m-1)}_{i_1,i_m}$, in which case the hyperdeterminant is the product of the conventional determinants of the matrices: $\text{Hyperdet}(T)=\prod_{j=1}^{m-1}\text{det}(A^{(j)})$. This is exactly the situation for the Laughlin's $\nu=1/(m-1)$ states with the open boundary condition, when the electron basis $\{\phi^e_{\alpha}\}$ is chosen to be the over-complete basis of coherent states (see Sec.\ref{sec:connection_jain}). 

Generally speaking, the tensor $T$ \emph{cannot} be decomposed as a product of matrices, and computing a generic hyperdeterminant is known to be a NP-hard problem \cite{hillar2013most}. Nevertheless, the crystalline momentum conservation leads to the selection rules in the bloch bases (see Eq.(\ref{eq:CG_selection_2},\ref{eq:CG_selection_1})), slightly reducing the computation complexity. Following the algorithm in Ref.\cite{barvinok1995new}, utilizing the selection rules, we have tested that on a laptop computer, computing one hyperdeterminant of a rank-4 tensor for $N=8$ electrons takes about two seconds. This allows us to perform variational Monte Carlo calculations for the projected FCI wavefunctions and compare them with the wavefunctions obtained from exact diagonalization (see Sec.\ref{sec:benchmark}).

\subsection{Connections with Jain's wavefunctions}\label{sec:connection_jain}
We have shown that the projected wavefunction $|\psi^g_e(\psi_{CF})\rangle$ is represented as the hyperdeterminant of a tensor. In this section, we first analytically show that under the open boundary condition and in the traditional LL context, these projected wavefunctions are identical to the ones obtained from Jain's construction. 

We will use the coherent state basis extensively in this section. For this purpose, we define the ladder operators satisfying $[a,a^\dagger]=1$ in the relevant single-particle Hilbert spaces:
\begin{align}
a_e&\equiv\frac{\mathcal R_{e,x}-i\mathcal R_{e,y}}{\sqrt{2} l_e}, &a_v&\equiv\frac{\mathcal R_{v,x}+i\mathcal R_{e,y}}{\sqrt{2} l_v},\notag\\
a_{\mathcal R}&\equiv\frac{\mathcal R_x-i\mathcal R_y}{\sqrt{2} l_{CF}}, &a_{\eta}&\equiv\frac{\mathcal \eta_{x}+i\eta_{y}}{\sqrt{2} l_{CF}}.\label{eq:ladder_op}
\end{align}
The relation Eq.(\ref{eq:CF_substitution}) between $e,v$ and CF $\mathcal R,\eta$ spaces becomes the bosonic Bogoliubov transformation between these ladder operators:
\begin{align}
a_{\mathcal R}&=\frac{1}{\sqrt{1-c^2}}(a_e-c a_v^\dagger),&a_{\eta}&=\frac{1}{\sqrt{1-c^2}}(-c a_e^\dagger+a_v)\notag\\
a_e&=\frac{1}{\sqrt{1-c^2}}(a_{\mathcal R}+c a_{\eta}^\dagger),&a_v&=\frac{1}{\sqrt{1-c^2}}(c a_{\mathcal R}^\dagger+a_{\eta})\label{eq:a_bogoliubov}
\end{align}

These operators and the magnetic translation operators defined in Eq.(\ref{eq:real_momentum},\ref{eq:D_v},\ref{eq:D_R_D_eta}) satisfy the algebra:
\begin{align}
D^\dagger_e(z_0) a_e D_e(z_0) &= a_e + \frac{\bar z_0}{\sqrt{2}l_e}\notag\\
D^\dagger_v(z_0) a_v D_v(z_0) &= a_v + \frac{ z_0}{\sqrt{2}l_v}\notag\\
D^\dagger_{\mathcal R}(z_0) a_{\mathcal R} D_{\mathcal R}(z_0) &= a_{\mathcal R} + \frac{\bar z_0}{\sqrt{2}l_{CF}}\notag\\
D^\dagger_{\eta}(z_0) a_{\eta} D_{\eta}(z_0) &= a_{\eta} + \frac{ z_0}{\sqrt{2}l_{CF}}
\end{align}
Let $|0_\alpha\rangle$ ($\alpha=e,v,\mathcal R,\eta$) be the coherent state annihilated by the $a_\alpha$ operator, the coherent state basis can be obtained via magnetic translation: $|z_\alpha\rangle\equiv D_\alpha(z)|0_\alpha\rangle$. In addition, the occupation number basis can also be defined: $|n_{\alpha}\rangle\equiv\frac{a^{\dagger n}_{\alpha}}{\sqrt{n!}}|0_{\alpha}\rangle$. For instance, the $n$-th CF LL corresponds to $|n_{\eta}\rangle$.

We will work \emph{in the symmetric gauge in this section}. The many-particle wavefunctions can be obtained using the coherent state basis. We focus on the CF space as a demonstration. 
Defining the position basis for CF $|\zeta_{CF}\rangle$ corresponding to a $\delta$-function located at $\zeta$, one may project it into the $n$-th CF LL. After choosing the appropriate normalization factor, it turns out that
\begin{align}
|n_{\eta}\rangle\langle n_{\eta}|\zeta_{CF}\rangle=(-1)^n |n_{\eta}\rangle \cdot D_{\mathcal R}(\zeta)|n_{\mathcal R}\rangle.
\end{align}
Therefore, for a single CF in the $n$-th LL: $|\phi_{CF}\rangle=|\phi_{\mathcal R}\rangle|n_{\eta}\rangle$, its wavefunction is:
\begin{align}
\langle \zeta_{CF} |\phi_{CF}\rangle=(-1)^n \langle n_{\mathcal R}|D_{\mathcal R}(\zeta)^\dagger|\phi_{\mathcal R}\rangle
\end{align}
If $n=0$, the wavefunction is identical to the overlap with the coherent state basis $\langle \zeta_{CF} |\phi_{CF}\rangle=\langle \zeta_{\mathcal R}|\phi_{\mathcal R}\rangle$.

Since the $e$ and $v$ particles only contain the guiding-center d.o.f., they may be viewed as if they are in the LLL:
\begin{align}
\psi_e(z_1,z_2,...,z_N)&=\langle z_{1,e},z_{2,e}... z_{N,e}|\psi_e\rangle,\notag\\
\psi_v(\omega_1,\omega_2,...,\omega_N)&=\langle \omega_{1,v},\omega_{2,v}... \omega_{N,v}|\psi_v\rangle
\end{align}
For instance,the Laughlin state $|\psi^{Laughlin}_{\nu=1/2s,v}\rangle$ in the vortex space is:
\begin{align}
\psi^{Laughlin}_{\nu=1/2s,v}(\omega_1,\omega_2,...,\omega_N)=& \prod_{i<j} (\bar\omega_i-\bar\omega_j)^{2s}  e^{-\frac{\sum_j|\omega_j|^2}{4l_v^2}}\notag\\
\equiv g_v^*(\{\omega_j\}) e^{-\frac{\sum_j|\omega_j|^2}{4l_v^2}},
\end{align}
where we introduced the polynomial $g_v(\{\omega_j\})=\prod_{i<j} (\omega_i-\omega_j)^{2s}$.

Next, we study the many-body CF wavefunction with $p$-filled CF LLs, which is a Slater determinant:
\begin{align}
&\psi^{p-LL}_{CF}(\zeta_1,\zeta_2,\cdots,\zeta_N)=\langle  \zeta_{1,CF},\zeta_{2,CF},\cdots,\zeta_{N,CF} |\psi^{p-LL}_{CF}\rangle\notag\\
&=\left|\begin{array}{llll} \bar\zeta_1^{p-1}&  \bar\zeta_2^{p-1} & ... &  \bar\zeta_N^{p-1}\\
\bar\zeta_1^{p-1}\zeta_1&  \bar\zeta_2^{p-1}\zeta_1 & ... &  \bar\zeta_N^{p-1}\zeta_N\\
... & ...& ...& ...\\
\bar\zeta_1^{p-1}\zeta_1^{M-1}& \bar\zeta_2^{p-1}\zeta_2^{M-1}& ...& \bar\zeta_N^{p-1}\zeta_N^{M-1}\\
\bar\zeta_1^{p-2}&  \bar\zeta_2^{p-2} & ... &  \bar\zeta_N^{p-2}\\
... & ...& ...& ...\\
... & ...& ...& ...\\
\bar\zeta_1^{2}\zeta_1^{M-1}& \bar\zeta_2^{2}\zeta_2^{M-1}& ...& \bar\zeta_N^{2}\zeta_N^{M-1}\\
\bar\zeta_1&  \bar\zeta_2 & ... &  \bar\zeta_N\\
\bar\zeta_1 \zeta_1&  \bar\zeta_2 \zeta_2 & ... &  \bar\zeta_N \zeta_N\\
... & ...& ...& ...\\
\bar\zeta_1 \zeta_1^{M-1}&  \bar\zeta_2 \zeta_2^{M-1} & ... &  \bar\zeta_N \zeta_N^{M-1}\\
1&  1 & ... &  1\\
\zeta_1&  \zeta_2 & ... &  \zeta_N\\
... & ...& ...& ...\\
 \zeta_1^{M-1}&   \zeta_2^{M-1} & ... &   \zeta_N^{M-1}\\
\end{array} \right|\cdot e^{-\frac{\sum_k|\zeta_k|^2}{4l_{CF}^2}}\notag\\
&\equiv g^{p-LL}_{CF}(\{\bar\zeta_k,\zeta_k\})e^{-\frac{\sum_k|\zeta_k|^2}{4l_{CF}^2}} ,\label{eq:p_2_slater}
\end{align}
where $M\equiv N/p$, and $g^{p-LL}_{CF}(\{\bar\zeta_k,\zeta_k\})$ is the polynomial part of the Slater determinant.

In order to perform the projection, it is crucial to compute the fusion coefficient in the $e,v$-coherent state bases $\langle z_e|\langle\omega_v|\zeta_{CF}\rangle$. Using properties of the transformation in Eq.(\ref{eq:a_bogoliubov}), it can be computed:
\begin{align}
\langle z_e|\langle\omega_v|\zeta_{CF}\rangle =& \sqrt{\frac{1+c}{1-c}}e^{-\frac{1}{4l_v^2}|\omega|^2+\frac{1}{2l_v^2}\bar\omega\big(-\frac{z}{c}+\frac{(1+c) \zeta}{c}\big)}\notag\\
&\cdot e^{\frac{-1}{4l_{CF}^2}\frac{1+c}{1-c}\left|\zeta\right|^2+\frac{1}{2l_{CF}^2}\frac{\bar\zeta z}{1-c}}\cdot e^{-\frac{|z|^2}{4l_e^2}}\label{eq:coherent_state_fusion}
\end{align}
We leave its derivation in Appendix \ref{app:coherent_state_fusion}.

Using the resolution of identity in coherent state basis, e.g., $\frac{1}{2\pi l_v^2}\int d\omega |\omega_v\rangle\langle\omega_v|=\mathbf 1$, following Eq.(\ref{eq:projection_ket}), we know that for an \emph{arbitrary} CF wavefunction:
\begin{align}
\psi_{CF}(\{\zeta_k\})\equiv\langle \{\zeta_{k,CF}\}|\psi_{CF}\rangle=g_{CF}(\{\bar\zeta_k,\zeta_k\})e^{-\frac{\sum_k|\zeta_k|^2}{4l_{CF}^2}},
\end{align}
the electronic projected wavefunction is given by:
\begin{align}
\psi_e(\{z_i\})&=\int \prod_j \frac{d\omega_j}{2\pi l_v^2} \prod_k \frac{d\zeta_k}{2\pi l_{CF}^2} \psi^{Laughlin \;*}_{\nu=1/2s,v}(\{\omega_j\})\notag\\
&\cdot {\psi}_{CF}(\{\zeta_k\})\cdot\langle z_{i,e}|\langle\omega_{i,v}|\zeta_{CF,i}\rangle,
\end{align}

One only needs to integrate out the complex variables $\{\omega_j\},\{\zeta_k\}$. Noticing the identity:
\begin{align}
\int \frac{d\omega}{2\pi l^2} \bar\omega^m \omega^n  e^{-\frac{|\omega|^2}{2l^2}+\frac{z\bar\omega}{2l^2}}=\left(2l^2\frac{d}{dz}\right)^m z^n,
\end{align}
one finds that:
\begin{align}
&\psi_e(\{z_i\})=e^{-\frac{1}{4l_e^2}\sum_i |z_i|^2}\notag\\
&\cdot \left.\left[g_v\left(\left\{\frac{1+c}{c}\zeta_i-\frac{z_i}{c}\right\}\right)g_{CF}\left(\left\{\frac{2l_e^2}{1+c}\frac{\partial}{\partial \zeta_i}, \zeta_i\right\}\right)\right]\right|_{\zeta_i=z_i}.\label{eq:integrating_out_rule}
\end{align}
Here, the derivatives $\frac{\partial}{\partial \zeta_i}$ should be moved to the leftmost of the polynomial expression in the second line. The identification $\omega_i=\frac{1+c}{c}\zeta_i-\frac{z_i}{c}$ is anticipated: from Eq.(\ref{eq:CF_substitution}), we know the operator $\zeta_{CF}=\frac{1}{1+c}\mathcal R_e+\frac{c}{1+c}\mathcal R_v$. $\mathcal R_e$ and $\mathcal R_v$ can be viewed as the position operators $z_e$ and $\omega_v$ projected into the LLL, leading to $\zeta=\frac{1}{1+c}\mathcal \omega +\frac{c}{1+c}\mathcal \zeta$.

However, in Jain's prescription, the electronic wavefunction is obtained as:
\begin{align}
&\psi^{\text{Jain}}_e(\{z_i\})=e^{-\frac{1}{4l_e^2}\sum_i |z_i|^2}\notag\\
&\quad\cdot \left.\left[g_v(\{\zeta_i\})g_{CF}\left(\left\{2l_e^2\frac{\partial}{\partial \zeta_i}, \zeta_i\right\}\right)\right]\right|_{\zeta_i=z_i}.\label{eq:integrating_out_rule_Jain}
\end{align}
which is apparently different from Eq.(\ref{eq:integrating_out_rule}). 

In fact, for a generic CF wavefunction $\psi_{CF}$ in the FCI systems, \emph{Jain's prescription and our prescription Eq.(\ref{eq:integrating_out_rule}) indeed give different electronic wavefunctions}! However, for the Galilean invariant CF wavefunction $g_{CF}=g_{CF}^{p-LL}$ in Eq.(\ref{eq:p_2_slater}), the two prescriptions give identical electronic wavefunctions, which we show below.

One may rewrite Eq.(\ref{eq:integrating_out_rule}) and Eq.(\ref{eq:integrating_out_rule_Jain}) in an equivalent fashion:
\begin{align}
&\psi_e(\{z_i\})=e^{-\frac{1}{4l_e^2}\sum_i |z_i|^2}\notag\\
&\cdot \left.\left[g_v(\{\omega_i\})g_{CF}\left(\left\{\frac{2l_e^2}{c}\left(\frac{\partial}{\partial \omega_i}+\frac{c}{1+c}\frac{\partial}{\partial \zeta_i}\right),\zeta_i\right\}\right)\right]\right|_{\zeta_i=\omega_i=z_i},\notag\\
&\psi^{\text{Jain}}_e(\{z_i\})=e^{-\frac{1}{4l_e^2}\sum_i |z_i|^2}\notag\\
&\cdot \left.\left[g_v(\{\omega_i\})g_{CF}\left(\left\{2l_e^2\left(\frac{\partial}{\partial \omega_i}+\frac{\partial}{\partial \zeta_i}\right), \zeta_i\right\}\right)\right]\right|_{\zeta_i=\omega_i=z_i}.
\end{align}
Due to the particular form of $g^{p-LL}_{CF}$ in Eq.(\ref{eq:p_2_slater}), it is easy to see that the derivatives $\frac{\partial}{\partial \zeta_i}$ can be neglected in both expressions. Consequently, up to the unimportant overall normalization factor, they give identical electronic wavefunctions. Essentially, any term involving $\frac{\partial}{\partial \zeta_i}$ would appear in the determinant as $(\frac{\partial}{\partial\omega_i})^m (\frac{\partial}{\partial\zeta_i})^n \zeta_i^l$. Varying $i$, these terms form a row in the determinant, which will be canceled by another row formed by $(\frac{\partial}{\partial\omega_i})^m (\frac{\partial}{\partial\zeta_i})^{n-1} \zeta_i^{l-1}$, unless $n=0$.

A nice feature of the current projective construction is the clarification of the Gaussian factor: in Jain's construction, the Gaussian factor $e^{-\frac{|z|^2}{4l_e^2}}$ is attached to the CF mean-field state, which is physically alarming since the CF should have the magnetic length $l_{CF}$, not $l_e$. \emph{In the current construction, the Gaussian factor is indeed $e^{-\frac{|\zeta|^2}{4l_{CF}^2}}$ for the CF mean-field state. Only after the projection the Gaussian factor $e^{-\frac{|z|^2}{4l_e^2}}$ emerges}.

Finally, we comment on the torus boundary condition. In this case, Jain's construction becomes more sophisticated \cite{pu2017composite}, and we do not pursue the analytical relationship with the present construction. However, Laughlin's wavefunctions are still well-known and represented using the Jacobi's $\vartheta$-function \cite{haldane1985periodic}. We have numerically tested that, for small system sizes with $N=3,4$ electrons, the projected $|\psi^g_e(\psi_{CF})\rangle$ is indeed identical to one of $m$-fold degenerate Laughlin's wavefunction. Here, a technical detail needs to be clarified: in order to define the CF LLL, one needs to define the coherent state $|0_{\eta}\rangle$ on a finite-size torus corresponding to $\mathcal H_{\eta}$. It is known that there exist different definitions of coherent states on a finite torus. We find that one needs to use the so-called \emph{continuous coherent state} \cite{fremling2014coherent} for $|0_{\eta}\rangle$ in order to reproduce Laughlin's wavefunctions on the torus after projection. We leave the discussion for the continuous coherent state on torus in Appendix \ref{app:ccs}.

\subsection{Connections with parton states for FCI systems}\label{sec:connection_eft}
Previously, there have been efforts to write down FCI wavefunctions using the parton construction \cite{lu2012symmetry}. For example, in order to construct a $\nu=1/3$ FCI state with the same topological order as the Laughlin's state, one splits the electron $c$ into three fermionic partons $f^{(\alpha)}$ ($\alpha=1,2,3$) in the real-space \cite{jain1989incompressible}:
\begin{align}
c_{\bm r}=f^{(1)}_{\bm r}f^{(2)}_{\bm r} f^{(3)}_{\bm r}. \label{eq:parton}
\end{align}
Each parton carries $1/3$ of the electron's charge and transform projectively under the crystalline symmetry group. It is then possible to have each $f^{(i)}$ to fill a Chern band with Chern number $C=-1$. The electron wavefunction after the identification Eq.(\ref{eq:parton}) is obviously a product of three Slater determinants $\psi_{f^{(i)}}$:
\begin{align}
\psi_e(\bm r_1,\bm r_2,... ,\bm r_N)=\prod_{\alpha=1}^3\psi_{f^{(\alpha)}}(\bm r_1,\bm r_2,... ,\bm r_N),\label{eq:parton_wavefunc}
\end{align}
where $\psi_{f^{(\alpha)}}=\text{det}[\phi^{(\alpha)}_i(\bm r_j)]$ is formed by the wavefunction of the filled parton orbitals $|\phi^{(\alpha)}_i\rangle$.

Although this parton construction is conceptually useful in classifying symmetry fractionalization, as focused in Ref.\cite{lu2012symmetry}, it has difficulty dealing with practical microscopics. One problem is that the electronic wavefunction $\psi_e$ in Eq.(\ref{eq:parton_wavefunc}) does not need to be within the electronic CB. In the regime where electronic band mixing can be neglected, such as the system in Eq.(\ref{eq:CB_H_e}), one would need another projection $\mathbf P_{CB}$ to project $\psi_e$ into the electronic Chern band. A related problem is that the construction Eq.(\ref{eq:parton}) involves the real-space wannier orbitals, which necessarily go beyond a single electronic CB. From the practical variational wavefunction viewpoint, this construction involves an unnecessarily large number of fictitious degrees of freedom. 

First, we would like to point out that, after projection to CB, $\mathbf P_{CB}|\psi_e\rangle$ is nothing but a hyperdeterminant. Introducing the fusion tensor:
\begin{align}
\langle \phi^e_i| \phi^{(1)}_j, \phi^{(2)}_k , \phi^{(3)}_l\rangle\equiv \int d\bm r \langle \phi^e_{i}|\bm r\rangle \langle \bm r| \phi^{(1)}_j\rangle\langle \bm r| \phi^{(2)}_k\rangle\langle \bm r| \phi^{(3)}_l\rangle,
\end{align}
where $ \phi^e_i$'s are a collection of electronic orbitals in the CB, one can easily show that the following overlap is the hyperdeterminant of this tensor:
\begin{align}
\langle\phi^e_{1},\phi^e_{2},...\phi^e_{N}|\mathbf P_{CB}|\psi_e\rangle=\text{Hyperdet}[\langle \phi^e_i| \phi^{(1)}_j, \phi^{(2)}_k , \phi^{(3)}_l\rangle]
\end{align}

Second, we want to mention that the current projective construction is \emph{not} equivalent to the usual parton construction for Jain's series in the absence of Galilean invariance. This is mostly easily seen in the disc geometry using the symmetric gauge, as discussed in Sec.\ref{sec:connection_jain}. In the usual construction, one would consider the $(2s+1)$ fermionic partons: $f^{(1)}_{\bm r}$ ($i=1,2,...,2s+1$). The first $(2s)$ partons each carry $p/(2ps+1)$ of the electron's charge at $\nu=1$, while the last parton carries $1/(2ps+1))$ of the electron's charge at $\nu=p$ \cite{jain1989incompressible}:\\
\indent $\bullet$ \emph{usual parton construction}:
\begin{align}
&c_{\bm r}=f^{(1)}_{\bm r}f^{(2)}_{\bm r}... f^{(2s)}_{\bm r} f^{(2s+1)}_{\bm r}.\label{eq:usual_parton}
\end{align}
Putting the first $(2s)$ partons each in the lowest LL, and putting the last parton in $p$ CF Chern bands, we have the wavefunction for each parton as:
\begin{align}
\psi_{f^{(\alpha)}}(\{\omega_j\})=&e^{-\frac{p\sum_j|\omega_j|^2}{4(2ps+1)l_e^2}} \prod_{i<j}(\omega_i-\omega_j), \;\;\alpha=1,2,...,2s.\notag\\
\psi_{f^{(2s+1)}}(\{\zeta_k\})=&e^{-\frac{\sum_k|\zeta_k|^2}{4(2ps+1)l_e^2}} g_{CF}(\{\bar\zeta_k,\zeta_k\})
\end{align}
The usual parton construction Eq.(\ref{eq:usual_parton}) identifies $\omega_i=\zeta_i=z_i$, where $z_i$'s are the coordinates of electrons. After projecting into the LLL of the electrons, one reproduces the Jain's prescription in Eq.(\ref{eq:integrating_out_rule_Jain}):
\begin{align}
\psi^{\text{usual-parton}}_e(\{z_i\})=\psi^{\text{Jain}}_e(\{z_i\}).
\end{align}
As discussed earlier, this is not the same wavefunction obtained by the current projective construction in Eq.(\ref{eq:integrating_out_rule}). This brings up an interesting question: Is there a real-space prescription similar to Eq.(\ref{eq:usual_parton}) to obtain the projected wavefunction in the current construction? 

It turns out that the projected wavefunction in Eq.(\ref{eq:integrating_out_rule}) can be obtained using the following real-space prescription:\\
\indent $\bullet$ \emph{current construction}:
\begin{align}
&c_{z}=\int d\zeta e^{\frac{-(1+c)}{4l_e^2}(\bar\zeta z-\bar z\zeta)} f^{(1)}_{\frac{1+c}{c}\zeta-\frac{z}{c} }f^{(2)}_{\frac{1+c}{c}\zeta-\frac{z}{c}}... f^{(2s)}_{\frac{1+c}{c}\zeta-\frac{z}{c}} f^{(2s+1)}_{\zeta}, \label{eq:current_parton}
\end{align}
where we used complex numbers to label the positions of particles. Again the identification $\omega_v=\frac{1+c}{c}\zeta_{CF}-\frac{1}{c}z_e$ is involved, where $\omega_v$ is the position of the first $(2s)$ partons that corresponds to the vortex. Different from the usual parton construction Eq.(\ref{eq:usual_parton}), the electron operator is \emph{not} the on-site combination of parton operators. 

\section{Benchmark results}\label{sec:benchmark}
\subsection{Models}
We will study two models in this section: a toy model describing Landau levels mixed due to a periodic potential with $C_4$ rotation symmetry and a realistic model for twisted bilayer MoTe$_2$ with $C_3$ rotation symmetry.
\subsubsection{Mixed Landau level (MLL) model}
Landau levels inherently possess a uniform distribution of Berry curvature with Chern number $C=-1$. One way to introduce the FCI physics into the system is to turn on a periodic potential $V_{\text{pp}}(\mathbf{r}_e)\equiv\sum_{\mathbf{G}_e} V_{\text{pp}}(\mathbf{G}_e)e^{i\mathbf{G}_e\cdot\mathbf{r}_e}$ between LLs; e.g. between $n=0$ and $n=1$ LLs. Such an LL-based model, named \emph{mixed Landau level model}, or MLL, is previously introduced in Ref.\cite{murthy2012hamiltonian}. The periodic potential has three effects: it mixes the LLs and thus modifies the Coulomb interaction projected into the lowest Chern band, which has a nonzero bandwidth, and leading to a non-constant distribution of Berry curvature. 

For illustrative purposes, we consider the square lattice potential and keep only the lowest harmonics of it, corresponding to $\pm G_{1,e},\pm G_{2,e}$ with the constant coefficient $V_{\text{pp}}(\mathbf{G}_e)=V_{10}/2$. The matrix element of periodic potential in the bloch basis between $n_1,n_2=0,1$ LLs then reads
\begin{align}
    &\frac{V_{10}}{2}\langle n_1|e^{i\mathbf{G}_e\cdot\eta_e}|n_2\rangle \langle\mathbf{k}_e|e^{i\mathbf{G}_e\cdot\mathbf{R}_e}|\mathbf{k}_e\rangle\notag\\
    &=\frac{V_{10}}{2}\rho_{n_2,n_1}(\mathbf{G}_e) e^{-\frac{i\ell_e^2}{2}G_{e,x} G_{e,y}} e^{i\ell_e^2 (G_{e,x}k_{e,y}-G_{e,y}k_{e,x})},
\end{align}
where the cyclotron part of the matrix element takes the form of (see, for example, the Appendix of Ref.\cite{murthy2003hamiltonian})
\begin{align*}
&\langle n_1|e^{i\mathbf{G}_e\cdot\eta_e}|n_2\rangle\equiv\rho_{n_2,n_1}(\mathbf{q}_e)\notag\\
&=e^{-\frac{q_e^2\ell_e^2}{4}}\sqrt{\frac{n_<!}{n_>!}} L_{n<}^{|n_1-n_2|}\left[\frac{q_e^2\ell_e^2}{2}\right]\begin{cases}
    (\tfrac{i\ell_e\bar z_{q_e}}{\sqrt{2}})^{|n_1-n_2|},&n_1>n_2\\
    (\tfrac{i\ell_e z_{q_e}}{\sqrt{2}})^{|n_1-n_2|}, &n_1\leq n_2\\
\end{cases}
\end{align*}
%with $z_{q,e}\equiv q_{e,x}+iq_{e,y}$ 
and the guiding-center matrix element follows from the guiding-center algebra Eq.\eqref{eq:e_guiding_center}. As a result, the MLL model is represented in the LL bloch basis \cite{murthy2012hamiltonian} as 
\begin{align}\label{eq:MLL effecitve Hamiltonian}
    &H=\notag\\
    &\begin{bmatrix}
        \frac{\widetilde{V}}{\sqrt\pi}(\cos k_{e,x}+\cos k_{e,y}) & \widetilde{V}(-i\sin k_{e,x}-\sin k_{e,y}) \\[0.5em]
        \widetilde{V}(i\sin k_{e,x}-\sin k_{e,y}) & \omega+\frac{\pi-1}{\sqrt{\pi}}\widetilde{V}(\cos k_{e,x}+\cos k_{e,y}) 
    \end{bmatrix},
\end{align}
where we define $\widetilde{V}\equiv e^{-\pi/2}\sqrt{\pi}V_{10}$, and the LL separation $\mathop{\mathrm{diag}}\{0, \omega\}$ is inserted. The MLL model Eq.\eqref{eq:MLL effecitve Hamiltonian} stays in a topological regime with Chern number $C=-1$, as long as the periodic potential satisfies $V_{10}\leq V_{10}^c\equiv\frac{e^{\pi/2}}{2\pi}\omega\approx0.766\omega$.

We consider the bare Coulomb interaction $V(\mathbf{q}_e)=\frac{2\pi e^2}{\epsilon |\mathbf{q}_e|}$. This interaction is projected to the lowest Chern band, and we fix the LL separation to be $\omega=2$ in unit of $\frac{e^2}{\epsilon l_e}$, leaving periodic potential $V_{10}$ as the only remaining tunable parameter. When $V_{10}=0$ the system returns to the traditional LL problem, whose many-body gap (magnetoroton gap) has been estimated $\sim 0.066$ in units of $\frac{e^2}{\epsilon l_e}$ \cite{balram2017positions}, in the thermodynamic limit. When $V_{10}$ is large enough, a gap-closing quantum phase transition is expected.

\subsubsection{TMD moiré (tMoTe$_2$) model}
A very recent experimental progress on searching for the zero-field FCI phase is the reported realization in R-stacked twisted $\mathrm{MoTe_2}$ \cite{cai2023signatures,zeng2023thermodynamic,park2023observation}. A realistic continuum model is to consider the $\mathbf{K}$ valley moir\'e Hamiltonian \cite{wu2019topological,wang2023fractional}
\begin{equation}\label{eq: tMoTe2 moiré Hamiltonian}
    H_\mathbf{K}=\begin{bmatrix}
        h_b(\bm r) & T(\bm r)\\
        T^\dagger(\bm r) & h_t(\bm r)
    \end{bmatrix},
\end{equation}
where $h_{b/t}(\bm r)\equiv-\hbar^2(\mathbf k_e-\mathbf K_{b/t})^2/2m^*+\Delta_{b/t}(\bm r)$ is the top/bottom layer Hamiltonian subject to the moir\'e potential $\Delta_{b/t}(\bm r)=2v\sum_{i=1,3,5}\cos(\mathbf g_i\cdot\bm r\pm\psi)$, and $T(\bm r)\equiv w(1+e^{-i\mathbf g_2\cdot\bm r}+e^{-i\mathbf g_3\cdot\bm r})$ is the interlayer tunnelings. Here $m^*\approx0.6m_e$ is the effective mass, $\mathbf g_i=\frac{4\pi}{\sqrt{3}a_M}(\cos\frac{\pi(i-1)}{3}, \sin\frac{\pi(i-1)}{3})$ are moir\'e reciprocal vectors with $a_M\simeq a/\theta=3.52\text{\AA}/3.89^\circ$, and $v=20.8$meV, $\psi=+107.7^\circ$, $w=-23.8$meV are parameters extracted from the large-scale DFT study of $\mathrm{tMoTe_2}$ \cite{wang2023fractional}, different from that obtained by fitting different stacking regions in Ref.\cite{wu2019topological}. The moir\'e Hamiltonian for $\mathbf{K'}$ valley can be obtained via time-reversal transformation: $H_{\mathbf{K'}}=[H_\mathbf{K}(\mathbf k_e\rightarrow -\mathbf k_e)]^*$. Due to the spin-orbit coupling, the spin and valley degrees of freedom are locked: spin up for $H_\mathbf{K}$ and spin down for $H_{\mathbf{K'}}$.

The observed FCI states appear at filling $\nu=-2/3,-3/5$ in the presence of ferromagnetic order. In the mean-field picture, this means that the topmost band for the minority spin is at filling $\nu=1/3,2/5$, while the majority spin bands are fully filled. This topmost Chern band has a bandwidth $\sim 9$meV using the above parameters in Eq.(\ref{eq: tMoTe2 moiré Hamiltonian}) in the absence of the Coulomb interaction introduced below. 

We consider the Coulomb interaction $\mathrm{tMoTe_2}$ model:
\begin{align}
H_U\equiv \frac{1}{2A}\sum_{\mathbf q_e} V(\mathbf q_e):\boldsymbol\rho(\mathbf q_e)\boldsymbol\rho(-\mathbf q_e):\label{eq:Coulomb}
\end{align}
where the electron density $\boldsymbol\rho$ is the summation from both valleys and both layers. $V(\mathbf q_e)$ is the dual-gate screening Coulomb interaction
\begin{equation}
    V(\mathbf q_e)=\frac{e^2\tanh(|\mathbf q_e|d)}{2\varepsilon_0\varepsilon_r|\mathbf q_e|}
\end{equation}
with a typical gate distance $d=300$\AA~and the relative dielectric constant $\varepsilon_r=15$, as is used in Ref.\cite{wang2023fractional}.

In our calculation, we assume the existence of ferromagnetism and project the Coulomb interaction Eq.\eqref{eq:Coulomb} into the spin minority topmost Chern band. Note that the Chern number $C=+1$ ($C=-1$) for this topmost band of $H_\mathbf{K}$ ($H_{\mathbf{K'}}$). To map to the $C=-1$ LLL, we simulate the case $H_{\mathbf{K'}}$ being partially filled (i.e., \emph{the minority spin is down spin}).

It is known that upon hole doping, the bandwidth of this Chern band is significantly renormalized from $\sim 9$meV to a smaller value due to the Coulomb interactions between this band and other filled bands, as shown in Ref.\cite{dong2023composite}. Another way to see this effect is to perform a particle-hole transformation, as done in Ref.\cite{wang2023fractional}.

Here, as a benchmarking exercise, we are motivated to investigate the effect of bandwidth in the FCI system. Therefore, we choose the bandwidth via a tuning parameter $\lambda$ instead of fixing a specific value. Precisely speaking, we simulate the model:
\begin{equation}
    H=\lambda P H_{\mathbf{K'}} P + P H_U P \label{eq:simulated_model}
\end{equation}
with the projection operation $P$ eleminates any fermion operator $c_k$ or $c^\dagger_k$ outside the partially filled Chern band. When $\lambda=0$ the CB is completely flat.

\subsection{Exact diagonalization, Hartree-Fock and time-dependent Hartree-Fock}
\subsubsection{CF mean-field ground states}

We construct both the MLL model and the $\mathrm{tMoTe_2}$ model on $6\times4$, $6\times6$, and $9\times 9$ samples at the same filling fraction $\nu=1/3$. The original models have trivial periodic boundary conditions. However, to map to the LLL, for the $9\times 9$ sample of the $\mathrm{tMoTe_2}$ model, twisted boundary conditions in the LLL $\varphi_{1,e}=\varphi_{2,e}=\pi$ are introduced, due to the identification of the operator algebra in Eq.(\ref{eq:CB_map_LLL_symmetry}). Other samples have trivial periodic boundary conditions after mapping to the LLL $\varphi_{1,e}=\varphi_{2,e}=0$. In the projected wavefunction simulations,  we always choose the vortex space $\mathcal H_v$ to have trivial periodic boundary conditions: $\varphi_{1,v}=\varphi_{2,v}=0$.

\begin{figure}[!htp]
    \centering
    \includegraphics[width=0.5\textwidth]{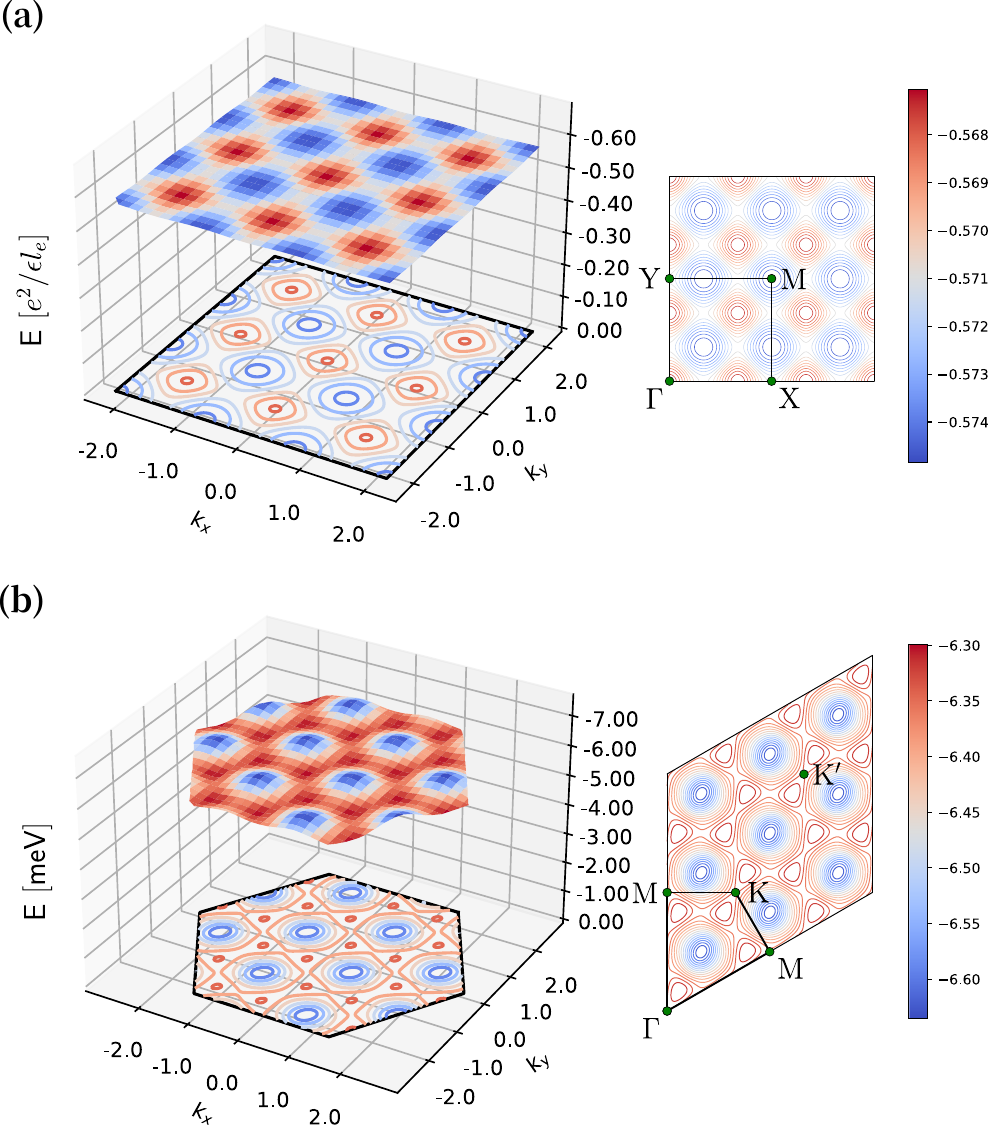}
    \caption{Filled CF band dispersion obtained using the Hartree-Forck approximation. The calculation is done for MLL model (a) with parameter $V_{10}=0.12$ units of $\frac{e^2}{\epsilon\ell_e}$ and $\mathrm{MoTe_2}$ model (b) with parameter $\lambda=0.6$ on a $9\times9$ lattice. These Hartree-Fock CF bands turn out to be \emph{nearly flat}. As is shown in Eq.\eqref{eq: translation fractionalization}, a three-fold periodicity of CF band dispersion emerges as a manifestation of the translation-symmetry fractionalization, as is seen in both subfigures on the right. Note that we have used the \emph{electron's Brillouin Zone} (BZ) to plot the CF dispersion for better visualization of the symmetry fractionalization (in the tMoTe$_2$ model the electron's BZ is the moiré BZ.). The CF's BZ should be 1/3 of the electron's BZ due to the enlarged real-space unit cell along the $\mathbf a_1$ direction.}
    \label{fig: CF dispersion}
\end{figure}

For both samples of $6\times 6$ and $9\times9$ unit cells, the system sizes are consistent with the $D_{\mathcal R}(\mathbf a_{i,e})$ projective translational symmetry in both directions as well as the projective $C_n$ rotation symmetry for the CF ($n=4$ for the MLL model and $n=3$ for the tMoTe$_2$ model). Particularly, Eq.\eqref{eq: translation fractionalization} tells that CF dispersion will display a $3$-fold periodicity in the CF Brillouin Zone (BZ), a well-known feature due to the translation-symmetry fractionalization. We perform Hartree-Fock self-consistent study for the $9\times9$ sample to obtain the composite-fermion band dispersion (see FIG.\ref{fig: CF dispersion} for the filled CF Chern band). 

\subsubsection{Overlap between projected wavefunctions and ED ground states}
\begin{figure}
    \centering
    \includegraphics[width=0.9\linewidth]{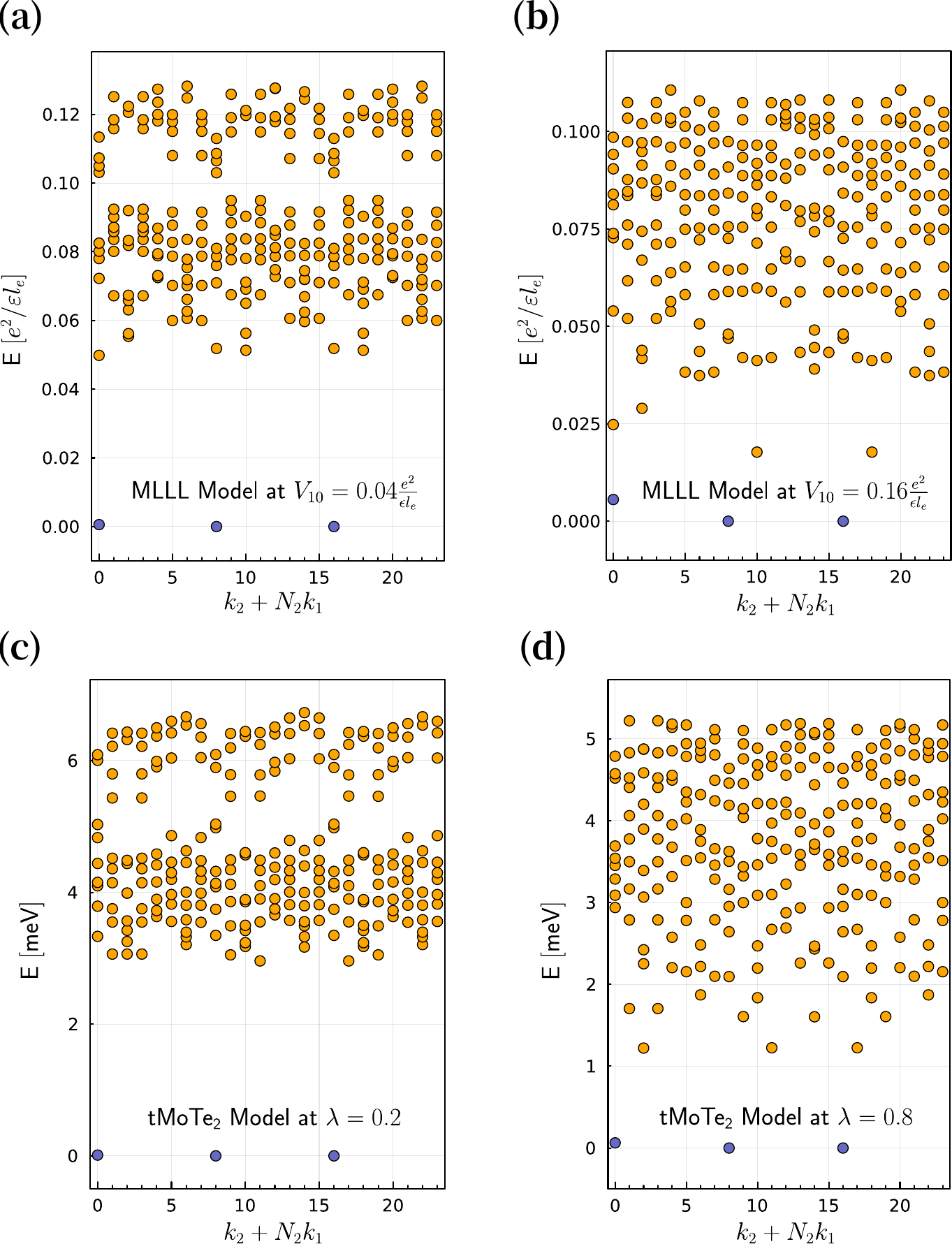}
    \caption{Selected many-body spectrum obtained from exact diagonalization on $6\times4$ unit cells. (a),(b) for MLL model and (c),(d) for $\mathrm{tMoTe_2}$ model. (a) and (c) are close to the flat band limit, while (b) and (d) have much smaller many-body gaps. The three nearly degenerate topological ground states are highlighted in dark purple.}
    \label{fig: ED}
\end{figure}

We perform exact diagonalization (ED) on the sample of $6\times4$ unit cells with the tuning parameter: the periodic potential $V_{10}$ for MLL model, and the band scaling factor $\lambda$ for $\mathrm{tMoTe_2}$ model. The many-body spectra for selected parameter values are shown in FIG.\ref{fig: ED}. As the parameter is large enough, we observe a gap-closing phase transition ($V_{10}\sim 0.18\frac{e^2}{\epsilon l_e}$ for the MLL model and $\lambda\sim 1.0$ for the tMoTe$_2$ model).

Both the Laughlin state and our proposed projected wavefunction (in the form of \emph{combinatorial hyperdeterminants}) can be obtained by projecting the CF states back to the electronic many-body Fock space. The only difference is that for Laughlin's state \emph{non-optimized} mean-field states are used (corresponding to fully filled CF LLL), while for our projective construction, the Hartree-Fock self-consistent mean-field states $|\Psi^{MF}_{CF}\rangle$ are used (corresponding to fully filled lowest-energy CF Chern band). 

Due to the $6\times4$ system size, the projected wavefunction $\mathbf P_g|\Psi^{MF}_{CF}\rangle$ is not translational symmetric along the $\mathbf a_{1,e}$ direction with $6$ unit cells. Namely, $\mathbf P_g|\Psi^{MF}_{CF}\rangle$ is a superposition of sectors with the center of mass (COM) crystalline momentum at $\Gamma$, $\frac{1}{3}\mathbf G_{1,e}$ and $\frac{2}{3}\mathbf G_{1,e}$. When we perform the overlap calculation with the three-fold ground states obtained from ED at these three COM momenta, we use the corresponding COM sector of the same projected wavefunction $\mathbf P_g|\Psi^{MF}_{CF}\rangle$. 

\begin{figure*}
    \centering
    \includegraphics[width=\textwidth]{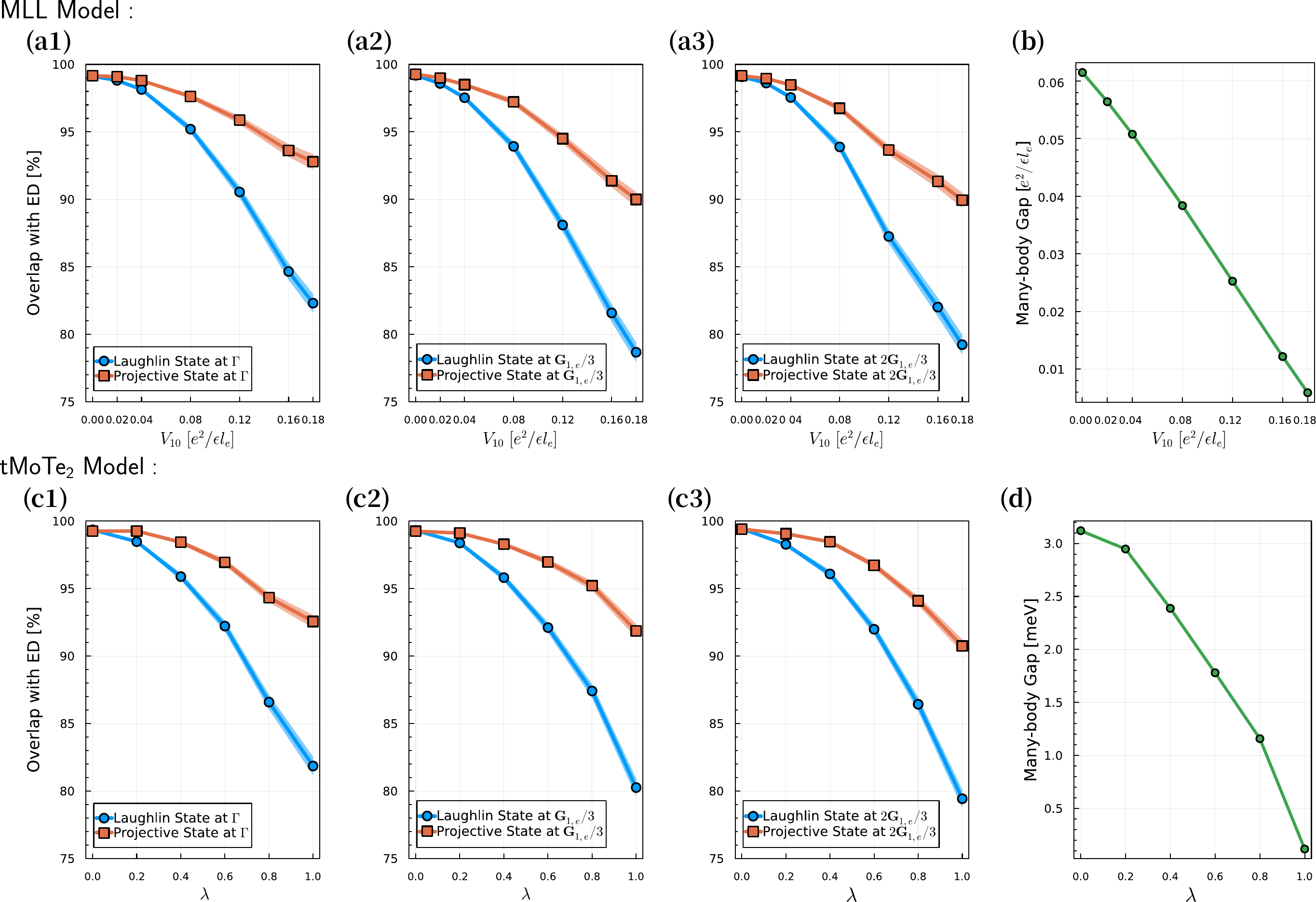}
    \caption{(a1)-(a3) and (c1)-(c3) exhibit the overlaps between our eletronic hyperdeterminant states and the ED ground states (red line), and the the overlaps between Laughlin's states and the ED ground states (blue line) for both MLL model and $\mathrm{tMoTe_2}$ model. Here, the overlap is defined as $|\langle \psi_{\text{ED}}|\psi_{\text{Proj}}\rangle|$ where the two wavefunctions are obtained from ED and the projective construction respectively. There exists three sectors of ground states carrying center-of-mass momentum $\Gamma, \frac{1}{3}\mathbf G_{1,e}$, and $\frac{2}{3}\mathbf G_{1,e}$ (the dark purple points in FIG.\ref{fig: ED}). The ribbon around each overlap curve represents the error bar due to variational Monte Carlo samplings. The many-body gap (i.e., the energy difference between the ground state manifold and the first excited state.) as a function of the tuning parameter is plotted in (b) and (d).}
    \label{fig: ED overlap}
\end{figure*}

It turns out that, \emph{this projective construction outperforms Laughlin's states across the entire parameter space for all the three COM sectors}, as is shown in Fig.\ref{fig: ED overlap}. Notice that our optimization is performed only for the CF mean-field ground states, \emph{not} on the level of the projected electronic wavefunctions. These benchmark results indicate the present projective construction can indeed capture the microscopics of the FCI states.

\subsubsection{Magnetoroton spectra and quantum numbers}
For the samples of $6\times 6$ and $9\times9$ unit cells, we obtain the magnetoroton spectra using the time-dependent Hartree-Fock (TDHF) approximation, where eigenmodes come into pairs $\pm\hbar\omega_a(\mathbf q_e)$, with $a$ labels the magnetoroton band. The positive bands correspond to excitations above the ground state. In our TDHF calculation a nearly dispersionless CF particle-hole (PH) continuum in both models is observed, consistent with the nearly flat mean-field CF bands. This PH continuum occurs at energy $\sim 0.23 \frac{e^2}{\epsilon l_e}$ for the MLL model at $V_{10}=0.04 \frac{e^2}{\epsilon l_e}$, and at energy $\sim 11.2$meV for the tMoTe$_2$ model at $\lambda=0.2$. 

\begin{figure*}[htp!]
    \centering
    \includegraphics[width=0.95\textwidth]{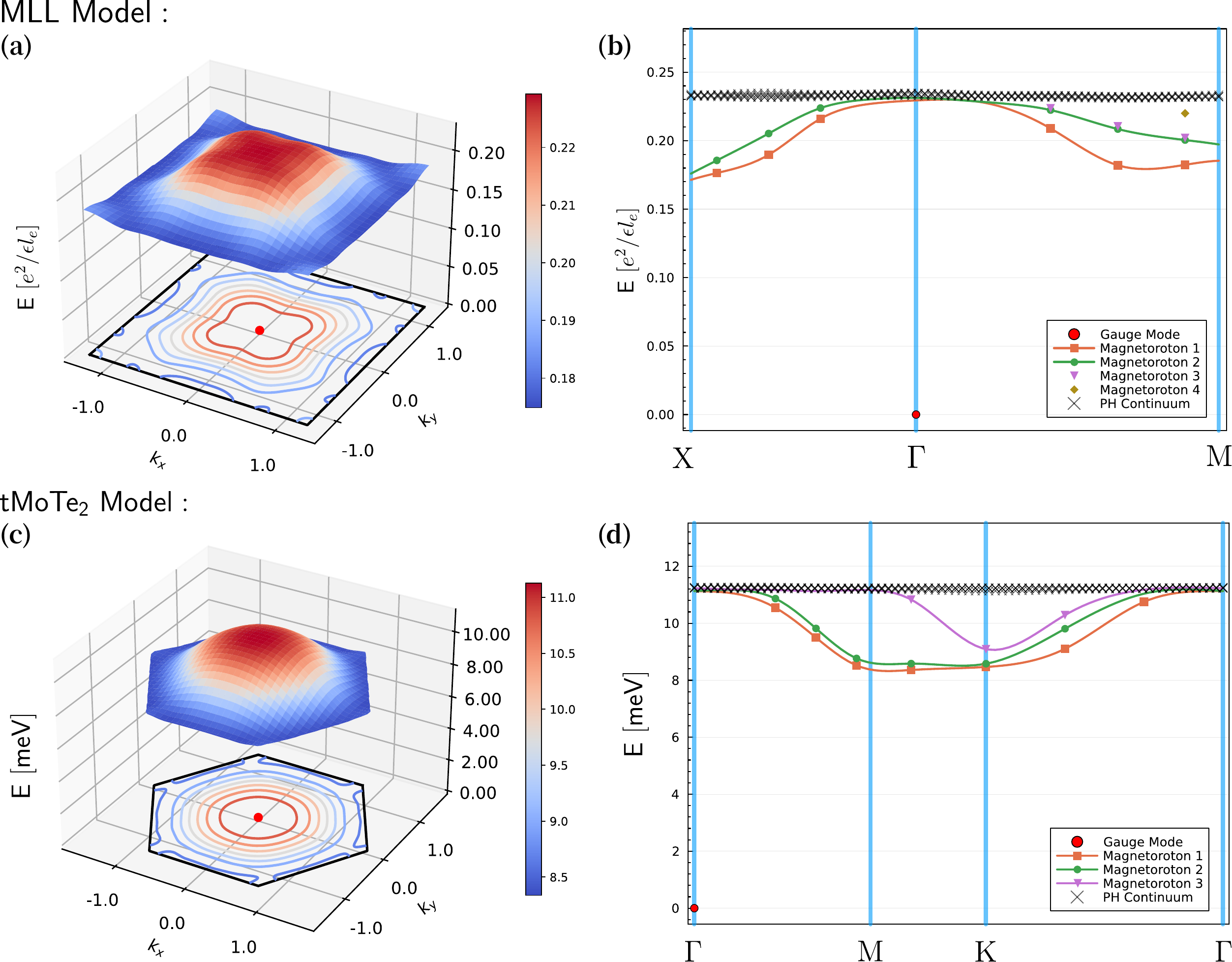}
    \caption{Magnetoroton bands for samples of $9\times9$ unit cells obtained using the TDHF approximation. (a), (b) are for the MLL model with the parameter $V_{10}=0.04$ units of $e^2/\epsilon l_e$, and (c), (d) are for the $\mathrm{tMoTe_2}$ model with the parameter $\lambda=0.2$. (a) and (c) are the fitted 2D surface contour of the lowest magnetoroton band (band-1), while (b) and (d) exhibit the full magnetoroton spectra, where scattering points represent the raw data along the path, and the higher horizontal crosses are from the particle-hole (PH) continuum. See Fig.\ref{fig: CF dispersion} for the definitions of the high-symmetry points. Note: there is no raw data point along the $\mathrm{X}$-$\mathrm{M}$ line in the MLL model due to the choice of $9\times9$ sample size. We provide the raw data for a smaller $6\times6$ MLL sample in the Appendix. \ref{app:extended_data} for comparison. We do not perform band fitting for the magnetoroton band-3 and band-4 for the MLL model due to the limited number of points outside the PH continuum. The appearance of the nearly zero-energy modes (red dots in all four figures) at the $\Gamma$-point is due to the $\mathcal R_v$ gauge degrees of freedom discussed in the main text near the end of Sec.\ref{sec:HF_TDHF}.}
    \label{fig: magnetoroton spectrum}
\end{figure*}

Below the PH continuum, we observe four (three) branches of magnetoroton bands for the MLL (tMoTe$_2$) model. We plot the magnetoroton bands $\omega_a(\mathbf q_e)$ from the TDHF calculation in FIG.\ref{fig: magnetoroton spectrum}. In both models, \emph{the lowest energy magnetorotons are found near the BZ boundary}. The high energy magnetoroton bands (i.e., band-3 and band-4 for the MLL model and band-3 for the tMoTe$_2$ model) are visible below the PH continuum only in a small region of the BZ. Even the lowest magnetoroton band (band-1) merges into the PH continuum near the $\Gamma$-point. The rotation eigenvalues for the high-symmetry points of the magnetoroton bands are computed in Table.\ref{tab: rotation eigvals}.

\begin{table}[htp!]
    \centering
    \begin{tabular}{|| c | c | p{1.5cm} | p{1.5cm} | p{1.5cm}}
        % \hline
        % MLL model & $\Gamma:i$ & $\text{X or Y}:-1$ & $\mathrm{M}:-i$ \\
        % \hline
        % $\mathrm{tMoTe_2}$ model & $\Gamma:e^{i2\pi/3}$ & $\mathrm{K}:1$ & $\mathrm{K'}:1$ \\
        % \hline\hline
        \hline
            & band \# & $\Gamma$      & $\mathrm{X}$ or $\mathrm{Y}$  & $\mathrm{M}$\\ 
        \hline
        MLL model & 1 & \diagbox[innerwidth=1.5cm, height=\line]{}{} & $-1$ & $-i$ \\
            & 2 & \diagbox[innerwidth=1.5cm, height=\line]{}{} & $1$ & $1$ \\
            & 3 & \diagbox[innerwidth=1.5cm, height=\line]{}{} & \diagbox[innerwidth=1.5cm, height=\line]{}{} & $-1$ \\
            & 4 & \diagbox[innerwidth=1.5cm, height=\line]{}{} & \diagbox[innerwidth=1.5cm, height=\line]{}{} & $i$ \\
        \hline\hline
            & band \# & $\Gamma$      & $\mathrm{K}$  & $\mathrm{K'}$            \\
        \hline
        $\mathrm{tMoTe_2}$ model & 1 & \diagbox[innerwidth=1.5cm, height=\line]{}{} & $1$           & $1$           \\
            & 2 & \diagbox[innerwidth=1.5cm, height=\line]{}{} & $e^{-i2\pi/3}$ & $e^{-i2\pi/3}$ \\
            & 3 & \diagbox[innerwidth=1.5cm, height=\line]{}{} & $e^{i2\pi/3}$ & $e^{i2\pi/3}$  \\
        \hline\hline
    \end{tabular}
    \caption{Rotation eigenvalues for high-symmetry points of TDHF bands. For MLL model, unlike $\Gamma$ or $\mathrm{M}$ points, the rotation eigenvalue for X and Y points are defined for $C_2$ rotation instead of $C_4$. For $\mathrm{tMoTe_2}$ model, all high symmetry points are defined for $C_3$ rotation, and the magnetorotons at $\mathrm{K}$ and $\mathrm{K'}$ points are related by the $C_{2y}\mathcal T$ symmetry \cite{wu2019topological} and thus have identical eigenvalues. Note: when the magnetoroton band merges into the particle-hole(PH) continuum, the corresponding rotation eigenvalue is not presented since one cannot separate the magnetoroton state from the PH continuum. See Fig.\ref{fig: magnetoroton spectrum} for details.}
    \label{tab: rotation eigvals}
\end{table}

We find that the energy scale of the magnetoroton excitations obtained using the TDHF approximation is larger than the excitation energy scale obtained from ED (by a factor $\sim 3$ in both models for the parameters chosen in Fig.\ref{fig: magnetoroton spectrum}). Performing the projection $\mathbf P_g$ is expected to improve the energetics significantly. Due to the complexity of the calculation, we leave computing the projected magnetoroton energies as a future project.

Finally, for the models studied here, no CF LL band inversion is observed. Namely, the FCI quantum phase remains adiabatically connected to the traditional Laughlin's wavefunctions. It would be very interesting to identify and simulate a model where CF band inversion actually occurs. We again leave this as a future direction.

\section{Discussion and Conclusions}\label{sec:conclusion}
In this paper, we present a general projective construction for the composite fermion states in a partially filled Chern band with Chern number $\pm 1$. In the context of the traditional fractional Quantum Hall liquids, the current construction clarifies a few physical puzzles and unifies several previous studies. In the context of FCI, the current construction paves a route to extract important and experimental relevant microscopic information for the FCI states, including magnetoroton spectrum, magnetoroton quantum numbers, and anyon quasiparticle band structures and  crystalline symmetry fractionalization pattern, the Fermi surface shape of the composite Fermi liquid, etc. Some of these seem difficult to access using other methods. We demonstrate how to apply our construction and extract microscopic information in some model systems, including the model for the twisted bilayer MoTe$_2$.

This work also leaves many open questions. A practical question is about the computation of a hyperdeterminant, which is known to be NP-hard. In the present work, we have used translational symmetry to slightly reduce the computational complexity. This allows us to compute hyperdeterminant exactly up to a system size comparable to those used in exact diagonalization. Is it possible to compute hyperdeterminants for larger systems? There may be two directions to proceed. First, instead of computing hyperdeterminant exactly, there may be algorithms to perform the projection approximately. Second, instead of considering the general hyperdeterminant wavefunctions, one may focus on a subclass of wavefunctions whose hyperdeterminants are easier to compute. 

On the conceptual side, one open question is about the non-abelian fractional quantum Hall states. The simplest state in this regard may be the Pfaffian state obtained via pairing on the composite Fermi surface \cite{moore1991nonabelions,lee2007particle,levin2007particle,son2015composite,read2000paired}. We expect the present construction, after moderate revision, can be applied to such states in the FCI context. The mathematically relevant object is the so-called hyperpfaffian \cite{aboud2016hyperpfaffians,jing2014quantum}, which is the natural generalization of pfaffian, but defined for tensors. 

Another crucial conceptual question that we did not answer in this work is the effective theories associated with the projected wavefunctions. We have demonstrated that in the context of the Galilean invariant traditional Fractional quantum Hall liquids, the projected wavefunctions in our construction are identical to those obtained by Jain's prescription, whose low energy Chern-Simons effective theories have been studied previously using various methods \cite{lopez1991fractional,zhang1992chern,lu2016classification,gromov2015framing,gromov2017bimetric,geracie2015spacetime,sohal2018chern,wen1999projective,barkeshli2010effective}. Even in this Galilean invariant case, finding the correct long-wavelength effective theories can be nontrivial. A remarkable example was established by Dong and Senthil recently \cite{dong2020noncommutative}, where they investigated the composite Fermi liquid for the $\nu=1$ bosonic system. This system has two apparently different theories: the Halperin-Lee-Read theory (HLR) theory \cite{halperin1993theory} and the Pasquier-Haldane-Read (PHR) theory \cite{pasquier1998dipole,read1998lowest}. The former theory is not within the LLL, leading to an effective theory with a Chern-Simons term. The latter theory is within the LLL, but apparently leads to an effective theory with no Chern-Simons term. Dong and Senthil showed that the effective theory of PHR is defined in a noncommutative space. After approximately mapping to a commutative field theory, the same Chern-Simon term as HLR emerges. 

The present construction includes the effects of the crystalline potential and generally applies to Jain's sequence and the composite Fermi liquid in FCI systems. Similar to the PHR theory, our construction is explicitly within the partially filled Chern band. The HLR theory, however, is parallel to the usual parton construction \emph{without} projecting into the Chern band (see Eq.(\ref{eq:usual_parton})). We leave the investigation of the long-wavelength effective theories for the proposed projected wavefunctions in future works. 

\begin{acknowledgements}
We thank Senthil Todadri, Ashvin Vishwanath, Yong Baek Kim and Yuan-Ming Lu for helpful discussions. We acknowledge the HPC resources from Andromeda cluster at Boston College.  D.X. acknowledges support from the Center on Programmable Quantum Materials, an Energy Frontier Research Center funded by DOE BES under award DE-SC0019443. 
\end{acknowledgements}

\vspace{1em}
\noindent\emph{Note added.---} After the completion of the manuscript, we became aware that Junren Shi \cite{shi2023quantum} generalizes the Pasquier-Haldane-Read’s construction for the $\nu=1$ bosonic composite Fermi liquid to the case of Galilean invariant $\nu=1/2$ fermionic composite Fermi liquid in the disc geometry, where a projection to the $\nu=1/2$ bosonic Laughlin state in the vortex space is used, and coincides with our construction in this case.

\appendix
\section{The rotation transformations of LLL bloch states and the CB-LLL mapping}\label{app:mag_rot}
Let us consider a 2D rotation symmetric sample with $e^{i\theta}\mathbf a_{i,e}= \sum_j r_{ij}\mathbf a_{j,e}$ and $N_{1,e}=N_{2,e}\equiv N$, where $r_{ij}\in\mathbb Z$ and $i,j\in\{1,2\}$. Choosing $z_0=\mathbf a_{i,e}$ in Eq.(\ref{eq:R_T_identity}), then
\begin{align}
&D_e(\mathbf a_{i,e})=[U_{R,e}(\theta)R_e(\theta)]^{-1} D_e(e^{i\theta}\mathbf a_{i,e})[U_{R,e}(\theta)R_e(\theta)]\notag\\
&\quad=[U_{R,e}(\theta)R_e(\theta)]^{-1} D_e(\sum_{ij}\mathbf a_{j,e})[U_{R,e}(\theta)R_e(\theta)].
\end{align}
Applying to the bloch basis $|\mathbf k_e\rangle_{\text{LLL}}$, and expanding $D_e(\sum_{ij}\mathbf a_j)$ using the GMP algebra
\begin{align}
D_e(\sum_j r_{ij}\mathbf a_{j,e})=e^{i\pi r_{i1}r_{i2}}D_e( r_{i1}\mathbf a_{1,e})D_e( r_{i2}\mathbf a_{2,e}),
\end{align}
we have
\begin{align}
e^{-i \mathbf k_e\cdot \mathbf a_{i,e}}=e^{i\pi r_{i1}r_{i2}} e^{-i (R_{\theta}\mathbf k_e)\cdot  (e^{i\theta}\mathbf a_{i,e})}.
\end{align}

Using the identity $(-1)^{r_{i1}+r_{i2}+1}=(-1)^{r_{i1}r_{i2}}$ (because $r_{i1},r_{i2}$ cannot be both even, since $\det(r_{ij})=1$), and noting that $e^{-i \mathbf k_e\cdot \mathbf a_{i,e}}\equiv e^{-i (e^{i\theta}\mathbf k_e)\cdot (e^{i\theta}\mathbf a_{i,e})}$, one can obtain the expression:
\begin{align}
e^{-i (R_{\theta}\mathbf k_e)\cdot  (e^{i\theta}\mathbf a_{i,e})}&=e^{-i (e^{i\theta}(\mathbf k_e-\mathbf K_e)+\mathbf K_e)\cdot (e^{i\theta}\mathbf a_{i,e})},\notag\\
\text{where }\mathbf K_e&\equiv \frac{\mathbf G_{1,e}}{2}+\frac{\mathbf G_{2,e}}{2}.
\end{align}

Therefore, generally speaking, the rotation should be viewed as about the $[\pi,\pi]$-point $\mathbf K_e$ of the BZ. For the $C_2$ and $C_4$ systems, the phase factor $e^{i\pi r_{i1}r_{i2}}$ is trivial and the rotation can also be viewed as about the $[0,0]$-point. However, for the $C_3$ and $C_6$ systems, this phase factor is nontrivial, and one does need to view the rotation as about the $[\pi,\pi]$-point (or momentum points differ by a reciprocal lattice vector). \emph{To have a uniform discussion, in this paper we always view the rotation as about the $[\pi,\pi]$-point in the LLL.}

Choosing $z_0=\frac{\mathbf a_{i,e}}{N}$ in Eq.(\ref{eq:R_T_identity}), we have:
\begin{align}
&[U_{R,e}(\theta)R_e(\theta)]\rho_e(\tfrac{\mathbf G_{1,e}}{N})[U_{R,e}(\theta)R_e(\theta)]^{-1}\notag\\
&\quad\quad=\rho_e(e^{i\theta}\tfrac{\mathbf G_{1,e}}{N})=e^{i\pi\frac{r_{21}r_{22}}{N^2}}\rho_e(r_{22}\tfrac{\mathbf G_{1,e}}{N})\rho_e(-r_{21}\tfrac{\mathbf G_{2,e}}{N})\notag\\
&[U_{R,e}(\theta)R_e(\theta)]\rho_e(\tfrac{\mathbf G_{2,e}}{N})[U_{R,e}(\theta)R_e(\theta)]^{-1}\notag\\
&\quad\quad=\rho_e(e^{i\theta}\tfrac{\mathbf G_{2,e}}{N})=e^{i\pi\frac{r_{11}r_{12}}{N^2}}\rho_e(-r_{12}\tfrac{\mathbf G_{1,e}}{N})\rho_e(r_{11}\tfrac{\mathbf G_{2,e}}{N})
\end{align}
Applying to Eq.(\ref{eq:rho_matrix_element}), one obtains:
\begin{align}
&e^{i\xi(\theta,\mathbf k_e+\frac{\mathbf G_{1,e}}{N})} e^{i\frac{\mathbf k_e\cdot\mathbf a_{2,e}}{N}}\notag\\
&\qquad=e^{i\pi\frac{r_{21}r_{22}}{N^2}}e^{ir_{22}\frac{(R_{\theta}\mathbf k_e-r_{21}\frac{\mathbf G_{2,e}}{N})\cdot\mathbf a_{2,e}}{N}}e^{i\xi(\theta,\mathbf k_e)}\notag\\
&e^{i\xi(\theta,\mathbf k_e+\frac{\mathbf G_{2,e}}{N})}\notag\\
&\qquad=e^{i\pi\frac{r_{11}r_{12}}{N^2}}e^{-ir_{12}\frac{(R_{\theta}\mathbf k_e+r_{11}\frac{\mathbf G_{2,e}}{N})\cdot\mathbf a_{2,e}}{N}}e^{i\xi(\theta,\mathbf k_e)}.
\end{align}
Namely:
\begin{align}
e^{i\xi(\theta,\mathbf k_e+\frac{\mathbf G_{1,e}}{N})}&=e^{-i\pi\frac{r_{21}r_{22}}{N^2}}e^{-i\frac{\mathbf k_e\cdot\mathbf a_{2,e}}{N}}e^{ir_{22}\frac{R_{\theta}\mathbf k_e\cdot\mathbf a_{2,e}}{N}}e^{i\xi(\theta,\mathbf k_e)}\notag\\
e^{i\xi(\theta,\mathbf k_e+\frac{\mathbf G_{2,e}}{N})}&=e^{-i\pi\frac{r_{11}r_{12}}{N^2}}e^{-ir_{12}\frac{R_{\theta}\mathbf k_e\cdot\mathbf a_{2,e}}{N}}e^{i\xi(\theta,\mathbf k_e)}.
\end{align}
These equations fully determine $e^{i\xi(\theta,\mathbf k_e)}$ up to an overall shift, which can be fixed by computing $e^{i\xi(\theta,\mathbf 0)}$. 

For instance, for a $C_4$ symmetric lattice, the matrix $(r_{ij})=i\sigma_y$, $R_{\theta}\mathbf{k}_e\equiv R_\theta(k_1 \frac{\mathbf G_{1,e}}{2\pi} + k_2 \frac{\mathbf G_{2,e}}{2\pi})=((-k_2+2\pi) \frac{\mathbf G_{1,e}}{2\pi} + k_1 \frac{\mathbf G_{2,e}}{2\pi})$, one finds:
\begin{align}
C_4:\; e^{i\xi(\frac{\pi}{2},\mathbf k_e)}=e^{-i \frac{k_1 k_2}{2\pi}}.
\end{align}
Using the BZ boundary condition Eq.(\ref{eq:BZ_BC}), the rotation eigenvalues are:
\begin{align}
&C_4 \text{ systems: }\notag\\
&\quad\quad C_4([(0,0)\text{ shifted by }(\pi,\pi)])=(-i), \notag\\
&\quad\quad C_4([(\pi,\pi)\text{ shifted by }(\pi,\pi)])=1,\notag\\
&\quad\quad C_2([(\pi,0)\text{ shifted by }(\pi,\pi)])=1.
\end{align}

For a $C_6$ symmetric lattice, we choose $\mathbf a_{2,e}=e^{i\frac{2\pi}{3}}\mathbf a_{1,e}$ and the $(r_{ij})$ matrix becomes $\begin{pmatrix} 1& 1 \\ -1 & 0\end{pmatrix}$. Consequently $\mathbf G_{2,e}=e^{i\frac{\pi}{3}}\mathbf G_{1,e}$. As is mentioned before, the rotation center is shifted to $[\pi,\pi]$, so $R_\theta\mathbf k_e\equiv R_{\theta} (k_1 \frac{\mathbf G_{1,e}}{2\pi} + k_2 \frac{\mathbf G_{2,e}}{2\pi})=(-(k_2-\pi)+\pi) \frac{\mathbf G_{1,e}}{2\pi} + ((k_1-\pi)+(k_2-\pi)+\pi) \frac{\mathbf G_{2,e}}{2\pi}=(-k_2+2\pi) \frac{\mathbf G_{1,e}}{2\pi} + (k_1+k_2-\pi) \frac{\mathbf G_{2,e}}{2\pi}$, and one finds:
\begin{align}
C_6:\; e^{i\xi(\frac{\pi}{3},\mathbf k_e)}=e^{-i\frac{\pi}{12}}e^{-i \frac{k_1 k_2+\frac{1}{2}k_2(k_2-2\pi)}{2\pi}}.
\end{align}
The rotation eigenvalues are:
\begin{align}
&C_6\text{ systems: }\notag\\
&\quad\quad C_6([(0,0)\text{ shifted by }(\pi,\pi)])=e^{-i\frac{\pi}{3}},\notag\\
&\quad\quad C_3([(\tfrac{2\pi}{3},\tfrac{2\pi}{3})\text{ shifted by }(\pi,\pi)])=1,\notag\\
&\quad\quad C_2([(\tfrac{\pi}{2},0)\text{ shifted by }(\pi,\pi)])=1.
\end{align}
The rotation transformation for $C_2$ and $C_3$ can be obtained by the square of the $C_4$ and $C_6$. These results show that in the LLL, the magnetic rotation eigenvalues are $e^{-i\theta}$ at the $\mathbf K_e=(\pi,\pi)$-point, and are trivial everywhere else.

In a general Chern band, The Chern number put a constraint on the rotation eigenvalues at these high-symmetry points \cite{fang2012bulk}:
\begin{align}
&C_2 \text{ systems }:\notag\\
& (-1)^C=C_2[(0,0)]\cdot C_2[(\pi,\pi)]\cdot C_2[(\pi,0)]\cdot C_2[(0,\pi)]\notag\\
&C_4 \text{ systems }:\notag\\
& e^{i\frac{\pi}{2}C}=(-1)^F C_4[(0,0)]\cdot C_4[(\pi,\pi)]\cdot C_2[(\pi,0)]\notag\\
&C_3 \text{ systems }:\notag\\
& e^{i\frac{2\pi}{3}C}=(-1)^F C_3[(0,0)]\cdot C_3[(\frac{2\pi}{3},\frac{2\pi}{3})]\cdot C_3[-(\frac{2\pi}{3},\frac{2\pi}{3})]\notag\\
&C_6 \text{ systems }:\notag\\
&e^{i\frac{\pi}{3}C}=(-1)^F C_6[(0,0)]\cdot C_3[(\frac{2\pi}{3},\frac{2\pi}{3})]\cdot C_2[(\frac{\pi}{2},0)],\label{eq:C_constraint}
\end{align}
where $(C_n)^n=(-1)^F$. Here, we have $C=-1$ and choose the convention that $(-1)^F=1$. It is straightforward to redefine the rotation operation to describe the case of $(-1)^F=-1$. 

Due to the mapping Eq.(\ref{eq:CB_LLL_map}), we know that if the CB has a rotation eigenvalue $e^{-i\theta}$ at the $\Gamma$-point and trivial everywhere else (coined ``\emph{the fundamental case}'' below), a smooth gauge satisfying Eq.(\ref{eq:CB_BZ_BC},\ref{eq:CB_rotation}) can be found following the prescription of Ref.\cite{jian2013crystal}. If the rotation eigenvalues do not match the fundamental case, one needs to redefine the rotation operation $R_{\text{CB}}(\theta)$ following two steps as below, without changing the algebra satisfied by $R_{\text{CB}}(\theta)$ and $T_{\text{CB}}(\mathbf a_{i,e})$.

In the first step, one redefines $R_{\text{CB}}(\theta)$ by multiplying a factor $e^{-im\theta}$ ($m\in\mathbb Z$): $R_{\text{CB}}(\theta)\rightarrow e^{-im\theta}R_{\text{CB}}(\theta)$, so that the eigenvalue $C_n[(0,0)]=e^{-i\theta}$, matching the fundamental case. This step induces a possible nontrivial Wen-Zee shift. After this step, the eigenvalues at the other high-symmetry points still may not match the fundamental case, in which case we need the second step. 

In the second step, we redefine $R_{\text{CB}}(\theta)$ by combining a translation. For example, for $C_4$ systems, after the first step, it is possible that $C_4[(\pi,\pi)]=C_2[(\pi,0)]=-1$. In this case, one redefine $R_{\text{CB}}(\theta)\rightarrow T_{\text{CB}}(\mathbf a_{1,e})R_{\text{CB}}(\theta)$, and the redefined rotation eigenvalues match the fundamental case. Physically, if $R_{\text{CB}}(\theta)$ is the $C_4$ rotation about a square lattice site, then $T_{\text{CB}}(\mathbf a_{1,e})R_{\text{CB}}(\theta)$ is the $C_4$ rotation about a plaquette center. Similar redefinitions can be made for $C_2$ (using either the link center or the plaquette center rotations) and $C_3$ systems (using the plaquette center rotation). For $C_6$ systems, the second step is not needed since one must have $C_3[(\frac{2\pi}{3},\frac{2\pi}{3})]=C_2[(\frac{\pi}{2},0)]=1$ after the first step. After these two steps of redefinition, a complete match with the fundamental case can always be made.

\section{Composite fermion substitution for the case of $\nu=\frac{1}{2s}$ composite Fermi liquid}\label{app:CFL_substitution}
In the case of $\nu=\frac{1}{2s}$, the bosonic vortex carries $q_v=-q_e$, and forms a $\nu=\frac{1}{2s}$ fractional quantum hall liquid. This corresponds to the $p\rightarrow\infty$ case of the Jain's sequence. In the disc geometry with the open boundary condition, $\mathcal R_e$ and $\mathcal R_v$ satisfies the algebra:
\begin{align}
[\mathcal R_{e,x},\mathcal R_{e,y}]&=-il_e^2,&[\mathcal R_{v,x},\mathcal R_{v,y}]&=il_e^2.
\end{align}
They can be used to construct the charge-neutral composite fermion variables:
\begin{align}
r_x&=\frac{1}{2}(\mathcal R_{e,x}+\mathcal R_{v,x})\notag\\
r_y&=\frac{1}{2}(\mathcal R_{e,y}+\mathcal R_{v,y})\notag\\
k_x&=\frac{-1}{l_e^2}(\mathcal R_{e,y}-\mathcal R_{v,y})\notag\\
k_y&=\frac{1}{l_e^2}(\mathcal R_{e,x}-\mathcal R_{v,x})
\end{align}
It is straightforward to check that these CF variables satisfy $[r_x,k_x]=[r_y,k_y]=i$, while all other commutators vanish. Note that $\bm k$ can be represented as:
\begin{align}
\bm k=\frac{1}{l_e^2}\hat z\times(\mathcal R_e-\mathcal R_v),
\end{align}
indicating that the CF's momentum is related to its electric dipole moment.

On a finite size system with $N_{1,e}\cdot N_{2,e}$ unit cells, one may choose either the real-space or momentum-space basis for the CF. For example, the momentum-space basis is given by the eigenstates of the translation operator:
\begin{align}
T_{CF}(z)=e^{-i\bm k\cdot z}=D_e(z)\cdot D_v(z)
\end{align}
The boundary-condition-allowed $z$ is given by $z=l_2\frac{\mathbf a_{1,e}}{N_{2,e}}-l_1\frac{\mathbf a_{2,e}}{N_{1,e}}$, $l_i\in \mathbb Z$. And the physically distinct $\bm k$ eigenvalues are:
\begin{align}
\bm k= (m_{1}+\frac{\varphi_{1,e}-\varphi_{1,v}}{2\pi})\frac{\mathbf G_{1,e}}{N_{1,e}}+(m_{2}+\frac{\varphi_{2,e}-\varphi_{2,v}}{2\pi})\frac{\mathbf G_{2}}{N_{2,e}},
\end{align}
where $m_i\in [0,N_{1,e}\cdot N_{2,e}-1]$ are integers. Since the number of fluxes $N_{\phi,e}=N_{\phi,v}=N_{1,e}\cdot N_{2,e}$, one finds that these $N_{\phi,e}^2$ number of momentum eigenstates exactly reproduce the dimension of the Hilbert space $\mathcal H_e\otimes \mathcal H_v$.

In the presence of crystalline potential, these momentum eigenstates will hybridize and form the CF band structure with the Brillion Zone characterized by $\mathbf G_{i,e}$, and each band has $N_{\phi,e}$ momentum points. On the mean-field level, the composite Fermi liquid is formed by filling the lowest (mean-field) energy band by the filling fraction $\frac{1}{2s}$. This CFL mean-field state can then be fed into the projector $\mathbf P_g$ to obtain the projected electronic wavefunction, which is still a hyperdeterminant.

\section{Density operator expectation values in Laughlin states on the torus}\label{app:density_expectation}
The discussion here largely follows Ref.\cite{wen1990ground}, apart from the numerical results. It is known that Laughlin's states at $\nu=1/m$ form a $m$-fold irreducible representation of the many-body magnetic translation algebra on a torus:
\begin{align}
\pmb{\bm D}_e(z_1)\pmb{\bm D}_e(z_2)&=e^{\frac{i}{2} \frac{z_1\times z_2}{l_e^2} N_e}\pmb{\bm D}_e(z_1+z_2)\notag\\
&=e^{i \frac{z_1\times z_2}{l_e^2} N_e} \pmb{\bm D}_e(z_2)\pmb{\bm D}_e(z_1)
\end{align}
For convenience of discussion below, we introduce the minimal translation displacement $\delta_1$ ($\delta_2$) along the $L_1$ ($L_1\tau$) direction of the sample that is consistent with the torus boundary condition. 
\begin{align}
\delta_1\equiv&\frac{L_{1}}{N_{\phi,e}},&\delta_2\equiv&\frac{L_{1}\tau}{N_{\phi,e}}, 
\end{align}
leading to
\begin{align}
\pmb{\bm D}_e(\delta_1)\pmb{\bm D}_e(\delta_2)=e^{i\frac{2\pi}{m}}\pmb{\bm D}_e(\delta_2)\pmb{\bm D}_e(\delta_1).
\end{align}
One can choose a gauge for the $m$-fold Laughlin's states $|\psi_i\rangle$ ($i \in [0,m-1]$) as the eigenstates of $\pmb{\bm D}_e(\delta_1)$, satisfying:
\begin{align}
\pmb{\bm D}_e(\delta_1)|\psi_i\rangle&=e^{i\phi_1}e^{i\frac{2\pi i}{m}}|\psi_i\rangle, & \pmb{\bm D}_e(\delta_2)|\psi_i\rangle&=|\psi_{i+1}\rangle,\label{eq:Laughlin_gauge}
\end{align}
where $|\psi_{i+m}\rangle\equiv e^{i\phi_2}|\psi_{i}\rangle$ and the phase factors $e^{i\phi_i}$ depend on the boundary condition. 

The many-particle density operator satisfies a relation with the magnetic translation operator:
\begin{align}
\pmb{\bm D}_e(z_0) \pmb{\boldsymbol\rho}_e(\mathbf q_e=\frac{iz_1}{l_e^2}) \pmb{\bm D}_e(z_0)^\dagger=e^{i\frac{z_0\times z_1}{l_e^2}}\pmb{\boldsymbol\rho}_e(\mathbf q_e=\frac{iz_1}{l_e^2}).
\end{align}
Plugging in $z_0=m\delta_1$ or $z_0=m\delta_2$, using Eq.(\ref{eq:Laughlin_gauge}), the above identity leads to:
\begin{align}
&\langle \psi_i|\pmb{\boldsymbol\rho}_e(\mathbf q_e)|\psi_j\rangle\neq 0 \notag\\
&\text{ only if } \mathbf q_e=\mathbf q_e(n_1,n_2)\equiv\frac{i (n_1 \frac{L_1}{m}+ n_2\frac{L_1\tau}{m})}{l_e^2},
\end{align}
where $n_i\in [0,m-1]$ are integers. 

Let's define the operators:
\begin{align}
\mathbf A(n_1,n_2)\equiv \pmb{\bm D}_e(n_1\delta_1+n_2\delta_2)^{-1} \pmb{\boldsymbol\rho}_e(\mathbf q_e(n_1,n_2)).\label{eq:A_def}
\end{align}
One can show that $\mathbf A(n_1,n_2)$ commutes with both $\pmb{\bm D}_e(\delta_1)$ and $\pmb{\bm D}_e(\delta_2)$, and consequently must be a constant in the ground state manifold.

\begin{table}
\centering
\vspace{2mm}
\begin{tabular}{||c | c | c | c | c||} 
 \hline
 $N_\phi$ & $\tau=i$ & $\tau=e^{i\frac{2\pi}{3}}$ \\ 
 \hline\hline
 2 & 1  & 1 \\\hline
 4 & -$\sqrt{2}$ & -1.156(1)\\\hline
 6 & 1.267(1) & 1.000(1) \\\hline
 8 & -0.8652(4) & -0.5591(7) \\\hline
 10 & 0.5658(7)& 0.3273(7) \\\hline
 12 & -0.3423(8) & -0.1831(9)\\\hline
 14 & 0.1993(8) & 0.0932(5) \\\hline
 16 & -0.1139(3) &  -0.0461(3)\\\hline
 18 & 0.0641(4) & 0.0228(2)\\\hline
 \hline
\end{tabular}
\caption{$\mathbf A(n_1=1,n_2=0)$ (see Eq.(\ref{eq:A_def})) for Laughlin's $\nu=1/2$ state, with the sample shape parameter $\tau=i$ and $\tau=e^{i\frac{2\pi}{3}}$, and boundary condition $\varphi_{1,e}=\varphi_{2,e}=0$ (see Eq.(\ref{eq:boundary_conditions})).}
\label{tb:rho_v_expectation}
\end{table}

We have checked numerically that $A(n_1,n_2)\sim e^{-c \cdot N_{\varphi,e}}$ in the ground state manifold exponentially decay in the thermodynamic limit ($c$ is a constant for a given $\tau$ and boundary condition.). For instance, in Table.\ref{tb:rho_v_expectation} we list the values of $\mathbf A(n_1=1,n_2=0)$ in the ground state manifold for Laughlin's $\nu=1/2$ states (electron is bosonic), computed via variational Monte Carlo.

\section{Time-dependent Hartree-Fock approximation in the presence of constraints}\label{app:TDHF}
Here we describe the general prescription to compute the excitation spectrum in the framework of TDHF in the presence of constraints and Lagrange multipliers. The original many-body Hamiltonian $\mathbf H=\mathbf H_0+\mathbf V$, where $\mathbf H_0$ is the two-body term. For simplicity, we consider $\mathbf V$ as the density-density interaction:
\begin{align}
\mathbf V=\frac{1}{2}\sum_{\mathbf q} V_{\mathbf q} \pmb{\boldsymbol\rho}(\mathbf q)\pmb{\boldsymbol\rho}(-\mathbf q),
\end{align}
where $\pmb{\boldsymbol\rho}(\mathbf q)$ is a fermion bilinear. We assume a collection of linearly independent Hermitian symmetry generators $\{\mathbf S_i\}$ that are fermion bilinears. They commute with $\mathbf H$, and form a closed algebra:
\begin{align}
[\mathbf S_i,\mathbf H]&=0,\;\forall i,\notag\\
[\mathbf S_i,\mathbf S_j]&=i\sum_{k} c_{ijk}\mathbf S_k\label{eq:S_algebra}
\end{align}
In the main text, the symmetry generators $\{\mathbf S_i\}$ are vortices density operators $\{\pmb{\boldsymbol\rho}_v(\mathbf q_v)\}$.

The mean-field \emph{free-fermion states} $|\psi\rangle$'s under consideration are those that satisfy the constraints:
\begin{align}
\langle \psi |\mathbf S_i| \psi\rangle =0.\;\forall i\label{eq:S_constraints}
\end{align}
$|\psi\rangle$ is completely captured by its single-body density matrix:
\begin{align}
\boldsymbol{\mathcal P}\equiv \sum_{\alpha,\beta}\langle \psi| f_{\beta}^\dagger f_{\alpha}|\psi\rangle f_{\alpha}^\dagger f_{\beta},
\end{align}
where $\alpha,\beta$ labels a basis in the single-particle Hilbert space.

For any single-body density matrix $\boldsymbol{\mathcal P}$, we define the Hartree-Fock approximated Hamiltonian:
\begin{align}
\mathbf H_{HF}(\boldsymbol{\mathcal P})\equiv \mathbf H_0+\mathbf V_{HF}(\boldsymbol{\mathcal P}),
\end{align}
where 
\begin{align}
\mathbf V_{HF}(\boldsymbol{\mathcal P}) &\equiv \frac{1}{2} \sum_{{\mathbf q}}V_{\mathbf q}\bigg[ \mathop{\mathrm{Tr}}[\pmb{\boldsymbol\rho}(\mathbf q)\boldsymbol{\mathcal P}]\pmb{\boldsymbol\rho}(-\mathbf q)+\pmb{\boldsymbol\rho}(\mathbf q)\mathop{\mathrm{Tr}}[\pmb{\boldsymbol\rho}(-\mathbf q)\boldsymbol{\mathcal P}]\notag\\
&\quad-\pmb{\boldsymbol\rho}(\mathbf q)\boldsymbol{\mathcal P}\pmb{\boldsymbol\rho}(-\mathbf q)-\pmb{\boldsymbol\rho}(-\mathbf q)\boldsymbol{\mathcal P}\pmb{\boldsymbol\rho}(\mathbf q) \bigg].
\end{align}

The standard static Hartree-Fock calculation boils down to finding $\boldsymbol{\mathcal P}_0$ that minimizes the variational energy $\langle\psi|\mathbf H|\psi\rangle$, subject to the constraints Eq.(\ref{eq:S_constraints}). One can show that under a small perturbation $\boldsymbol{\mathcal P}_0\rightarrow \boldsymbol{\mathcal P}_0+\delta\boldsymbol{\mathcal P}$, the linear order change of variational energy is:
\begin{align}
0=\delta\langle\psi|\mathbf H|\psi\rangle=\mathop{\mathrm{Tr}}[\mathbf H_{HF}(\boldsymbol{\mathcal P}_0)\delta\boldsymbol{\mathcal P}].
\end{align}
Quite generally, such a small perturbation can be parameterized by a small unitary rotation $\boldsymbol{\mathcal P}_0\rightarrow \mathbf U\boldsymbol{\mathcal P}_0\mathbf U^\dagger$ where $\mathbf U=e^{i\boldsymbol\phi}$, $\boldsymbol\phi$ is a small fermion bilinear operator. To the leading order, $\delta\boldsymbol{\mathcal P}=i[\boldsymbol\phi,\boldsymbol{\mathcal P}_0]$, so
\begin{align}
0=i\mathop{\mathrm{Tr}}[\mathbf H_{HF}(\boldsymbol{\mathcal P}_0)[\boldsymbol\phi,\boldsymbol{\mathcal P}_0]]=-i\mathop{\mathrm{Tr}}[\boldsymbol\phi[\mathbf H_{HF}(\boldsymbol{\mathcal P}_0),\boldsymbol{\mathcal P}_0]],\label{eq:P0_condition}
\end{align}
where we have used the trace identity $\mathop{\mathrm{Tr}}[\mathbf A[\mathbf B,\mathbf C]]=\mathop{\mathrm{Tr}}[\mathbf B[\mathbf C,\mathbf A]]$.

At this point, it is helpful to introduce the symplectic structure of the space of the fermion bilinear operators. We can separate any fermion bilinear operator $\mathbf A$ into two parts:
\begin{align}
\mathbf A&=\{\mathbf A\}_{phys}+\{\mathbf A\}_{unphys},\;\;\text{where}\notag\\
\{\mathbf A\}_{phys}&\equiv [[\mathbf A,\boldsymbol{\mathcal P}_0],\boldsymbol{\mathcal P}_0]\notag\\
&=(\mathbf 1-\boldsymbol{\mathcal P}_0)\mathbf A\boldsymbol{\mathcal P}_0+\boldsymbol{\mathcal P}_0\mathbf A(\mathbf 1-\boldsymbol{\mathcal P}_0).
\end{align}
Using the fact that $\boldsymbol{\mathcal P}_0$ is a projector, one can easily show that $[\mathbf A,\boldsymbol{\mathcal P}_0]=[\{\mathbf A\}_{phys},\boldsymbol{\mathcal P}_0]$. Namely, to consider the small unitary rotation above, it is sufficient to consider the linear space spanned by $\{\mathbf A\}_{phys}$, which we denote as $\boldsymbol{\mathcal W}$. Note that for one has $[\mathbf A,\boldsymbol{\mathcal P}_0]\in \boldsymbol{\mathcal W},\; \forall \text{ fermion bilinear } \mathbf A$, and there is a useful identity:
\begin{align}
\mathop{\mathrm{Tr}}[\boldsymbol{\mathcal P}_0[\mathbf A,\mathbf B]]=\mathop{\mathrm{Tr}}[\boldsymbol{\mathcal P}_0[\{\mathbf A\}_{phys},\{\mathbf B\}_{phys}]]
\end{align}

In $\boldsymbol{\mathcal W}$, we can define two different inner products. The first (single angle bracket) is a conventional one while the second (double angle bracket) is a symplectic one. 
\begin{align}
\langle \{\mathbf A\}_{phys},\{\mathbf B\}_{phys}\rangle &\equiv\mathop{\mathrm{Tr}}[ \{\mathbf A^\dagger\}_{phys}\cdot \{\mathbf B\}_{phys}]\notag\\
\llangle\{\mathbf A\}_{phys},\{\mathbf B\}_{phys}\rrangle &\equiv\mathop{\mathrm{Tr}}[ \{\mathbf A^\dagger\}_{phys}\cdot [ \{\mathbf B\}_{phys},\boldsymbol{\mathcal P}_0]]\notag\\
&=\mathop{\mathrm{Tr}}[\boldsymbol{\mathcal P}_0 [ \{\mathbf A^\dagger\}_{phys},\{\mathbf B\}_{phys}]].
\end{align}

The condition that $\delta \boldsymbol{\mathcal P}$ does not change the constraint relations Eq.(\ref{eq:S_constraints}) can also be written as $-i\mathop{\mathrm{Tr}}[\boldsymbol{\mathcal P}_0[\mathbf S_i,\boldsymbol{\phi}]]=0$, or in terms of the symplectic inner product introduced above:
\begin{align}
\llangle\{\mathbf S_i\}_{phys},\{\boldsymbol\phi\}_{phys}\rrangle=0.
\end{align}
We may denote the subspace in $\boldsymbol{\mathcal W}$ spanned by $\{\mathbf S_i\}_{phys}$ as $\boldsymbol{\mathcal W}_{\mathbf S}$. The above condition means that $\{\boldsymbol\phi\}_{phys}\in \overline{\boldsymbol{\mathcal W}_{\mathbf S}}$, where $\overline{\boldsymbol{\mathcal W}_{\mathbf S}}$ is the symplectic complement of $\boldsymbol{\mathcal W}_{\mathbf S}$. Drastically different from the conventional complement subspace, here we have:
\begin{align}
\boldsymbol{\mathcal W}_{\mathbf S} \subset \overline{\boldsymbol{\mathcal W}_{\mathbf S}},
\end{align}
which is a consequence of Eq.(\ref{eq:S_algebra}). The variational minimization problem now becomes finding $\boldsymbol{\mathcal P}_0$ so that Eq.(\ref{eq:P0_condition}) is satisfied for all $\boldsymbol\phi\in\overline{\boldsymbol{\mathcal W}_{\mathbf S}}$. 

If the objective was to find $\boldsymbol{\mathcal P}_0$ so that Eq.(\ref{eq:P0_condition}) is satisfied for all $\boldsymbol\phi$ in the \emph{entire space} $\boldsymbol{\mathcal W}$, then it would lead to the well-known self-consistent condition $[\mathbf H_{HF}(\boldsymbol{\mathcal P}_0),\boldsymbol{\mathcal P}_0]=0$. However, since $\overline{\boldsymbol{\mathcal W}_{\mathbf S}}$ is smaller than $\boldsymbol{\mathcal W}$, as long as $[\mathbf H_{HF}(\boldsymbol{\mathcal P}_0),\boldsymbol{\mathcal P}_0]\in \overline{\boldsymbol{\mathcal W}_{\mathbf S}}^\perp$, $\boldsymbol{\mathcal P}_0$ is a legitimate optimal solution to satisfy Eq.\eqref{eq:P0_condition} satisfied. Here $\overline{\boldsymbol{\mathcal W}_{\mathbf S}}^\perp$ is the conventional complement subspace of $\overline{\boldsymbol{\mathcal W}_{\mathbf S}}$ (Here and below we always use ${\;\cdot\;}^\perp$ to denote the conventional complement and $\overline{\;\cdot\;}$ to denote the symplectic complement).

One can show that $\overline{\boldsymbol{\mathcal W}_{\mathbf S}}^\perp$ is actually the symplectic dual of the subspace $\boldsymbol{\mathcal W}_{\mathbf S}$. Namely, they are orthogonal to each other w.r.t. the conventional inner product, have the same dimension, and the linear map $\mathbf x\mapsto [\mathbf x,\boldsymbol{\mathcal P}_0]$ is a one-to-one mapping between the two subspaces. One way to see this is to decompose $\overline{\boldsymbol{\mathcal W}_{\mathbf S}}$ into the direct sum of two mutually orthogonal subspaces (w.r.t. the conventional inner product): $\overline{\boldsymbol{\mathcal W}_{\mathbf S}}=\boldsymbol{\mathcal W}_{\mathbf S}\oplus \boldsymbol{\mathcal V}$. It follows that, by definition, $\boldsymbol{\mathcal W}$ is the direct sum of three mutually orthogonal (w.r.t. the conventional inner product) subspaces:
\begin{align}
\boldsymbol{\mathcal W}=\boldsymbol{\mathcal W}_{\mathbf S}\oplus\overline{\boldsymbol{\mathcal W}_{\mathbf S}}^\perp\oplus\boldsymbol{\mathcal V}\label{eq:W_decomposition}
\end{align}
Now by choosing an arbitrary $\mathbf x\in \boldsymbol{\mathcal W}_{\mathbf S}$, from above decomposition we know $\llangle \boldsymbol{\mathcal W}_{\mathbf S},\mathbf x\rrangle=\llangle \boldsymbol{\mathcal V},\mathbf x\rrangle=0$, i.e., $\langle \boldsymbol{\mathcal W}_{\mathbf S},[\mathbf x,\boldsymbol{\mathcal P}_0]\rangle=\langle \boldsymbol{\mathcal V},[\mathbf x,\boldsymbol{\mathcal P}_0]\rangle=0$. Thus we must have $[\mathbf x,\boldsymbol{\mathcal P}_0]\in \overline{\boldsymbol{\mathcal W}_{\mathbf S}}^\perp$. Noting that $[[\mathbf x,\boldsymbol{\mathcal P}_0],\boldsymbol{\mathcal P}_0]=\mathbf x$, we showed that $\mathbf x\mapsto [\mathbf x,\boldsymbol{\mathcal P}_0]$ is a one-to-one mapping between $\boldsymbol{\mathcal W}_{\mathbf S}$ and $\overline{\boldsymbol{\mathcal W}_{\mathbf S}}^\perp$. This mapping also sends $\boldsymbol{\mathcal V}$ back to $\boldsymbol{\mathcal V}$.

Therefore, there exist a collection of Lagrange multipliers $\lambda_i$, so that $[\mathbf H_{HF}(\boldsymbol{\mathcal P}_0),\boldsymbol{\mathcal P}_0]=-[\sum_i\lambda_i\mathbf S_i,\boldsymbol{\mathcal P}_0]$. This is equivalent to the condition:
\begin{align}
[\mathbf{\tilde H}_{HF}(\boldsymbol{\mathcal P}_0), \boldsymbol{\mathcal P}_0]\equiv\bigg[\mathbf H_{HF}(\boldsymbol{\mathcal P}_0)+\sum_i\lambda_i\mathbf S_i, \boldsymbol{\mathcal P}_0\bigg]=0.
\end{align}
This is the well-known prescription: one can introduce Lagrange multipliers so that the ground state of $\mathbf{\tilde H}_{HF}(\boldsymbol{\mathcal P}_0)$ satisfies the constraints Eq.(\ref{eq:S_constraints}), and perform the self-consistent calculation as usual. 

Now we are ready to study the time-evolution of the single-body density matrix near $\boldsymbol{\mathcal P}_0$: 
\begin{align}
[\mathbf{\tilde H}_{HF}(\boldsymbol{\mathcal P}), \boldsymbol{\mathcal P}]=i\hbar\dot{\boldsymbol{\mathcal P}}
\end{align}
To the linear order of $\boldsymbol\phi$, this leads to:
\begin{align}
&[\mathbf {\tilde H}_{HF}(\boldsymbol{\mathcal P}_0),[\boldsymbol\phi,\boldsymbol{\mathcal P}_0]]+[\mathbf V_{HF}([\boldsymbol\phi,\boldsymbol{\mathcal P}_0]),\boldsymbol{\mathcal P}_0]\notag\\
&\quad+(-i)\left[\sum_i\delta\lambda_i(\boldsymbol\phi) \mathbf S_i,\boldsymbol{\mathcal P}_0\right]=i\hbar [\dot{\boldsymbol\phi},\boldsymbol{\mathcal P}_0].
\end{align}
$\delta\lambda_i(\boldsymbol\phi)\propto \boldsymbol\phi$ is the adjustment of the Lagrange multipliers due to $\boldsymbol\phi$, so that the ground state of $\mathbf{\tilde H}_{HF}(\boldsymbol{\mathcal P})$ satisfies the constraints Eq.(\ref{eq:S_constraints}).
Equivalently, we can define the operator $\boldsymbol{\mathcal H}$:
\begin{align}
\boldsymbol{\mathcal H} \cdot \{\boldsymbol\phi\}_{phys} &\equiv [[\mathbf {\tilde H}_{HF}(\boldsymbol{\mathcal P}_0),\{\boldsymbol\phi\}_{phys}],\boldsymbol{\mathcal P}_0]\notag\\
&\quad+[\mathbf V_{HF}([\{\boldsymbol\phi\}_{phys},\boldsymbol{\mathcal P}_0]),\boldsymbol{\mathcal P}_0], \label{eq:H_eigen}
\end{align}
and introduce the linear operator $\boldsymbol{\mathcal L}$ to represent the eigen equation (using Jacobi identity and static condition)
\begin{align}
\boldsymbol{\mathcal L}\cdot \{\boldsymbol\phi\}_{phys}&\equiv [\boldsymbol{\mathcal H} \cdot \{\boldsymbol\phi\}_{phys},\boldsymbol{\mathcal P}_0]+(-i)\sum_i\delta\lambda_i(\boldsymbol\phi) \{\mathbf S_i\}_{phys}\notag\\
&=i\hbar\{\dot{\boldsymbol\phi}\}_{phys}=\hbar\omega\{\boldsymbol\phi\}_{phys} .\label{eq:L_eigen}
\end{align}

One can show that if $\{\boldsymbol\phi\}_{phys}\in \overline{\boldsymbol{\mathcal W}_{\mathbf S}}$, then $\boldsymbol{\mathcal L}\cdot \{\boldsymbol\phi\}_{phys}\in \overline{\boldsymbol{\mathcal W}_{\mathbf S}}$ as well. To see this, it is sufficient to show $[\boldsymbol{\mathcal H} \cdot \{\boldsymbol\phi\}_{phys},\boldsymbol{\mathcal P}_0]\in \overline{\boldsymbol{\mathcal W}_{\mathbf S}}$, or equivalently $\boldsymbol{\mathcal H} \cdot \{\boldsymbol\phi\}_{phys}\in \boldsymbol{\mathcal W}_{\mathbf S}^{\perp}$.

In fact, one can show that the operator $\boldsymbol{\mathcal H}$ is Hermitian (w.r.t the conventional inner product) in the full space $\boldsymbol{\mathcal W}$. It the follows that $\forall i$ and $\{\boldsymbol\phi\}_{phys}\in \overline{\boldsymbol{\mathcal W}_{\mathbf S}}$,
\begin{align}
\langle \{\mathbf S_i\}_{phys},\boldsymbol{\mathcal H} \cdot \{\boldsymbol\phi\}_{phys}\rangle=\langle \{\boldsymbol\phi\}_{phys},\boldsymbol{\mathcal H} \cdot \{\mathbf S_i\}_{phys} \rangle^*=0
\end{align}
This is because 
\begin{align}
\boldsymbol{\mathcal H} \cdot \{\mathbf S_i\}_{phys} \in \overline{\boldsymbol{\mathcal W}_{\mathbf S}}^\perp,\label{eq:H_S}
\end{align}
as a consequence of the symmetry, which we will explain next.

TDHF is known to be a conserving approximation. For instance, the Goldstone mode computed in TDHF is gapless. This can be demonstrated explicitly. A symmetry generator $\mathbf S_i$ should satisfy both $[\mathbf S_i,\mathbf H_0]=0$ and $[\mathbf S_i,\mathbf V]=0$. The latter condition leads to an important identity:
\begin{align}
[\mathbf S_i,\mathbf V_{HF}(\boldsymbol{\mathcal P})]=\mathbf V_{HF}([\mathbf S_i,\boldsymbol{\mathcal P}])
\end{align}
Therefore, if $\boldsymbol{\mathcal P}_0$ is a static Hartree-Fock solution with Lagrange multipliers $\lambda_j$, then $e^{i\epsilon \mathbf S_i}\boldsymbol{\mathcal P}_0 e^{-i\epsilon \mathbf S_i}$ is automatically another static Hartree-Fock solution with Lagrange multipliers unitary rotated by $e^{i\epsilon \mathbf S_i}$. One finds that
\begin{align}
\boldsymbol{\mathcal L}\cdot\{\mathbf S_i\}_{phys}=0, \text{ with } \delta\lambda_j(\mathbf S_i)=\sum_k c_{kij}\lambda_k
\end{align}
Namely, each $\mathbf S_i$ corresponds to an exact zero mode -- the Goldstone mode. This result in turn tells that $[\boldsymbol{\mathcal H} \cdot \{\boldsymbol\phi\}_{phys},\boldsymbol{\mathcal P}_0]\in\boldsymbol{\mathcal W}_{\mathbf S}$, which, under the one-to-one correspondence $\mathbf x\mapsto[\mathbf x,\boldsymbol{\mathcal P}], \boldsymbol{\mathcal W}_{\mathbf S}\rightarrow\overline{\boldsymbol{\mathcal W}_{\mathbf S}}^\perp$, also establishes the validity of Eq.(\ref{eq:H_S}).

We are now ready to find all the eigen modes in TDHF. Note the decomposition of $\boldsymbol{\mathcal W}$ in Eq.(\ref{eq:W_decomposition}). We should solve the eigenproblem of $\boldsymbol{\mathcal L}$ in 
$\overline{\boldsymbol{\mathcal W}_{\mathbf S}}=\boldsymbol{\mathcal W}_{\mathbf S}\oplus \boldsymbol{\mathcal V}$, and the subspace $\boldsymbol{\mathcal W}_{\mathbf S}$ is the null space of $\boldsymbol{\mathcal L}$. One then only needs to consider the operator $\boldsymbol{\mathcal L}$ in the subspace $\boldsymbol{\mathcal V}$, where the eigenvalues are generically nonzero. Introducing the projector $\mathbf P_{\boldsymbol{\mathcal V}}$ into the subspace $\boldsymbol{\mathcal V}$, we need to solve the eigenproblem for the operator $\boldsymbol{\mathcal L}_{\boldsymbol{\mathcal V}}$:
\begin{align}
&\boldsymbol{\mathcal L}_{\boldsymbol{\mathcal V}}\cdot\boldsymbol{\phi}=[\boldsymbol{\mathcal H}_{\boldsymbol{\mathcal V}}\cdot\boldsymbol{\phi},\boldsymbol{\mathcal P}_0],\notag\\
&\text{where }\boldsymbol{\mathcal L}_{\boldsymbol{\mathcal V}}\equiv \mathbf P_{\boldsymbol{\mathcal V}}\cdot\boldsymbol{\mathcal L}\cdot\mathbf P_{\boldsymbol{\mathcal V}}\text{ and }\boldsymbol{\mathcal H}_{\boldsymbol{\mathcal V}}\equiv \mathbf P_{\boldsymbol{\mathcal V}}\cdot\boldsymbol{\mathcal H}\cdot\mathbf P_{\boldsymbol{\mathcal V}}.
\end{align}
If $\boldsymbol{\mathcal L}_{\boldsymbol{\mathcal V}}\cdot \boldsymbol\phi =\hbar\omega \boldsymbol\phi $ with $\omega\neq 0$ for $\boldsymbol\phi\in \boldsymbol{\mathcal V}$, one can always extend $\boldsymbol\phi$ to $\overline{\boldsymbol{\mathcal W}_{\mathbf S}}$ by adding a unique component in $\boldsymbol{\mathcal W}_{\mathbf S}$ so that Eq.(\ref{eq:L_eigen}) holds.

The eigenproblem of $\boldsymbol{\mathcal L}_{\boldsymbol{\mathcal V}}$ can be shown to be equivalent to diagonalizing a free boson Hamiltonian via the bosonic Bogoliubov transformation (i.e., symplectic transformation): the eigenvalues are real and appear as $\pm \hbar\omega$ pairs. This is because $\boldsymbol{\mathcal H}_{\boldsymbol{\mathcal V}}$ satisfies the following conditions:
\begin{align}
\boldsymbol{\mathcal H}_{\boldsymbol{\mathcal V}}\cdot \boldsymbol{\phi}^\dagger=(\boldsymbol{\mathcal H}_{\boldsymbol{\mathcal V}}\cdot \boldsymbol{\phi})^\dagger,
\end{align}
which can be easily seen from Eq.(\ref{eq:H_eigen}) using $[\mathbf A^\dagger,\mathbf B^\dagger]\equiv-[\mathbf A,\mathbf B]^\dagger$. Consequently, if $\boldsymbol{\mathcal L}_{\boldsymbol{\mathcal V}}\boldsymbol{\phi}=\hbar\omega \boldsymbol{\phi}$, then $\boldsymbol{\mathcal L}_{\boldsymbol{\mathcal V}}\boldsymbol{\phi}^\dagger=-\hbar\omega \boldsymbol{\phi}^\dagger$.

Let's summary some main results here. Let the dimension of the linear space $\boldsymbol{\mathcal W}$ be $D_{\boldsymbol{\mathcal W}}$. In the energy eigenbasis of $\tilde {\mathbf H}_{HF}(\boldsymbol{\mathcal P}_0)$, $\boldsymbol{\mathcal W}$ is spanned by the fermion bilinears $c_{\alpha}^{\dagger} d_{i}$ and $d_{i}^{\dagger} c_{\alpha}$, where $i$ labels the filled single-particle orbitals and $\alpha$ labels the empty single-particle orbitals. Only these bilinears have nontrivial commutator with $\boldsymbol{\mathcal P}_0$. Therefore $D_{\boldsymbol{\mathcal W}}=2\cdot N_{filled}\cdot N_{empty}$, where $N_{filled}$ ($N_{empty}$) is the number of filled (empty) single-particle orbitals. 

In the presence of $N_c$ constraints, the perturbations corresponding to violation of the constraints span a subspace $\overline{\boldsymbol{\mathcal W}_{\mathbf S}}^\perp$, which is $N_c$ dimensional. The exact zero energy Goldstone modes, span a subspace $\boldsymbol{\mathcal W}_{\mathbf S}$, which is also $N_c$ dimensional. The nonzero energy modes can be found by studying the subspace $\boldsymbol {\mathcal V}$, which is $D_{\boldsymbol{\mathcal W}}-2\cdot N_c$ dimensional. $\boldsymbol{\mathcal W}$ has the important decomposition Eq.(\ref{eq:W_decomposition}).

\section{Derivation of Eq.(\ref{eq:coherent_state_fusion})}\label{app:coherent_state_fusion}
We will first compute a simpler fusion coefficient:
\begin{align}
A(\zeta)\equiv\langle\zeta_{CF}| 0_e\rangle|0_v\rangle.
\end{align}
Notice that the bosonic Bogoliubov transformation Eq.(\ref{eq:a_bogoliubov}) is generated by the unitary:
\begin{align}
U(c)\equiv e^{\text{arctanh}(c) (a_e^\dagger a_v^\dagger-a_ea_v)}=e^{\text{arctanh}(c) (a_{\mathcal R}^\dagger a_{\eta}^\dagger-a_{\mathcal R}a_{\eta})}.
\end{align}
$U(c)$ satisfies $U(c)^\dagger=U(-c)$ and:
\begin{align}
U(c) a_e U(c)^\dagger&=a_{\mathcal R},&U(c) a_v U(c)^\dagger&=a_{\eta},
\end{align}
We thus have the relation between the coherent states and the occupation number basis:
\begin{align}
|0_e\rangle|0_v\rangle=U(-c)|0_{\mathcal R}\rangle|0_\eta\rangle=\sqrt{1-c^2}\sum_{n=0}^{\infty} (-c)^n|n_{\mathcal R}\rangle|n_\eta\rangle,
\end{align}
leading to:
\begin{align}
A(\zeta)=\sqrt{1-c^2}\sum_n (-c)^n\langle\zeta_{CF}| n_{\mathcal R}\rangle|n_{\eta}\rangle
\end{align}
It is known that the relevant wavefunctions in the CF space are the Laguerre polynomials:
\begin{align}
\langle\zeta_{CF}| n_{\mathcal R}\rangle|n_{\eta}\rangle=(-1)^nL_n\left(\frac{|\zeta|^2}{2l_{CF}^2}\right) e^{-\frac{|\zeta|^2}{4l_{CF}^2}}.
\end{align}
In addition, the Laguerre polynomials have the generating function:
\begin{align}
\sum_n t^n L_n(x)=\frac{1}{1-t}e^{-\frac{tx}{1-t}}.
\end{align}

In the current situation, $t=c$, and one has:
\begin{align}
A(\zeta)=&\sqrt{1-c^2}\frac{1}{1-c} \exp\left[-\frac{c}{1-c}\frac{|\zeta|^2}{2l_{CF}^2}-\frac{|\zeta|^2}{4l_{CF}^2}\right]\notag\\
=&\sqrt{\frac{1+c}{1-c}}\exp\Big[-\Big(\frac{1+c}{1-c}\Big)\frac{|\zeta|^2}{4l_{CF}^2}\Big]
\end{align}

Next, we compute the complex conjugate of Eq.(\ref{eq:coherent_state_fusion}):
\begin{align}
B(z,\omega,\zeta)\equiv\langle\zeta_{CF}|z_e\rangle|\omega_v\rangle=\langle\zeta_{CF}|D_{e}(z) D_{v}(\omega)|0_e\rangle|0_v\rangle.\label{eq:B_definition}
\end{align}
Using Eq.(\ref{eq:D_e_D_v_to_D_R_D_eta}), one can show:
\begin{align}
D_{e}(z) D_{v}(\omega)=D_{\mathcal R}(X)D_\eta(Y).
\end{align}
where
\begin{align}
X&\equiv\frac{z-c^2\omega}{1-c^2}, &Y&\equiv\frac{c(\omega-z)}{1-c^2}.
\end{align}
Introducing:
\begin{align}
s&\equiv\frac{X+Y}{2},& d&\equiv\frac{X-Y}{2},
\end{align}
one has:
\begin{align}
&D_{e}(z) D_{v}(\omega)\notag\\
&=\Theta\left(\frac{d}{\sqrt{2}l_{CF}},\frac{2s}{\sqrt{2}l_{CF}}\right) D_{\mathcal R}(d)D_{\eta}(-d)D_{\mathcal R}(s)D_{\eta}(s),\label{eq:D_e_D_v}
\end{align}
where we have defined the phase factor involved in the magnetic translation algebra:
\begin{align}
\Theta(\alpha,\beta)\equiv e^{\frac{\alpha\bar\beta-\bar\alpha\beta}{2}},
\end{align}
whose exponent is bilinear in $\alpha,\beta$ and satisfies:
\begin{align}
\Theta(\alpha,\beta)&=\bar \Theta(\beta,\alpha)\notag\\
\Theta(\bar \alpha,\bar\beta)&=\bar \Theta(\alpha,\beta).
\end{align}
For instance:
\begin{align}
D_e(z_0)D_e(z_1)&=\Theta\left(\frac{\bar z_0}{\sqrt 2l_e},\frac{\bar z_1}{\sqrt 2l_e}\right)D_e(z_0+z_1)\notag\\
D_v(\omega_0)D_v(\omega_1)&=\Theta\left(\frac{\omega_0}{\sqrt 2l_v},\frac{\omega_1}{\sqrt 2l_v}\right)D_v(\omega_0+\omega_1).
\end{align}

Finally, in the symmetric gauge:
\begin{align}
\eta_x&=\frac{\hat x}{2}-l^2_{CF} \hat k_y,&\eta_y&=\frac{\hat y}{2}+l^2_{CF} \hat k_x,\notag\\
\mathcal R_x&=\frac{\hat x}{2}+l^2_{CF} \hat k_y,&\mathcal R_y&=\frac{\hat y}{2}-l^2_{CF} \hat k_x.\notag\\
\end{align}
leading to the operator identities:
\begin{align}
D_{\mathcal R}(z) D_{\eta}(-z)&=e^{\frac{i}{l_{CF}^2}(x\hat y-y\hat x)}\notag\\
D_{\mathcal R}(z) D_{\eta}(z)&=\hat T(2z),
\end{align}
where $z=x+iy$ and $\hat T(z)=e^{-i(x \hat k_x+y\hat k_y)}$ is the usual translation operator in the CF space. Therefore,
\begin{align}
D_{\mathcal R}(-d)D_{\eta}(d)|\zeta_{CF}\rangle &=\Theta\left(\frac{d}{\sqrt{2}l_{CF}},\frac{2\zeta}{\sqrt{2}l_{CF}}\right)|\zeta_{CF}\rangle\notag\\
D_{\mathcal R}(-s)D_{\eta}(-s)|\zeta_{CF}\rangle&=|(\zeta-2s)_{CF}\rangle
\end{align}

Plugging these results into Eq.(\ref{eq:D_e_D_v},\ref{eq:B_definition}), one finds:
\begin{align}
B(z,\omega,\zeta)=\Theta\left(\frac{d}{\sqrt{2}l_{CF}},\frac{2s-2\zeta}{\sqrt{2}l_{CF}}\right)A(\zeta-2s).
\end{align}
After some basic manipulations and taking the complex conjugate, Eq.(\ref{eq:coherent_state_fusion}) is established.

\section{Continuous coherent state and the projective construction for Laughlin's states on the torus}\label{app:ccs}
The mean-field CF picture for the Laughlin's state corresponds to fully fill the CF LLL. On a finite torus, one needs to define the coherent state $|0_{\eta}\rangle$ in the cyclotron space of the CF. One natural definition for such a coherent state is to project the $\delta$-function at the origin $|\zeta_{CF}=0\rangle$ to the CF LLL, which has been termed as the continuous coherent state in Ref.\cite{fremling2014coherent}. We have numerically tested for a small number of electrons $N=3,4$, the projected CF wavefunction is identical to one of the $m$-fold degenerate Laughlin's states on the torus.

This choice of $|0_{\eta}\rangle$ naturally preserves the magnetic rotation symmetry. Namely, when the sample size and the boundary conditions are consistent with the magnetic rotation symmetry, the projected wavefunction obtained with this prescription is a rotational eigenstate.

\section{Extended Data}\label{app:extended_data}
\begin{figure}
    \centering
    \includegraphics[width=0.95\linewidth]{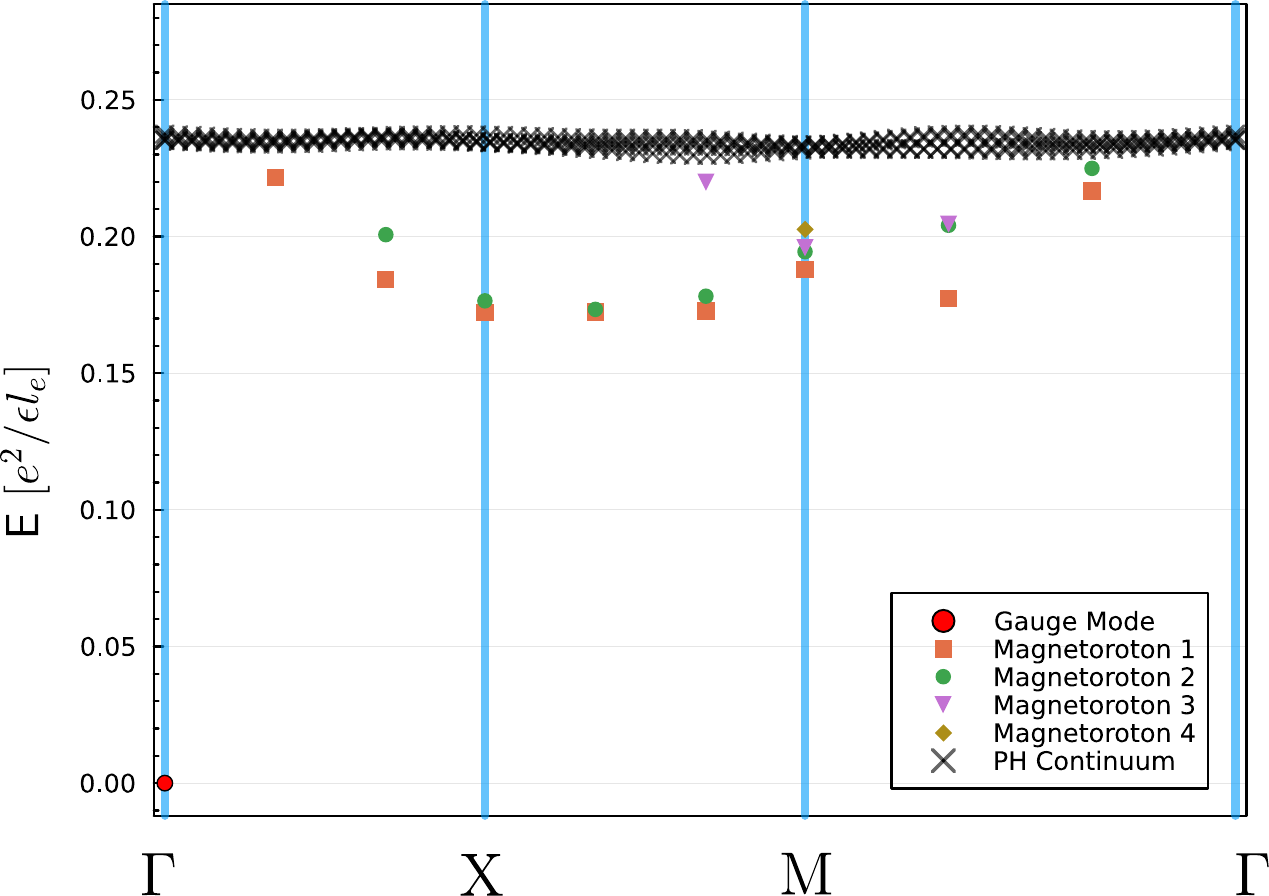}
    \caption{Magnetoroton bands for the $6\times6$ MLL model.}
    \label{fig: supplemental tdhf bands}
\end{figure}

In Fig.\ref{fig: supplemental tdhf bands}, we provide the magnetoroton bands raw data of the $6\times6$ sample for the MLL model, as a supplement to Fig.\ref{fig: magnetoroton spectrum} in the main text. Still, the points mixed with the particle-hole continuum are removed.

\bibliography{FCI_ProjWfc}
\bibliographystyle{apsrev} % apsrev is format for PRL of APS, comment this for appearance of long reference title
\end{document}